\documentclass[epj]{svjour}
\usepackage{graphicx}
\usepackage{color}

\def\mb#1{\setbox0=\hbox{$#1$}\kern-.025em\copy0\kern-\wd0
\kern-0.05em\copy0\kern-\wd0\kern-.025em\raise.0233em\box0}

\begin{document}
   \title{Nonlinear mean field Fokker-Planck equations. Application to the chemotaxis of biological populations }

 \author{P.H. Chavanis}

\institute{ Laboratoire de Physique Th\'eorique, Universit\'e Paul
Sabatier, 118 route de Narbonne 31062 Toulouse, France\\
\email{chavanis@irsamc.ups-tlse.fr}}

\titlerunning{Nonlinear mean field Fokker-Planck equations.}

   \date{To be included later }

   \abstract{We study a general class of nonlinear mean field
   Fokker-Planck equations in relation with an effective generalized
   thermodynamical (E.G.T.) formalism. We show that these equations describe
   several physical systems such as: chemotaxis of bacterial
   populations, Bose-Einstein condensation in the canonical ensemble,
   porous media, generalized Cahn-Hilliard equations, Kuramoto model,
   BMF model, Burgers equation, Smoluchowski-Poisson system for
   self-gravitating Brownian particles, Debye-H\"uckel theory of
   electrolytes, two-dimensional turbulence...  In particular, we show
   that nonlinear mean field Fokker-Planck equations can provide
   generalized Keller-Segel models for the chemotaxis of
   biological populations. As an example, we introduce a new model of
   chemotaxis incorporating both effects of anomalous diffusion and
   exclusion principle (volume filling). Therefore, the notion of
   generalized thermodynamics can have applications for concrete
   physical systems. We also consider nonlinear mean field
   Fokker-Planck equations in phase space and show the passage from
   the generalized Kramers equation to the generalized Smoluchowski
   equation in a strong friction limit. Our formalism is simple and
   illustrated by several explicit examples corresponding to Boltzmann,
   Tsallis, Fermi-Dirac and Bose-Einstein entropies among others.   }

   \maketitle
%

\section{Introduction}
\label{sec_introduction}

The study of Fokker-Planck equations 
(Fokker (1914) \cite{fokker},
Planck (1917) \cite{planck}) 
is of considerable interest in physics
since the pioneering work of Einstein (1905) \cite{einstein} on the
Brownian motion. The simplest Fokker-Planck equation is the
Smoluchowski (1915) \cite{smoluchowski} equation. This is a
drift-diffusion equation describing the diffusion of particles in
physical space in the presence of an external potential (e.g. the
sedimentation of colloids in a gravitational field). A more general
Fokker-Planck equation is the Kramers (1940) \cite{kramers} equation,
previously introduced by Klein (1921) \cite{klein}, which takes into
account inertial effects and describes the diffusion of particles in
phase space when they experience a friction force. The Smoluchowski
equation is recovered from the Kramers equation in a strong friction
limit where inertial effects are negligible. In this sense, the
Smoluchowski equation describes an overdamped evolution. These
Fokker-Planck equations \cite{risken} are consistent with usual
thermodynamics in the canonical ensemble. They monotonically decrease
the Boltzmann free energy ($H$-theorem) and relax towards the
equilibrium Boltzmann distribution.

Recently, several researchers have tried to extend the usual concepts
of thermodynamics and kinetic theory in order to describe complex
systems that are characterized by non-Boltzmannian distributions. In
that respect, some generalized forms of entropic functionals
\footnote{We recall that the Boltzmann entropy can be obtained from a {\it
combinatorial analysis} assuming that all the accessible microstates
are equiprobable. This is the basic postulate of statistical
mechanics. Non-standard entropies can be relevant for complex systems
where this postulate breaks down, i.e. when the accessible microstates
are {\it not} equiprobable. This happens when the system prefers some
regions of phase space better than others or when the particles are
subjected to exclusion/inclusion principles or fine-grained
constraints \cite{cras}.} have been introduced. One of the most
popular ``generalized entropy'' is the Tsallis (1988) \cite{tsallis}
entropy, but other entropies have been presented by Abe (1997)
\cite{abe}, Borges \& Roditi (1998)
\cite{br}, Kaniadakis (2001) \cite{k1}, Naudts (2004) \cite{naudts}, and 
Kaniadakis {\it et al.} (2005) \cite{k2}. It was later realized
that these entropic functionals are special cases of the one-parameter
family of entropies introduced earlier by Harvda \& Charvat (1967)
\cite{hc} or of the two-parameters family of entropies introduced by
Mittal (1975) \cite{mittal} and Sharma \& Taneja (1975)
\cite{st}. Other famous forms of entropies have been presented 
by Reyni (1970) \cite{reyni} and Sharma \& Mittal (1975) \cite{sm}. We
refer to Kaniadakis \& Lissia
\cite{kl} for a very interesting discussion of these historical
aspects, starting from the early works of Euler in 1779. 

Following these developments, some researchers have tried to develop
out-of-equilibrium theories associated to a generalized
thermodynamical framework. In particular, it has been first shown by
Plastino \& Plastino (1995) \cite{pp} that the Tsallis
$q$-distributions are the steady states of a nonlinear Fokker-Planck
equation taking into account anomalous diffusion. This type of
equations had been previously considered by mathematicians to describe
porous media \cite{spohn}. The seminal work of Plastino
\& Plastino \cite{pp} has been further developed by 
Tsallis \& Bukman (1996) \cite{bukman}, Stariolo (1997)
\cite{stariolo}, Borland (1998) \cite{borland} and Nobre {\it et al.}  
(2004) \cite{nobre} among others. On the other hand, Kaniadakis
\& Quarati (1994)
\cite{kq} have introduced nonlinear Fokker-Planck equations whose steady states
are the Fermi-Dirac
\footnote{A generalized Fokker-Planck equation leading to the
Fermi-Dirac statistics has also been introduced by Chavanis {\it et
al.} (1996) \cite{csr} in the context of the violent relaxation of
collisionless stellar systems described by the Vlasov equation.  This
is based on the Lynden-Bell's form of entropy (1968)
\cite{lb} which becomes similar to the Fermi-Dirac entropy in the
two-levels approximation of the theory.} and Bose-Einstein
statistics. These kinetic equations take into account an
exclusion (fermions) or inclusion (bosons) principle leading to
quantum-like statistics at equilibrium. The case of intermediate
statistics, interpolating between fermions and bosons, has also been
considered in
\cite{kq}. Recently, the bosonic Kramers equation has been studied in \cite{bose} and was shown to reproduce the
phenomenology of the Bose-Einstein condensation in the canonical
ensemble.

The above-mentioned nonlinear Fokker-Planck (NFP) equations are
associated with {\it special} forms of entropic functionals (Tsallis,
Fermi-Dirac, Bose-Einstein). More recently, Martinez {\it et al.} 
(1998) \cite{martinez},  Kaniadakis (2001)
\cite{k1}, Frank (2002) \cite{frank1} and Chavanis (2003) 
\cite{gen}  have studied generalized forms of NFP equations associated
with an almost {\it arbitrary} entropic functional. They can be viewed
as generalized Kramers and Smoluchowski equations where the
coefficients of diffusion, friction and drift explicitly depend on the
local density of particles. Physically, this can take into account
microscopic constraints (exclusion volume effects, steric hindrance,
non-extensive effects...) that modify the dynamics of the particles at
small scales and lead to non-standard equilibrium distributions
\footnote{Generalized Kramers and Smoluchowski equations describe {\it
dissipative} systems where the temperature is fixed instead of the
energy. They are therefore associated with the canonical ensemble. The
appropriate thermodynamical potential is the free energy $F=E-TS$
which decreases monotonically with time at fixed mass. Generalized
Boltzmann and Landau equations describing {\it conservative} systems
where the energy is fixed have been introduced by Kaniadakis (2001)
\cite{k1} and Chavanis (2004) \cite{gen2}. They are associated with
the microcanonical ensemble. The proper thermodynamical potential is
the entropy $S$ which increases monotonically with time at fixed mass
and energy.}. Martinez {\it et al.} \cite{martinez} determined the NFP
equation in order to recover, as a steady state, the equilibrium state
produced by minimizing a generalized form of free energy at fixed
mass. Kaniadakis \cite{k1} obtained the NFP equation from the Master
equation (see also Curado \& Nobre
\cite{cn}) by allowing the transition probabilities to depend on
the concentration of particles in the initial and arrival
states. Frank \cite{frank1} derived the NFP equation from a
generalized free energy functional by using the linear thermodynamics
of Onsager. Chavanis \cite{gen} obtained the NFP equation by using a
form of Maximum Entropy Production Principle (MEPP). This corresponds
to a variational version of the linear thermodynamics of Onsager. We
refer to the book of Frank
\cite{frank} for a first survey on nonlinear Fokker-Planck equations.

Another topic of active research in statistical physics concerns the
dynamics and thermodynamics of systems with long-range interactions
\cite{houches}. Several Hamiltonian systems with long-range interactions 
have been studied in the microcanonical ensemble such as
self-gravitating systems, two-dimensional vortices, the HMF model, the
free-electron laser etc... For such systems the mean field
approximation provides a very good description of the system and
becomes exact in a proper thermodynamic limit $N\rightarrow
+\infty$. The microcanonical ensemble is the correct description of
{\it isolated} systems evolving at fixed energy. On the other hand,
some authors have introduced a canonical version of these models so as
to treat systems that are {\it dissipative}. This leads to the notion
of Brownian systems with long-range interactions. Their dynamics is
described by mean field Fokker-Planck equations where the temperature
is fixed (instead of the energy).  These mean field Fokker-Planck
equations were introduced early by Kuramoto (1984) \cite{kuramoto} to
describe the synchronization of globally coupled nonlinear oscillators
and more recently by Marzel \& Aslangul (2001) \cite{ma}, Chavanis
(2006)
\cite{hb} and Frank \cite{frank} in a more general context. 
Some specific studies have been
made for self-gravitating Brownian particles \cite{crs,sc} and for the
BMF model \cite{cvb} which is the canonical version of the HMF model
\cite{antoni}.

In view of the importance of these two topics: generalized
thermodynamics and long-range interactions, we have introduced in
\cite{gen} a class of nonlinear mean field Fokker-Planck equations
(see Eq. (81) of \cite{gen}) that incorporate {\it both} a generalized
free energy functional and a long-range potential of interaction. As
an illustration, we studied with C. Sire a model of self-gravitating
Langevin particles \cite{lang} (see also \cite{shiino}) that combines
self-gravity (long-range interactions) and anomalous diffusion
(generalized thermodynamics) related to the Tsallis entropy and to the
polytropic equation of state. In other words, this model couples the
NFP equation introduced by Plastino \& Plastino \cite{pp} to the
gravitational Poisson equation. Unfortunately, this model of
self-gravitating Langevin particles has no clear application in
astrophysics (because self-gravitating systems are generally not
overdamped and not dissipative) and was introduced essentially as an
interesting dynamical model with rich mathematical
properties. However, it was noted in
\cite{lang} that this model could have applications in unexpected
area, like in the chemotaxis of bacterial populations...

The name chemotaxis refers to the motion of organisms induced by
chemical signals \cite{murray}. In some cases, the biological
organisms (bacteria, amoebae, endothelial cells, ants...) secrete a
substance (pheromone, smell, food, ...) that has an attractive effect
on the organisms themselves. Therefore, in addition to their diffusive
motion, they move preferentially along the gradient of concentration
of the chemical they secrete (chemotactic flux). When attraction
prevails over diffusion, the chemotaxis can trigger a
self-accelerating process until a point at which aggregation takes
place. This is the case for the slime mold {\it Dictyostelium
discoideum} and for the bacteria {\it Escherichia coli}. This is
referred to as chemotactic collapse. A model of slime mold aggregation
has been introduced by Patlak (1953) \cite{patlak} and Keller \& Segel
(1971) \cite{ks} in the form of two coupled differential equations.
The first equation is a drift-diffusion equation describing the
evolution of the concentration of bacteria and the second equation is
a diffusion equation with terms of source and degradation describing
the evolution of the concentration of the chemical. In the simplest
model, the diffusion coefficient $D$ and the mobility $\chi$ of the
bacteria are constant. This forms the standard Keller-Segel
model. However, the original Keller-Segel model allows these
coefficients to depend on the concentration of the bacteria and of the
chemical. The case where these coefficients depend on the
concentration of the chemical $c({\bf r},t)$, but not on the
concentration of bacteria $\rho({\bf r},t)$, has been considered by
Othmer \& Stevens \cite{os}. This leads to ordinary mean field
Fokker-Planck equations (with respect to $\rho({\bf r},t)$) with space
and time dependent coefficients. On the other hand, if we assume that
the diffusion coefficient and the mobility of the bacteria depend on
their concentration $\rho({\bf r},t)$, but not on the concentration
$c({\bf r},t)$ of the secreted chemical, the original Keller-Segel
model takes the form of a generalized mean field Fokker-Planck
equation. {\it Therefore, the Keller-Segel model represents a
fundamental example of nonlinear mean field Fokker-Planck equation
with physical applications in biology, thereby justifying a notion of
generalized thermodynamics.} The analogy between the Keller-Segel
model and NFP equations was first pointed out in
\cite{gen} and further developed in subsequent papers (see, e.g.,
\cite{csbio}). This analogy makes possible to interprete results of
chemotaxis in terms of nonlinear Fokker-Planck equations and
generalized thermodynamics, which has not been considered so far by
applied mathematicians working on this domain \cite{horstmann}. It
thus allows to make a bridge between two different communities. In
this analogy, the model of self-gravitating Langevin particles
introduced by Chavanis \& Sire \cite{lang} also provides a generalized
Keller-Segel model of chemotaxis taking into account anomalous
diffusion (the application of this model to chemotaxis has been
emphasized in \cite{csmass}). More generally, we can use the numerous
results accumulated in the context of generalized thermodynamics to
propose new forms of generalized Keller-Segel models with potential
applications in biology.

The aim of this paper is to develop a simple and rich formalism that
allows to deal with nonlinear mean field Fokker-Planck equations. We
shall illustrate this formalism on several examples and show the
inter-connections between different topics. The paper is organized as
follows. In Sec. \ref{sec_phys}, we consider NFP equations in physical
space. This corresponds to overdamped models where inertial effects
are neglected. We review and complete the basic properties of these
equations in relation with an effective generalized thermodynamical
(E.G.T.) formalism. In Sec. \ref{sec_gle}, we show that they can be
obtained from generalized Langevin equations. In Sec. \ref{sec_h}, we
show that they admit an $H$-theorem (in the canonical ensemble) for a
generalized free energy. We stress that the Legendre structure of the
free energy and the Einstein relation are preserved in this
generalized thermodynamical framework. In Sec. \ref{sec_sta}, we
determine the steady states of these nonlinear mean field
Fokker-Planck equations and show that they are solutions of an
integrodifferential equation. In Sec. \ref{sec_min} (and in Appendix
\ref{sec_conn}), we show that a steady state of a nonlinear mean field
Fokker-Planck equation is linearly dynamically stable if and only if
(iff) it is a (local) minimum of the free energy at fixed mass. In
Sec. \ref{sec_gs}, we show that a NFP equation in physical space with
a constant mobility and a density-dependent diffusion coefficient can be
written in the form of a generalized Smoluchowski equation
incorporating a barotropic equation of state.  In Secs. \ref{sec_phen}
and \ref{sec_kin}, we show the correspondance between the
phenomenological derivations of the NFP equations given by Frank
\cite{frank1} and Chavanis \cite{gen} and the kinetic derivation given
by Kaniadakis \cite{k1}.  In Sec. \ref{sec_ex}, we present several
explicit examples of NFP equations and mention their potential
applications to the problem of chemotaxis (see Sec. \ref{sec_ks}). In
particular, we introduce a new model of chemotaxis that incorporates
both effects of anomalous diffusion and exclusion principle (volume
filling).  The corresponding generalized entropy is expressed in the
form of integrals that can be explicited in particular cases. In
Sec. \ref{sec_phase}, we consider NFP equations in phase space taking
into account inertial effects. In Secs. \ref{sec_gk}-\ref{sec_eos},
we list their main properties. In Sec. \ref{sec_strong}, we consider
the strong friction limit and derive the generalized Smoluchowski
equation from the generalized Kramers equation. We use a method of
moments that is simpler than the Chapman-Enskog expansion presented in
\cite{lemou}. In Sec. \ref{sec_ee}, we consider explicit examples
corresponding to the Boltzmann, Tsallis and Fermi-Dirac entropies. The
Appendices contain important results that complete the basic
properties of the NFP equations discussed in the text. In Appendix
\ref{sec_iso}, we show that a generalized isotropic BGK operator has
properties similar to those possessed by a nonlinear Kramers operator.
In Appendix \ref{sec_conn}, we establish a simple relation showing the
equivalence between linear dynamical stability (exponential damping of
the perturbation) and generalized thermodynamical stability (minimum
of free energy at fixed mass). In Appendix
\ref{sec_bif}, we study the stability of a spatially homogeneous
solution of the nonlinear mean field Fokker-Planck equation and
evidence a critical point. In Appendices
\ref{sec_passage} and \ref{sec_equi}, we show that a distribution
function $f$ in phase space is a minimum of the free energy $F[f]$ at
fixed mass iff the corresponding distribution $\rho$ in physical space
is a minimum of the corresponding free energy $F[\rho]$ at fixed
mass. This implies that a distribution function $f$ in phase space is
a stable steady state of the generalized Kramers equation iff the
corresponding distribution $\rho$ in physical space is a stable steady
state of the corresponding generalized Smoluchowski equation.  In
Appendix \ref{sec_eddington}, we extend to $d$ dimensions the Eddington
formula that allows to obtain the distribution function
$f=f(\epsilon)$ in phase space from the knowledge of the barotropic
equation of state $p=p(\rho)$ in physical space. In Appendix
\ref{sec_ht}, we derive the $H$-theorems associated with the NFP equations
and in Appendix \ref{sec_pol} we derive the polytropic equation of
state associated with the Tsallis statistics.

\section{Nonlinear mean field Fokker-Planck equations in physical space}
\label{sec_phys}

We first describe nonlinear mean field Fokker-Planck equations in
physical space where the inertia of the particles is neglected. They
can be viewed as models describing an overdamped dynamics.

\subsection{Generalized Langevin equations}
\label{sec_gle}

We consider a system of $N$ particles whose individual dynamics is
described by the stochastic Ito-Langevin equations
\begin{eqnarray}
\label{gle1} \frac{d{\bf
r}_i}{dt}=-\chi(\rho_i)\nabla\Phi_i+\sqrt{2D(\rho_i)}{\bf R}_i(t),
\end{eqnarray}
where ${\bf R}_i(t)$ is a white noise satisfying $\langle {\bf
R}_i(t)\rangle={\bf 0}$ and $\langle
R_{i,\alpha}(t)R_{j,\beta}(t')\rangle
=\delta_{ij}\delta_{\alpha,\beta}\delta(t-t')$ where $i=1,...,N$ label
the particles and $\alpha=1,...,d$ label the coordinates of space. We have noted $\rho_i=\rho({\bf r}_{i}(t),t)$ and $\Phi_i=\Phi({\bf r}_{i}(t),t)$. In
ordinary models, the mobility $\chi$ and the diffusion coefficient $D$
are constant. In that case, the statistical equilibrium state is the
Boltzmann distribution $\rho\sim e^{-\Phi/T}$ where the temperature
$T=1/\beta$ is given by the Einstein relation $T=D/\chi$. In the
present study, we shall consider more general situations and allow the
mobility $\chi(\rho)$ and the diffusion coefficient $D(\rho)$ to
depend on the local concentration of particles $\rho({\bf
r},t)=\langle \sum_{i=1}^{N}\delta({\bf r}-{\bf
r}_{i}(t))\rangle$. This is an heuristic approach to take into account
microscopic constraints that affect the dynamics of particles at small
scales and lead to non-Boltzmannian equilibrium distributions. Indeed,
it is not surprising that the mobility or the diffusive properties of
a particle depend on its environment. For example, in a dense medium,
its motion can be hampered by the presence of the other particles so
that its mobility is reduced.

On the other hand, in ordinary models, the particles move in a fixed
external potential $\Phi_{ext}({\bf r})$. In the present study, we
want to be more general and take into account the possibility that the
potential $\Phi({\bf r},t)$ is created self-consistently by the
particles themselves. In this paper, we shall neglect statistical
correlations and use a mean field description (for more general models
taking into account statistical correlations see, e.g.,
\cite{hb,new}). Therefore, we assume that the potential is given by a
relation of the form
\begin{eqnarray}
\label{gle2} \Phi({\bf r},t)=\int \rho({\bf r}',t)u(|{\bf r}-{\bf
r}'|)d{\bf r}',
\end{eqnarray}
where $u(|{\bf r}-{\bf r}'|)$ is a binary potential of interaction and
$\rho({\bf r},t)$ is the smooth distribution of particles. In general,
the mean field approximation gives a very good description of systems
with weak long-range binary interactions and it becomes exact in a proper 
thermodynamic limit $N\rightarrow +\infty$ \cite{hb}. In
Eq. (\ref{gle2}), the potential is expressed as a convolution product:
$\Phi=u* \rho$. Of course, the potential can be due to the combined
effect of the self-interaction and an external field, in which case
$\Phi=\Phi_{ext}+u*\rho$.  We shall also consider the case where it is
determined by an equation of the form
\begin{eqnarray}
\label{gle3}
\epsilon\frac{\partial \Phi}{\partial t}=\Delta \Phi-k^{2}\Phi-\lambda\rho,
\end{eqnarray}
where $\epsilon$ and $\lambda$ are positive constants. For
$\epsilon=0$, Eq. (\ref{gle3}) becomes the screened Poisson equation
\begin{eqnarray}
\label{gle4}
\Delta \Phi-k^{2}\Phi=\lambda\rho. 
\end{eqnarray}
Therefore, we can identify $k^{-1}$ as the screening length. If we
assume furthermore that $k=0$, we get the Poisson equation
\begin{eqnarray}
\label{gle5}
\Delta \Phi=\lambda\rho. 
\end{eqnarray}
These last two equations can be put in the form of
Eq. (\ref{gle2}). Note also that, in the stationary state,
Eq. (\ref{gle3}) reduces to Eq. (\ref{gle4}).

\subsection{Drift-diffusion equations}
\label{sec_dde}

For the stochastic process (\ref{gle1}), the evolution of the smooth
density of particles $\rho({\bf r},t)$ is governed by the nonlinear
mean field Fokker-Planck equation \cite{gen,hb}:
\begin{eqnarray}
\label{dde1}
\frac{\partial\rho}{\partial t}=\nabla\cdot \left\lbrack \nabla(D(\rho)\rho)+\chi(\rho)\rho\nabla\Phi\right\rbrack,
\end{eqnarray}
coupled to Eq. (\ref{gle2}) or (\ref{gle3}). Let us introduce the
notations
\begin{eqnarray}
\label{dde2} Dh(\rho)=\frac{d}{d\rho}(\rho D(\rho)), \qquad \chi
g(\rho)=\rho\chi(\rho),
\end{eqnarray}
where $D$ and $\chi$ are positive constants and $h(\rho)$ and
$g(\rho)$ are positive functions. These notations are chosen such that the
usual stochastic equations with constant diffusion $D(\rho)=D$ and
constant mobility $\chi(\rho)=\chi$ are recovered for $h(\rho)=1$ and
$g(\rho)=\rho$.  With these notations, the nonlinear Fokker-Planck
equation (\ref{dde1}) can be rewritten
\begin{eqnarray}
\label{dde3} \frac{\partial\rho}{\partial t}=\nabla\cdot \left ( D
h(\rho)\nabla\rho+\chi
 g(\rho)\nabla\Phi\right ).
\end{eqnarray}
It can be put  in the conservative form
\begin{eqnarray}
\label{dde4}
\frac{\partial\rho}{\partial t}=-\nabla\cdot {\bf J},
\end{eqnarray}
where
\begin{eqnarray}
\label{dde5}
{\bf J}=-\left (Dh(\rho)\nabla\rho+\chi g(\rho)\nabla\Phi\right ),
\end{eqnarray}
is a diffusion current. This structure guarantees the conservation of
mass $M=\int \rho d{\bf r}$ provided that the normal component of the
current at the boundary vanishes. 

\subsection{Relation to some known models}
\label{sec_known}

Known models can be recovered as particular cases of
Eq. (\ref{dde1}). When $\Phi_{ext}$ is an external potential and when
$D(\rho)=D$ and $\chi(\rho)=\chi$ are constant, we recover the
Smoluchowski equation $\partial_{t}\rho=\nabla\cdot
(D\nabla\rho+\chi\rho\nabla\Phi_{ext})$  describing the Brownian
motion of colloidal suspensions in a fixed gravitational field
\cite{smoluchowski}. When $\rho D(\rho)=K\rho^{\gamma}$ and $\Phi_{ext}=0$,
we recover the porous medium equation $\partial_{t}\rho=K\Delta
\rho^{\gamma}$ \cite{spohn}, and when $\Phi_{ext}\neq 0$, we recover the nonlinear Fokker-Planck equation $\partial_{t}\rho=\nabla\cdot
(K\nabla\rho^{\gamma}+\chi\rho\nabla\Phi_{ext})$ introduced by
Plastino \& Plastino \cite{pp} in connection to the Tsallis statistics
\cite{tsallis}.  When $D(\rho)=D$ and $\chi(\rho)=\chi$ are constant
and $u=-\frac{k}{2\pi}\cos(\theta-\theta')$ we obtain the Brownian
Mean Field (BMF) model \cite{cvb}, which is the canonical (fixed $T$)
version of the microcanonical (fixed $E$) Hamiltonian Mean Field (HMF)
model \cite{antoni}. This is closely related to the Kuramoto model
\cite{kuramoto} which describes the synchronization of globally
coupled nonlinear oscillators. When $D(\rho)=D$ and $\chi(\rho)=\chi$
are constant and $\Phi$ is the gravitational potential satisfying
$\Delta \Phi=4\pi G\rho$, Eqs. (\ref{dde1})-(\ref{gle5}) become the
Smoluchowski-Poisson system describing a model of self-gravitating
Brownian particles studied by Chavanis \& Sire
\cite{crs,sc,post,tcoll,virial1,virial2}. When $D(\rho)=D$ and $\chi(\rho)=\chi$ are
constant and $\Phi$ is the self-consistent electrostatic potential, we
recover the equations introduced by Debye
\& H\"uckel \cite{dh} in their model of electrolytes. 
Models of the form (\ref{dde1})-(\ref{gle4}) including a time
dependent temperature $\beta(t)$ assuring the conservation of energy
have been introduced by Robert \& Sommeria \cite{rs} and Chavanis
\cite{physicaD} to describe the violent relaxation of two-dimensional
vortices in geophysical and astrophysical flows. In the Quasi
Geostrophic (QG) model, $k^{-1}$ represents the Rossby length. Similar
equations have been proposed in
\cite{csr} to model the violent relaxation of stellar systems.  On the
other hand, for short range interactions, we can expand the potential
in the form $\Phi({\bf r},t)=a\rho({\bf
r},t)+\frac{b}{2}\Delta\rho({\bf r},t)$ and Eq. (\ref{dde1}) leads to
a generalized form of the Cahn-Hilliard equation (see
\cite{lemou,csbio,new} for details). As a particular case, for $D=0$ and
$u=a \delta$ (leading to $\Phi({\bf r},t)=a\rho({\bf r},t)$), we get
$\partial_{t}\rho=\chi a\nabla (g(\rho)\nabla\rho)$ and for
$g(\rho)=\rho$, we get the porous medium equation
$\partial_{t}\rho=\frac{1}{2}\chi a\Delta\rho^{2}$. Finally, in $d=1$,
assuming $D(\rho)=D$, $\chi(\rho)=\chi$ and
$u'=-\frac{1}{2\chi}\delta$, we get the Burgers equation
$\partial_{t}\rho+\rho\partial_{x}\rho=D\partial_{xx}\rho$
\cite{burgers}. Therefore, the class of nonlinear mean field Fokker-Planck equations (\ref{dde1})-(\ref{gle2}) introduced in \cite{gen} can
find physical applications in different areas \cite{cras}.

\subsection{Generalized Keller-Segel model of chemotaxis}
\label{sec_ks}

In addition to the previous examples, nonlinear mean field
Fokker-Planck equations can find important applications in the context
of chemotaxis \cite{murray}. The original Keller-Segel model \cite{ks}
describing the chemotaxis of bacterial populations consists in two
coupled differential equations
\begin{eqnarray}
\label{ks1}
\frac{\partial\rho}{\partial t}=\nabla\cdot \left (D_{2}\nabla\rho)-\nabla\cdot (D_{1}\nabla c\right ),
\end{eqnarray}
\begin{eqnarray}
\label{ks2}
\epsilon\frac{\partial c}{\partial t}=D_{c}\Delta c-k(c)c+f(c)\rho, 
\end{eqnarray}
that govern the evolution of the density of bacteria $\rho({\bf r},t)$
and the evolution of the secreted chemical $c({\bf r},t)$. The
bacteria diffuse with a diffusion coefficient $D_{2}$ and they also
move in a direction of a positive gradient of the chemical
(chemotactic drift). The coefficient $D_{1}$ is a measure of the
strength of the influence of the chemical gradient on the flow of
bacteria. On the other hand, the chemical is produced by the bacteria
with a rate $f(c)$ and is degraded with a rate $k(c)$. It also
diffuses with a diffusion coefficient $D_{c}$. In the general
Keller-Segel model, $D_{1}=D_{1}(\rho,c)$ and $D_{2}=D_{2}(\rho,c)$
can both depend on the concentration of the bacteria and of the
chemical. This takes into account microscopic constraints, like
close-packing effects, that can hinder the movement of bacteria.
If we assume that the quantities only depend on the concentration of bacteria
\footnote{In this paper, we shall restrict ourselves to this situation. The original Keller-Segel model (\ref{ks1})-(\ref{ks2}) where  $D_{1}$ and $D_{2}$ depend on both  $\rho({\bf r},t)$ and $c({\bf r},t)$ does not seem to possess the nice ``thermodynamical'' properties of the reduced Keller-Segel (\ref{ks3})-(\ref{ks4}) such as Legendre structure of the free energy functionals, canonical $H$-theorem, Einstein relation etc.} and write $D_{2}=Dh(\rho)$, $D_{1}=\chi g(\rho)$, $k(c)=k^{2}$, $f(c)=\lambda$ and $D_{c}=1$, we obtain 
\begin{eqnarray}
\label{ks3} \frac{\partial\rho}{\partial t}=\nabla\cdot \left ( D
h(\rho)\nabla\rho-\chi g(\rho)\nabla c\right ),
\end{eqnarray}
\begin{eqnarray}
\label{ks4}
\epsilon\frac{\partial c}{\partial t}=\Delta c-k^{2}c+\lambda\rho. 
\end{eqnarray}
These equations are isomorphic to the nonlinear mean field
Fokker-Planck equations (\ref{dde3})-(\ref{gle3}) provided that we make the
correspondance $\Phi({\bf r},t)=-c({\bf r},t)$: the potential of
interaction is played by the concentration of the secreted chemical.

It is important to note that the Keller-Segel model is a {\it mean field}
model. If we come back to the exact microscopic equations
\begin{eqnarray}
\label{c1}
\frac{d{\bf r}_{i}}{dt}=\chi\nabla c_{d}({\bf r}_{i}(t),t)+\sqrt{2D}{\bf R}_{i}(t),
\end{eqnarray}
\begin{eqnarray}
\label{c2}
\epsilon\frac{\partial c_{d}}{\partial t}=\Delta c_{d}-k^2 c_{d}+\lambda \sum_{i=1}^{N}\delta({\bf r}-{\bf r}_{i}(t)),
\end{eqnarray}
governing the evolution of each particle (bacteria, cells, social
insects,...), the mean field approximation leading to the standard Keller-Segel
model (\ref{ks3})-(\ref{ks4}) with $h(\rho)=1$ and $g(\rho)=\rho$
amounts to neglecting statistical correlations; see
\cite{stevens,ng,crrs,hb,virial2,kinbio,new} for details. Let us first
assume that the system is Markovian and possesses no intrinsic memory
in the sense that $c_{d}({\bf r},t)=-\int u(|{\bf r}-{\bf
r}'|)\rho_{d}({\bf r}',t)d{\bf r}'$ where $\rho_{d}({\bf r},t)=
\sum_{i=1}^{N}\delta({\bf r}-{\bf r}_{i}(t))$ is the exact distribution
of particles.  When the particles interact via a weak long-range
binary potential $u(|{\bf r}-{\bf r}'|)$, it can be shown that the
mean field approximation becomes exact in a proper thermodynamic limit
$N\rightarrow +\infty$
\cite{hb}.  For example, in Eq. (\ref{c2}) these assumptions correspond
to $\epsilon=0$ and $k\ll 1$. By contrast, when memory is not zero and
the interaction is short range, corresponding to $\epsilon\neq 0$ and
$k\gg 1$ in Eq. (\ref{c2}), the mean field approximation can lead to
wrong results. In the context of chemotaxis, the differences between
mean field and non mean field models have been discussed by Grima
\cite{grima} who showed the failure of the mean field approximation
for the prediction of the width of the aggregate sizes (the
disagreement becomes very severe close to the critical point where we
know that mean field approximation breaks down in general).  On the
other hand, the mean field approximation assumes that the number of
particles $N\rightarrow +\infty$. In stellar systems and plasmas, for
example, this is always the case. However, for biological systems, the
number of interacting bacteria or cells is frequently less than a few
thousands so it may be relevant to return to a microscopic description
of the bacteria or cells' movement in terms of $N$-body stochastic
equations like (\ref{c1})-(\ref{c2}) as discussed in
\cite{kinbio}. In this paper, however, we shall exclusively consider
mean field models. As we have seen, mean field approximation works
well for Markovian systems ($\epsilon=0$) with weak long-range
interactions if (i) $N\rightarrow +\infty$ and (ii) if we are not too
close from a critical point.  Mean field approximation breaks down:
(i) for non-Markovian systems ($\epsilon\neq 0$) (ii) close to a critical point (iii) for
small values of $N$.

\subsection{Generalized free energy and H-theorem}
\label{sec_h}

When $\Phi_{ext}$ is an external potential, we define the energy by
\begin{eqnarray}
\label{h1}
E=\int \rho \Phi_{ext} \, d{\bf r}.
\end{eqnarray} 
When $\Phi$ is determined by Eq. (\ref{gle2}), the self-interaction energy 
is given by
\begin{eqnarray}
\label{h2}
E=\frac{1}{2}\int \rho \Phi \, d{\bf r}.
\end{eqnarray} 
Finally, when  $\Phi$ is determined by Eq. (\ref{gle3}), the energy is 
\begin{eqnarray}
\label{h3}
E=\frac{1}{2\lambda}\int \left\lbrack (\nabla \Phi)^{2}+k^{2} \Phi^{2}\right \rbrack
\, d{\bf r}+\int \rho \Phi \, d{\bf r}.
\end{eqnarray} 
For $\epsilon=0$, the expression (\ref{h3}) reduces to Eq. (\ref{h2}).
On the other hand, we define the temperature by
\begin{eqnarray}
\label{h4}
T=\frac{D}{\chi}.
\end{eqnarray} 
Therefore, the Einstein relation is preserved in the generalized
thermodynamical framework. We also set $\beta=1/T$. We introduce the
generalized entropic functional
\begin{eqnarray}
\label{h5}
S=-\int C(\rho)\, d{\bf r},
\end{eqnarray}
where $C(\rho)$ is a convex function ($C''\ge 0$) defined by
\begin{eqnarray}
\label{h6}
C''(\rho)=\frac{h(\rho)}{g(\rho)}.
\end{eqnarray}
Note that the relation (\ref{h6}) defines the function $C(\rho)$ up to a term
of the form $A\rho+B$. After integration over the domain, the first term
is proportional to the mass which is a conserved quantity and the
second term is just a constant. Therefore, these terms play no physical 
role. However, we can adapt the values of the constants $A$ and $B$ in
order to obtain convenient expressions of the entropy.

Since the system is dissipative, the energy is not conserved. What is
fixed instead of the energy is the temperature defined by the Einstein
relation (\ref{h4}). Since $D\propto T$, the temperature measures the
strength of the stochastic force in Eq. (\ref{gle1}). This corresponds to a
canonical description where the system is in contact with a heat
bath. Note that the heat bath is completely disconnected from the
long-range potential of interaction; it corresponds to short-range
interactions modelled by the stochastic term in Eq. (\ref{gle1}). For an
isolated system described by the microcanonical ensemble the proper
thermodynamical potential is the entropy. Alternatively, for a
dissipative system described by the canonical ensemble, the relevant
thermodynamical potential is the free energy. We introduce the
generalized free energy
\begin{eqnarray}
\label{h7}
F=E-TS.
\end{eqnarray}
The definition of the free energy (Legendre transform) is preserved in
the generalized thermodynamical framework.

When the energy is given by Eqs. (\ref{h1}) or (\ref{h2}), a
straightforward calculation (see Appendix \ref{sec_ht}) shows that
\begin{eqnarray}
\label{h8}
\dot F=-\int \frac{1}{\chi g(\rho)}(Dh(\rho)\nabla\rho+\chi g(\rho)\nabla \Phi)^{2}d{\bf r}.
\end{eqnarray}
When the energy is given by Eq. (\ref{h3}), we obtain
\begin{eqnarray}
\label{h9}
\dot F=-\frac{1}{\lambda\epsilon}\int (\Delta \Phi-k^{2}\Phi-\lambda\rho)^{2} d{\bf r}\nonumber\\
-\int \frac{1}{\chi g(\rho)}(Dh(\rho)\nabla\rho+\chi g(\rho)\nabla \Phi)^{2}d{\bf r}.
\end{eqnarray}
Therefore, $\dot F\le 0$. This forms an $H$ theorem in the canonical ensemble. It is sometimes useful to introduce the Massieu function 
\begin{eqnarray}
\label{h10}
J=S-\beta E,
\end{eqnarray}
which is related to the free energy by $J=-\beta F$. Clearly, we have
$\dot J\ge 0$. We can now consider particular cases: if $D=0$ (leading
to $T=0$), we get $F=E$ so that $\dot E\le 0$. If $\chi=0$ (leading to
$\beta=0$), we have $J=S$ so that $\dot S\ge 0$.

\subsection{Stationary solution}
\label{sec_sta}

The steady state of Eq. (\ref{dde3}) satisfies $\dot F=0$ leading to
${\bf J}={\bf 0}$ or explicitly
\begin{eqnarray}
\label{sta1}
Dh(\rho)\nabla\rho+\chi g(\rho)\nabla \Phi={\bf 0}.
\end{eqnarray}
Using Eqs. (\ref{h4}) and (\ref{h6}), we get
\begin{eqnarray}
\label{sta2}
C''(\rho)\nabla\rho+\beta\nabla \Phi={\bf 0},
\end{eqnarray}
which can be integrated into
\begin{eqnarray}
\label{sta3}
C'(\rho)=-\beta \Phi-\alpha,
\end{eqnarray}
where $\alpha$ is a constant of integration.  Since $C$ is convex, this
equation can be reversed to give
\begin{eqnarray}
\label{sta4}
\rho({\bf r})=F(\beta \Phi({\bf r})+\alpha),
\end{eqnarray}
where $F(x)=(C')^{-1}(-x)$ is a monotonically decreasing
function. Thus, in the steady state, the density is a monotonically
decreasing function $\rho=\rho(\Phi)$ of the potential.  We have the identity
\begin{eqnarray}
\label{sta5}
\rho'(\Phi)=-\frac{\beta}{C''(\rho)}.
\end{eqnarray}
Substituting Eq. (\ref{gle2}) in Eq. (\ref{sta3}), we find that the
density profile is determined by an integro-differential equation of
the form
\begin{eqnarray}
\label{sta6}
C'(\rho)=-\beta\int \rho({\bf r}')u(|{\bf r}-{\bf
r}'|)d{\bf r}' -\alpha.
\end{eqnarray}
More specifically, when the potential is given by Eq. (\ref{gle4}), we
obtain a mean field equation of the form
\begin{eqnarray}
\label{sta7}
\Delta \Phi-k^{2}\Phi=\lambda F(\beta \Phi+\alpha).
\end{eqnarray}
The constant of integration $\alpha$ is determined by the total mass
$M$ (which is a conserved quantity). Finally, we note that the
generalized entropy (\ref{h5}) is related to the distribution
(\ref{sta4}) by  \cite{super}:
\begin{eqnarray}
\label{sta7b}
C(\rho)=-\int^{\rho}F^{-1}(x)dx.
\end{eqnarray}
Equation (\ref{sta4}) determines the distribution from the entropy and
Eq. (\ref{sta7b}) determines the entropy from the distribution.

\subsection{Minimum of free energy}
\label{sec_min}

The critical points of free energy at fixed mass are determined by the variational problem
\begin{eqnarray}
\label{min1}
\delta F+T\alpha\delta M=0,
\end{eqnarray}
where $\alpha$ is a Lagrange multiplier. We first consider the case
where $\Phi$ is given by Eq. (\ref{gle2}). Therefore, the free energy that we consider is
\begin{eqnarray}
\label{min6} F[\rho]={1\over 2}\int \rho\Phi d{\bf  r}+T\int C(\rho) d{\bf r}.
\end{eqnarray} 
After straightforward
calculations, we find that Eq. (\ref{min1}) leads to
\begin{eqnarray}
\label{min2}
C'(\rho)=-\beta \Phi-\alpha.
\end{eqnarray}
Therefore, comparing with Eq. (\ref{sta3}), we find that a stationary
solution of Eq. (\ref{dde3}) is a critical point of $F$ at fixed
mass. Furthermore, it is shown in Appendix \ref{sec_conn} that a
steady state of Eq. (\ref{dde3}) is linearly dynamically stable iff it is a
{\it minimum} (at least local) of $F$ at fixed mass.  This property also 
results from Lyapunov's direct method \cite{frank}. Indeed, we have established that
\begin{eqnarray}
\label{min2add}
\dot F\le 0, \qquad \dot F=0 \leftrightarrow \partial_{t}\rho=0.
\end{eqnarray}
This implies that $\rho({\bf r})$ is linearly dynamically stable iff
it is a (local) minimum of $F$ at fixed mass (maxima or saddle points
of $F$ are dynamically unstable). In this sense, dynamical and
generalized thermodynamical stability in the canonical ensemble
coincide. Furthermore, if $F$ is bounded from below
\footnote{There are important cases, like the system of 
self-gravitating Brownian particles, where the free energy is not
bounded from below. In that case, the system can either relax towards
a local minimum of $F$ at fixed mass (when it exists) or collapse to a
Dirac peak \cite{post}, leading to a divergence of the free energy
$F(t)\rightarrow -\infty$.}, we can conclude from Lyapunov's theory
that the system will converge to a stable steady state for
$t\rightarrow +\infty$ which is a (local) minimum of $F[\rho]$ at fixed
mass.  If several local minima exist, the choice of the final steady
state will depend on a complicated notion of {\it basin of
attraction}.

In conclusion, a steady solution of the nonlinear mean field Fokker-Planck
equation (\ref{dde3}) is stable iff it satisfies the minimization
problem:
\begin{eqnarray}
\label{min3}
\min_{\rho}\quad \lbrace F[\rho]\quad |\quad M[\rho]=M\rbrace.
\end{eqnarray}
Taking the second variations of $F$ and using Eq. (\ref{sta5}),  the
condition of stability can be written
\begin{eqnarray}
\label{min4} \delta^{2}F[\delta\rho]=-{1\over 2}\biggl\lbrace \int {(\delta
\rho)^{2}\over \rho'(\Phi)}d{\bf  r}-\int \delta\rho\delta\Phi
d{\bf  r}\biggr\rbrace \ge 0,
\end{eqnarray}
for all perturbations $\delta\rho$ that conserve mass. If
$\Phi_{ext}({\bf r})$ is an external potential, the second term in
Eq. (\ref{min4}) vanishes. Therefore, the second variations of the
free energy are always positive $\delta^{2}F=-T\delta^{2}S=(1/2)T\int
C''(\rho){(\delta \rho)^{2}}\ge 0$ so that a critical point of $F$ at
fixed mass is necessarily a minimum. If the potential is given by
Eq. (\ref{gle3}), the free energy is a functional $F[\rho,\Phi]$ of
the density $\rho$ and potential $\Phi$. The cancellation of the first
order variations of $F$ with respect to $\delta\rho$ and $\delta\Phi$
yields Eqs. (\ref{min2}) and (\ref{gle4}), respectively. The condition
of stability can be written
\begin{eqnarray}
\label{min5} \delta^{2}F[\delta\rho,\delta\Phi]=-{1\over 2}\int {(\delta
\rho)^{2}\over \rho'(\Phi)}d{\bf  r}+\int \delta\rho\delta\Phi
d{\bf  r}\nonumber\\
-\frac{1}{2\lambda}\int (\Delta\delta\Phi-k^{2}\delta\Phi)\delta\Phi d{\bf r} \ge 0,
\end{eqnarray}
for all perturbations $\delta\rho$ and $\delta\Phi$ that conserve
mass. From now on, we shall only consider the case where the potential
is given by Eq. (\ref{gle2}) since the case of Eq. (\ref{gle3}) can be treated similarly
by following the lines sketched above. 

\subsection{Particular cases}
\label{sec_pc}

If we take $h(\rho)=1$ and  $g(\rho)=1/C''(\rho)$, we get
\begin{eqnarray}
\label{pc1}
\frac{\partial\rho}{\partial t}=\nabla\cdot \left ( D \nabla\rho+
\frac{\chi}{C''(\rho)}  \nabla \Phi\right ).
\end{eqnarray}
In that case, we have a constant diffusion $D(\rho)=D$ and a variable
mobility $\chi(\rho)=\chi/(\rho C''(\rho))$. If we take
$g(\rho)=\rho$ and $h(\rho)=\rho C''(\rho)$, we get
\begin{eqnarray}
\label{pc2}
\frac{\partial\rho}{\partial t}=\nabla\cdot \left ( D\rho C''(\rho) \nabla\rho+{\chi}\rho  \nabla \Phi\right ).
\end{eqnarray}
In that case, we have a constant mobility $\chi(\rho)=\chi$ and a
variable diffusion $D(\rho)=D\rho \lbrack C(\rho)/\rho\rbrack'$. Note
that the condition $D(\rho)\ge 0$ requires that $\lbrack
C(\rho)/\rho\rbrack'\ge 0$. This gives a constraint on the possible
forms of $C(\rho)$.

\subsection{Generalized Smoluchowski equation}
\label{sec_gs}

The NFP equation (\ref{pc2}) can be put in the form of a generalized
Smoluchowski (GS) equation
\begin{eqnarray}
\label{gs1}
\frac{\partial\rho}{\partial t}=\nabla\cdot \left \lbrack  \chi (\nabla p+
\rho  \nabla \Phi )\right \rbrack,
\end{eqnarray}
with a barotropic equation of state $p(\rho)$ given by
\begin{eqnarray}
\label{gs2}
p'(\rho)=T\rho C''(\rho).
\end{eqnarray}
Since $C$ is convex, we have $p'(\rho)\ge 0$. 
On the other hand, integrating Eq. (\ref{gs2}) twice, we get
\begin{eqnarray}
\label{gs3}
TC(\rho)=\rho \int^{\rho}\frac{p(\rho')}{\rho^{'2}}d\rho'.
\end{eqnarray}
Therefore, the free energy (\ref{min6}) can be rewritten
\begin{eqnarray}
\label{gs4}
F[\rho]=\frac{1}{2}\int\rho\Phi d{\bf r}+\int  \rho \int^{\rho}\frac{p(\rho')}{\rho^{'2}}d\rho'd{\bf r}.
\end{eqnarray}
With these notations, the $H$-theorem (\ref{h8}) becomes
\begin{eqnarray}
\label{gs5}
\dot F=-\int \frac{\chi}{\rho}(\nabla p+\rho\nabla \Phi)^{2}d{\bf r}\le 0.
\end{eqnarray}
The stationary solutions of the GS equation (\ref{gs1}) satisfy the relation
\begin{eqnarray}
\label{gs6}
\nabla p+\rho\nabla \Phi={\bf 0},
\end{eqnarray}
which is similar to a condition of hydrostatic equilibrium. Since $p=p(\rho)$, this relation can be integrated to give $\rho=\rho(\Phi)$ through
\begin{eqnarray}
\label{gs7}
\int^{\rho}\frac{p'(\rho')}{\rho'}d\rho'=-\Phi.
\end{eqnarray}
This is equivalent to
\begin{eqnarray}
\label{gs8}
\frac{p'(\rho)}{\rho}=-\frac{1}{\rho'(\Phi)}.
\end{eqnarray}
This relation can also be obtained from Eqs. (\ref{gs2}) and
(\ref{sta5}). Therefore, we recover the fact that, in the steady
state, $\rho=\rho(\Phi)$ is a monotonically decreasing function of
$\Phi$. We also note the identity
\begin{eqnarray}
\label{gs9}
p(\rho)=\frac{1}{\chi}D(\rho)\rho=T\rho^{2} \left \lbrack \frac{C(\rho)}{\rho} \right\rbrack'=T\lbrack C'(\rho)\rho-C(\rho)\rbrack.\nonumber\\ 
\end{eqnarray}

Finally, we note that the relation (\ref{gs7}), equivalent to the
condition of hydrostatic equilibrium (\ref{gs6}), can also be obtained
by extremizing the free energy (\ref{gs4}) at fixed mass writing
$\delta F-\alpha\delta M=0$. More precisely, we have the important
result: {\it a steady solution of the generalized Smoluchowski
equation (\ref{gs1}) is linearly dynamically stable iff it is a
(local) minimum of the free energy $F[\rho]$ at fixed mass
$M[\rho]=M$.} This corresponds to the minimization problem
(\ref{min3}) with Eq. (\ref{gs4}). Note that the GS equation has been
introduced here in a very general context. In this sense, it is valid
for arbitrary value of $\chi$ (not necessarily small). Alternatively,
we shall see in Sec. \ref{sec_strong} that a generalized Smoluchowski
equation with $\chi=1/\xi\rightarrow 0$ can be derived from a
generalized Kramers equation in the strong friction limit
$\xi\rightarrow +\infty$.

\subsection{Phenomenological derivation of the nonlinear Fokker-Planck equation}
\label{sec_phen}

The form of the diffusion current appearing in the NFP equation
(\ref{dde3}) can be obtained by different phenomenological procedures.

\subsubsection{Functional derivative}
\label{sec_func}

For a given free energy functional $F[\rho]$, we can introduce
phenomenologically a dynamical model by writing the evolution of the
density as a continuity equation $\partial_{t}\rho=-\nabla \cdot {\bf
J}$ where the current is proportional to the gradient of the
functional derivative of the free energy \cite{frank}, i.e.
\begin{eqnarray}
\label{func1}
\frac{\partial\rho}{\partial t}=\nabla\cdot \left \lbrack \chi({\bf r},t)\rho \nabla\frac{\delta F}{\delta\rho}\right \rbrack.
\end{eqnarray}
For the free energy (\ref{min6}), we have 
\begin{eqnarray}
\label{func2}
\frac{\delta F}{\delta\rho}=TC'(\rho)+\Phi,
\end{eqnarray}
so that
\begin{eqnarray}
\label{func3}
\frac{\partial\rho}{\partial t}=\nabla\cdot \left \lbrack \chi({\bf r},t)\left (T\rho C''(\rho) \nabla\rho+\rho  \nabla \Phi\right )\right\rbrack.
\end{eqnarray}
This equation is more general than Eq. (\ref{dde3}). It shows that,
for a given free energy, we can introduce an infinite class of NFP
equations where $\chi({\bf r},t)$ is an {\it arbitrary} positive
function of position and time. In particular, $\chi({\bf r},t)$ can be
a positive function of the density $\rho({\bf r},t)$. If we set
$\chi({\bf r},t)=\chi g(\rho)/\rho$ we recover Eq. (\ref{dde3}) as a
particular case. We can also write Eq. (\ref{func3}) in the
alternative form
\begin{eqnarray}
\label{func4}
\frac{\partial\rho}{\partial t}=\nabla\cdot \left \lbrack \tilde{\chi}({\bf r},t)\left (T\nabla\rho+\frac{1}{C''(\rho)}  \nabla \Phi\right )\right\rbrack,
\end{eqnarray}
where $\tilde{\chi}({\bf r},t)$ is an {\it arbitrary} positive
function of position and time.  If we set $\tilde{\chi}({\bf
r},t)=\chi h(\rho)$ we recover Eq. (\ref{dde3}). These two alternative
forms (\ref{func3}) and (\ref{func4}) were given in \cite{gen}.

On the other hand, the formulation (\ref{func1}) ensures that the free
energy decreases monotonically provided that $\chi({\bf r},t)$ is
positive. Indeed,
\begin{eqnarray}
\label{func5}
\dot F=\int \frac{\delta F}{\delta\rho}\frac{\partial\rho}{\partial t}\, d{\bf r}=-\int \frac{\delta F}{\delta\rho} \nabla\cdot {\bf J}\, d{\bf r}\nonumber\\
=\int {\bf J}\cdot \nabla \frac{\delta F}{\delta\rho}\, d{\bf r}=-\int \chi({\bf r},t)\rho\left (\nabla \frac{\delta F}{\delta\rho}\right )^{2}\, d{\bf r}\le 0.
\end{eqnarray}
Furthermore, a steady state of Eq. (\ref{func1}) satisfies $\dot F=0$, i.e $\nabla (\delta F/\delta\rho)=0$ leading to 
\begin{eqnarray}
\label{func6}
\frac{\delta F}{\delta\rho}=-T\alpha,
\end{eqnarray}
where $\alpha$ is a constant of integration. This is equivalent to the
extremization of $F$ at fixed mass $M$ if we write the first
variations as $\delta F+T\alpha\delta M=0$. Therefore, a steady state
extremizes the free energy at fixed mass. Using Eq. (\ref{func2}), we find that
the equilibrium density profile satisfies
\begin{eqnarray}
\label{func7}
C'(\rho)=-\beta \Phi-\alpha.
\end{eqnarray}
Finally, using Lyapunov's direct method, one can show that a steady
state $\rho({\bf r})$ of Eq. (\ref{func1}) is linearly dynamically
stable iff it is a (local) minimum of $F$ at fixed mass.  We again
emphasize that the stationary states of Eq. (\ref{func3}), and the
$H$-theorem, only depend on the form of the free energy $F[\rho]$ and
are independent on the positive function $\chi({\bf r},t)$.  These
properties are therefore valid for the whole class of NFP equations
associated with a given free energy functional. Therefore, for a given
free energy specified by $C(\rho)$, we can construct an infinite class
of NFP equations with arbitrary $\chi({\bf r},t)$ that  have the
same equilibrium states (\ref{func7}) but a different dynamics.
 
\subsubsection{Onsager's linear thermodynamics}
\label{sec_onsager}

The previous approach is equivalent to Onsager's linear
thermodynamics. Indeed, at equilibrium, we expect that the
distribution $\rho({\bf r})$ minimizes the free energy $F$ at fixed
mass. This leads to Eq. (\ref{func6}) or (\ref{func7}). Noting that the
chemical potential
\begin{eqnarray}
\label{onsager1}
\lambda({\bf r},t)\equiv \frac{\delta F}{\delta\rho}=TC'(\rho)+\Phi,
\end{eqnarray}
is uniform at equilibrium, the linear thermodynamics of Onsager
suggests that, close to equilibrium, the current is proportional to
the gradient of the chemical potential. If we write
\begin{eqnarray}
\label{onsager2}
{\bf J}=-\chi({\bf r},t)\rho \nabla\lambda({\bf r},t),
\end{eqnarray}
the linear thermodynamics of Onsager leads to Eq. (\ref{func1}).  

\subsubsection{Maximum Free Energy Dissipation Principle}
\label{sec_mepp}

The same results can be obtained from a variational formulation,
called the Maximum Free Energy Dissipation (MFED) principle which is
the canonical ensemble version of the Maximum Entropy Production (MEP)
principle \cite{gen}.  At equilibrium, the optimal distribution
$\rho({\bf r})$ minimizes the free energy $F[\rho]$ at fixed mass
$M$. Out of equilibrium, we may expect that the optimal current ${\bf
J}$ maximizes the rate of free energy dissipation $\dot F[{\bf J}]$
under some constraints. This can be viewed as a variational
formulation of Onsager's linear thermodynamics. The rate of
dissipation of free energy is given by
\begin{eqnarray}
\label{mepp1}
\dot F=\int \frac{\delta F}{\delta\rho}\frac{\partial\rho}{\partial t}\, d{\bf r}=-\int \frac{\delta F}{\delta\rho} \nabla\cdot {\bf J}\, d{\bf r}
=\int {\bf J}\cdot \nabla \frac{\delta F}{\delta\rho}\, d{\bf r}.\nonumber\\
\end{eqnarray}
We shall determine the optimal current ${\bf J}_{*}$ which maximizes
the rate of dissipation of free energy $\dot F$ under the constraint
$J^{2}\le C({\bf r},t)$ putting a physical bound on $|{\bf J}|$. It
can be shown that the bound is always reached so that we can replace
the inequality by an equality. Thus, we write the variational problem
as
\begin{eqnarray}
\label{mepp2}
\delta \dot F+\delta \left (\int \frac{{\bf J}^{2}}{2\rho \chi({\bf r},t)}\, d{\bf r}\right )=0,
\end{eqnarray}
where $\chi({\bf r},t)$ is a local Lagrange multiplier. 
Performing the variations on ${\bf J}$, we obtain 
\begin{eqnarray}
\label{mepp3}
{\bf J}_{*}=-\chi({\bf r},t)\rho \nabla \frac{\delta F}{\delta\rho},
\end{eqnarray}
 which returns
Eq. (\ref{func1}). Note that if we introduce the ``dissipation'' function
\begin{eqnarray}
\label{mepp4}
E_{d}\equiv \int \frac{{\bf J}^{2}}{2\rho \chi({\bf r},t)}\, d{\bf r},
\end{eqnarray}
we have
\begin{eqnarray}
\label{mepp5}
E_{d}[{\bf J}_{*}]=-\frac{1}{2}\dot F[{\bf J}_{*}]. 
\end{eqnarray}
On the other hand, since $\delta^{2}(\dot F+E_{d})=-\int \frac{(\delta
J)^{2}}{2\rho\chi}d{\bf r}\le 0$, the optimal current (\ref{mepp3})
{\it maximizes} the dissipation of free energy under the constraint
$J^{2}\le C({\bf r},t)$.

\subsection{Kinetic derivation of the nonlinear Fokker-Planck equation}
\label{sec_kin}

Nonlinear Fokker-Planck equation where the diffusion coefficient and
the mobility explicitly depend on the local concentration of particles
can be derived from a kinetic theory, starting from the master
equation, and assuming that the probabilities of transition explicitly
depend on the occupation numbers (concentrations) of the initial and
arrival states. We briefly summarize the approach developed by
Kaniadakis \cite{k1} and make the link with the phenomenological
equations studied previously.

We introduce a stochastic dynamics by defining the probability of
transition of a particle from position ${\bf r}$ to position ${\bf
r}'$. Following Kaniadakis \cite{k1}, we assume the following
form
\begin{eqnarray}
\label{kin1}
\pi({\bf r}\rightarrow {\bf r}')=w({\bf r},{\bf r}-{\bf r}')a\lbrack\rho({\bf r},t)\rbrack b\lbrack\rho({\bf r}',t)\rbrack.
\end{eqnarray}
Usual stochastic processes correspond to $a(\rho)=\rho$ and
$b(\rho)=1$: the probability of transition is proportional to the
density of the initial state and independent on the density of the
final state.  They lead to the ordinary Fokker-Planck equation
(\ref{sm1}) as will be shown below. Here, we assume a more general
dependence on the occupancy in the initial and arrival states. This
can account for microscopic constraints like close-packing effects
that can inhibitate the transition. Quite generally, the evolution of
the density satisfies the master equation
\begin{eqnarray}
\label{kin2}
\frac{\partial\rho}{\partial t}=\int \left\lbrack \pi({\bf r}'\rightarrow {\bf r})-\pi({\bf r}\rightarrow {\bf r}')\right\rbrack d{\bf r}'.
\end{eqnarray}
Assuming that the evolution is sufficiently slow, and local, such that the dynamics only permits values of ${\bf r}'$ close to ${\bf r}$, one can develop the term in brackets in Eq. (\ref{kin2}) in powers of  ${\bf r}-{\bf r}'$. Proceeding along the lines of \cite{k1}, we obtain a Fokker-Planck-like equation 
\begin{equation}
\label{kin3}
\frac{\partial\rho}{\partial t}=\frac{\partial}{\partial x_{i}}\left\lbrack\left (\zeta_{i}+\frac{\partial\zeta_{ij}}{\partial x_{j}}\right )\gamma(\rho)+\gamma(\rho)\frac{\partial\ln \kappa(\rho)}{\partial\rho}\zeta_{ij}\frac{\partial\rho}{\partial x_{j}}\right\rbrack,
\end{equation}
with
\begin{equation}
\label{kin4}
\gamma(\rho)=a(\rho)b(\rho),\qquad \kappa(\rho)=\frac{a(\rho)}{b(\rho)},
\end{equation}
and
\begin{equation}
\label{kin5}
\zeta_{i}({\bf r})=-\int y_{i}w({\bf r},{\bf y})d{\bf y},
\end{equation}
\begin{equation}
\label{kin6}
\zeta_{ij}({\bf r})=\frac{1}{2}\int y_{i}y_{j}w({\bf r},{\bf y})d{\bf y}.
\end{equation}
The moments $\zeta_{i}$ and $\zeta_{ij}$ are fixed by the ordinary Langevin equation 
\begin{eqnarray}
\label{kin7} \frac{d{\bf
r}}{dt}=-\chi\nabla\Phi+\sqrt{2D}{\bf R}(t),
\end{eqnarray}
where $\chi$ and $D$ are constant.  Assuming isotropy
$\zeta_{i}=J_{i}$, $\zeta_{ij}=D\delta_{ij}$, the kinetic equation (\ref{kin3})
becomes
\begin{equation}
\label{kin8}
\frac{\partial\rho}{\partial t}=\nabla\cdot \left\lbrack ({\bf J}+\nabla D)\gamma(\rho)+\gamma(\rho)\frac{\partial\ln \kappa(\rho)}{\partial\rho}D\nabla \rho\right\rbrack.
\end{equation}
Now, according to the Langevin equation (\ref{kin7}), $D$ is independent on ${\bf r}$ and ${\bf J}=\chi\nabla \Phi$. Thus, we get
\begin{equation}
\label{kin9}
\frac{\partial\rho}{\partial t}=\nabla\cdot \left\lbrack D\gamma(\rho)\frac{\partial\ln \kappa(\rho)}{\partial\rho}\nabla\rho+\chi\gamma(\rho)\nabla \Phi \right\rbrack.
\end{equation}
If we define
\begin{equation}
\label{kin10}
h(\rho)=\gamma(\rho)\frac{\partial\ln \kappa(\rho)}{\partial\rho}, \qquad g(\rho)=\gamma(\rho),
\end{equation}
the foregoing equation can be written
\begin{eqnarray}
\label{kin11a}
\frac{\partial\rho}{\partial t}=\nabla\cdot \left \lbrack Dh(\rho)\nabla\rho+\chi g(\rho)\nabla \Phi\right \rbrack,
\end{eqnarray}
and it coincides \footnote{In Sec. \ref{sec_gle}, we have obtained
generalized Fokker-Planck equations by using ordinary Master equations (based on usual transition
probabilities $a(\rho)=\rho$ and $b(\rho)=1$) and
generalized Langevin equations where the diffusion coefficient and the
mobility depend on the density. In this section, we have obtained
generalized Fokker-Planck equations by using generalized Master equations (based on density dependent
transition probabilities) and ordinary Langevin equations
with constant coefficients.}  with the phenomenological equation
(\ref{dde3}). We note that
\begin{eqnarray}
\label{kin11b}
\ln\kappa(\rho)=C'(\rho).
\end{eqnarray}
We also have the relations
\begin{eqnarray}
\label{kin12}
a(\rho)=\sqrt{\gamma(\rho)\kappa(\rho)}=\sqrt{g(\rho)}e^{C'(\rho)/2},
\end{eqnarray}
\begin{eqnarray}
\label{kin13}
b(\rho)=\sqrt{\frac{\gamma(\rho)}{\kappa(\rho)}}=\sqrt{g(\rho)}e^{-C'(\rho)/2}.
\end{eqnarray}
Inversely
\begin{eqnarray}
\label{kin14}
g(\rho)=a(\rho)b(\rho), \qquad C'(\rho)=\ln\left\lbrack \frac{a(\rho)}{b(\rho)}\right\rbrack,
\end{eqnarray}
\begin{eqnarray}
\label{kin15}
h(\rho)=b(\rho)a'(\rho)-a(\rho)b'(\rho).
\end{eqnarray}

It seems natural to assume that the transition probability is
proportional to the density of the initial state so that
$a(\rho)=\rho$. In that case, we obtain an equation of the form
\begin{equation}
\label{kin16}
\frac{\partial\rho}{\partial t}=\nabla\cdot \left ( D\left\lbrack b(\rho)-\rho b'(\rho)\right \rbrack \nabla\rho+\chi\rho b(\rho)\nabla \Phi\right ). 
\end{equation}
Note that the coefficients of diffusion and mobility are not
independent since they are both expressed in terms of
$b(\rho)$. Choosing $b(\rho)=1$, i.e. a probability of transition which
does not depend on the population of the arrival state, leads to the
standard Fokker-Planck equation (\ref{sm1}).  If, now, we assume that
the transition probability is blocked (inhibited) if the concentration
of the arrival state is equal to $\sigma_0$, then it seems natural to
take $b(\rho)=1-\rho/\sigma_{0}$. In that case, we obtain
\begin{equation}
\label{kin17}
\frac{\partial\rho}{\partial t}=\nabla\cdot \left ( D\nabla\rho+\chi\rho (1-\rho/\sigma_{0})\nabla \Phi\right ),
\end{equation}
which will be considered in Sec. \ref{sec_ff}.  Inversely, we can
wonder what the general form of the mobility will be if we assume a
normal diffusion $h(\rho)=1$. This leads to $b(\rho)-\rho b'(\rho)=1$
which is integrated in $b(\rho)=1+K\rho$ where $K$ is a
constant. Interestingly, we find that this condition selects the class
of fermions ($K=-1$) and bosons ($K=+1$) and intermediate statistics
(arbitrary $K$). The corresponding NFP equation is
\begin{equation}
\label{kin18}
\frac{\partial\rho}{\partial t}=\nabla\cdot \left ( D\nabla\rho+\chi\rho (1+K\rho)\nabla \Phi\right ),
\end{equation}
which will be considered in Sec. \ref{sec_fbi}.  

\section{Examples of nonlinear Smoluchowski equations and generalized Keller-Segel models}
\label{sec_ex}

In this section, we give several explicit examples of nonlinear mean
field Fokker-Planck equations. Some correspond to well-known forms of
entropies, and others are new. We emphasize that these equations can
have applications in different domains of physics as discussed in
Sec. \ref{sec_known}. Importantly, they can provide generalized
Keller-Segel models of chemotaxis (see Sec. \ref{sec_ks}). Most of
these models have not been considered before in biology because the
connection with generalized thermodynamics was not made. This is why
we give a relatively detailed description of these models since their
applications in biology are new.

\subsection{Standard model: Boltzmann entropy}
\label{sec_sm}

If we take $h(\rho)=1$ and $g(\rho)=\rho$, we get the ordinary
Smoluchowski equation
\begin{eqnarray}
\label{sm1}
\frac{\partial\rho}{\partial t}=\nabla\cdot \left ( D \nabla\rho+
\chi \rho \nabla \Phi\right ).
\end{eqnarray}
It corresponds to an ordinary diffusion $D(\rho)=D$ and a constant
mobility $\chi(\rho)=\chi$.  The associated stochastic process is
\begin{eqnarray}
\label{sm2}
\frac{d{\bf r}}{dt}=-\chi\nabla \Phi+\sqrt{2D}{\bf R}(t).
\end{eqnarray}
The entropy is the Boltzmann entropy 
\begin{eqnarray}
\label{sm3}
S=-\int \rho\ln\rho\, d{\bf r},
\end{eqnarray}
and the stationary solution of Eq. (\ref{sm1}) is the Boltzmann
distribution
\begin{eqnarray}
\label{sm4}
\rho=e^{-\beta \Phi-\alpha-1}.
\end{eqnarray}
The pressure law is  
\begin{eqnarray}
\label{sm5}
p(\rho)=\rho T.
\end{eqnarray}
This is similar to the equation of state for an isothermal gas with
constant temperature $T$. When the Fokker-Planck equation (\ref{sm1})
is coupled to the Poisson equation (\ref{gle5}), we obtain the
Smoluchowski-Poisson system describing a gas of self-gravitating
Brownian particles \cite{crs,sc,post,tcoll,virial1,virial2}. When the Fokker-Planck
equation (\ref{sm1}) is coupled to the field Eq. (\ref{gle3}), we
obtain the standard Keller-Segel model describing the chemotactic
aggregation of biological populations
\cite{horstmann}.

\subsection{Power law diffusion: Tsallis entropy}
\label{sec_pld}

If we take $h(\rho)=q\rho^{q-1}$ and $g(\rho)=\rho$, we obtain the NFP equation
\begin{eqnarray}
\label{pld1}
\frac{\partial\rho}{\partial t}=\nabla\cdot \left ( D \nabla\rho^{q}+
\chi \rho \nabla \Phi\right ).
\end{eqnarray}
It corresponds to a power law diffusion $D(\rho)=D\rho^{q-1}$ and
a constant mobility $\chi(\rho)=\chi$. This equation was introduced by
Plastino \& Plastino \cite{pp}.  The associated stochastic process, introduced by Borland \cite{borland}, is
\begin{eqnarray}
\label{pld2}
\frac{d{\bf r}}{dt}=-\chi\nabla \Phi+\sqrt{2D}\rho^{\frac{q-1}{2}}{\bf R}(t).
\end{eqnarray}
This model can take into
account effects of non-ergodicity and nonextensivity.  It leads to a
situation of anomalous diffusion related to the Tsallis statistics.
For $q=1$, we recover the standard Brownian model with a constant
diffusion coefficient, corresponding to a pure random walk. In that
case, the sizes of the random kicks are uniform and do not depend on
where the particle happens to be. For $q\neq 1$, the size of the
random kicks changes, depending on the distribution of the particles
around the ``test'' particle. A particle which is in a region that is
highly populated [large $\rho({\bf r},t)$] will tend to have larger
kicks if $q>1$ and smaller kicks if $q<1$. Since the microscopics
depends on the actual density in space, this creates a bias in
the ergodic behavior of the system.  Then, the dynamics has a fractal
or multi-fractal phase space structure
\cite{borland}. The generalized entropy associated to Eq. (\ref{pld1}) is the
Tsallis entropy
\begin{eqnarray}
\label{pld3}
S=-\frac{1}{q-1}\int (\rho^{q}-\rho)\, d{\bf r},
\end{eqnarray}
and the stationary solution is the Tsallis distribution
\begin{eqnarray}
\label{pld4}
\rho=\left (\frac{1}{q}\right )^{\frac{1}{q-1}}\left\lbrack 1-(q-1)(\beta\Phi+\alpha)\right\rbrack_{+}^{1/(q-1)}.
\end{eqnarray}
The pressure law is  
\begin{eqnarray}
\label{pld5}
p(\rho)=T\rho^{q}.
\end{eqnarray}
This is similar to a polytropic gas with an equation of state
$p=K\rho^{\gamma}$ where $K=T$ plays the role of a polytropic
temperature and $q=\gamma$ is the polytropic index (we also set
$\gamma=1+1/n$). Note that the Tsallis entropy can be written
\begin{eqnarray}
\label{pld6}
S=-\int \rho \ln_{(q)}\rho \, d{\bf r},
\end{eqnarray}
where we have introduced  the $q$-logarithm
\begin{eqnarray}
\label{pld7}
\ln_{(q)}(x)=\frac{1}{q-1}(x^{q-1}-1).
\end{eqnarray}
The stationary solution can be written
\begin{eqnarray}
\label{pld8}
\rho=\left (\frac{1}{q}\right )^{\frac{1}{q-1}}e_{(q)}^{-\beta\Phi-\alpha}
\end{eqnarray}
with  the $q$-exponential
\begin{eqnarray}
\label{pld9}
e_{(q)}(x)=\lbrack 1+(q-1)x\rbrack^{\frac{1}{q-1}}.
\end{eqnarray}
For $q=1$, we recover the standard model (\ref{sm1}). For $q=2$, we
have some simplifications. In that case, the NFP equation (\ref{pld1}) becomes
\begin{eqnarray}
\label{pld10}
\frac{\partial\rho}{\partial t}=\nabla\cdot \left ( D \nabla\rho^{2}+\chi \rho \nabla \Phi\right ).
\end{eqnarray}
The entropy is the quadratic functional
\begin{eqnarray}
\label{pld11}
S=-\int \rho^{2} d{\bf r},
\end{eqnarray}
and the stationary solution is 
\begin{eqnarray}
\label{pld12}
\rho=-\frac{1}{2}(\beta \Phi+\alpha),
\end{eqnarray}
corresponding to a linear relation between the density and the
potential. In that case, the differential equation (\ref{sta7})
determining the steady state reduces to the Helmholtz
equation. Finally, the pressure is
\begin{eqnarray}
\label{pld13}
p(\rho)=T\rho^{2},
\end{eqnarray}
corresponding to a polytrope with index $n=1$.  When the NFP equation
(\ref{pld1}) is coupled to the Poisson equation (\ref{gle5}), we
obtain the polytropic Smoluchowski-Poisson system describing
self-gravitating Langevin particles. When the NFP equation
(\ref{pld1}) is coupled to the field Eq. (\ref{gle3}), we obtain a
generalized Keller-Segel model of chemotaxis taking into account
anomalous diffusion. These models have been introduced by Chavanis
\& Sire \cite{lang,csbio,csmass}.

\subsection{Logotropic distributions: log-entropy}
\label{sec_log}

If we take $h(\rho)=1/\rho$ and $g(\rho)=\rho$, we obtain
\begin{eqnarray}
\label{log1}
\frac{\partial\rho}{\partial t}=\nabla\cdot \left ( D \nabla\ln\rho+
\chi \rho \nabla \Phi\right ).
\end{eqnarray}
The generalized entropy associated to Eq. (\ref{log1}) is the log-entropy
\begin{eqnarray}
\label{log2}
S=\int \ln\rho \, d{\bf r},
\end{eqnarray}
and the stationary solution is 
\begin{eqnarray}
\label{log3}
\rho=\frac{1}{\alpha+\beta \Phi}. 
\end{eqnarray}
For a quadratic potential $\Phi_{ext}=r^{2}/2$, this corresponds to
the  Lorentzian  function. The pressure is
\begin{eqnarray}
\label{log4}
p(\rho)=T\ln\rho.
\end{eqnarray}
This is similar to a logotropic equation of state \cite{pudritz}. This
is also connected to a polytropic equation of state (or Tsallis
distribution) with $\gamma=q=0$. Indeed, the logotropic model
(\ref{log1}) can be deduced from Eq. (\ref{pld1}) by writing $D\nabla\rho^{q}=Dq\rho^{q-1}\nabla\rho$, taking $q=0$ and
re-defining $Dq\rightarrow D$. When the NFP equation (\ref{log1}) is
coupled to the Poisson equation (\ref{gle5}), we obtain the logotropic
Smoluchowski-Poisson system. When the NFP equation (\ref{log1}) is coupled to
the field Eq. (\ref{gle3}), we obtain a generalized Keller-Segel model of
chemotaxis. These models have been introduced by Chavanis
\& Sire \cite{logotrope}.

\subsection{Power law diffusion and drift: Tsallis entropy}
\label{sec_pldd}

We introduce here a new model generalizing the polytropic model
(\ref{pld1}). If we take $h(\rho)=q\rho^{q+\mu-1}$ and
$g(\rho)=\rho^{\mu+1}$, we obtain
\begin{eqnarray}
\label{pldd1}
\frac{\partial\rho}{\partial t}=\nabla\cdot \left ( Dq\rho^{q+\mu-1} 
\nabla\rho+
\chi \rho^{\mu+1} \nabla \Phi\right ).
\end{eqnarray}
This corresponds to a power law diffusion
$D(\rho)=\frac{Dq}{q+\mu}\rho^{q+\mu-1}$ and a power law mobility
$\chi(\rho)=\chi \rho^{\mu}$.  The associated 
stochastic process is
\begin{eqnarray}
\label{pldd2}
\frac{d{\bf r}}{dt}=-\chi \rho^{\mu}\nabla\Phi+\sqrt{\frac{2Dq}{q+\mu}}\rho^{\frac{q+\mu-1}{2}}{\bf R}(t).
\end{eqnarray}
Since $\rho^{\mu}$ can be put in factor of the diffusion current in
Eq. (\ref{pldd1}), this model belongs to the infinite family of NFP
equations associated to the
Tsallis entropy with index $q$ (see discussion in Sec. \ref{sec_func}).

For $\mu=0$, we recover Eq. (\ref{pld1}) with a constant mobility and a power
law diffusion. For $(\mu,q)=(0,0)$, we recover the logotropic
Smoluchowski equation (\ref{log1}) provided that we make the transformation
$Dq\rightarrow D$. For $\mu=1-q$, we have a normal diffusion and a
power law mobility
\begin{eqnarray}
\label{pldd3}
\frac{\partial\rho}{\partial t}=\nabla\cdot \left ( Dq 
\nabla\rho+
\chi \rho^{2-q} \nabla \Phi\right ).
\end{eqnarray}
For $q=2$, we get
\begin{eqnarray}
\label{pldd4}
\frac{\partial\rho}{\partial t}=\nabla\cdot \left (2 D 
\nabla\rho+
\chi  \nabla \Phi\right ),
\end{eqnarray}
which has the same equilibrium states as Eq. (\ref{pld10}). If we
assume furthermore that $\Phi$ is given by the Poisson equation
(\ref{gle5}), Eq. (\ref{pldd4}) reduces to the linear equation
\begin{eqnarray}
\label{pldd5}
\frac{\partial\rho}{\partial t}=2 D 
\Delta\rho+
\chi \lambda \rho.
\end{eqnarray}
Finally, for $q=0$ (making the transformation $qD\rightarrow D$), we
obtain
\begin{eqnarray}
\label{pldd6}
\frac{\partial\rho}{\partial t}=\nabla\cdot \left ( D
\nabla\rho+
\chi \rho^{2} \nabla \Phi\right ),
\end{eqnarray}
which has the same equilibrium states as Eq. (\ref{log1}).  When the
NFP equation (\ref{pldd1}) is coupled to the field equation (\ref{gle3}),
we obtain a generalized Keller-Segel model of chemotaxis taking into
account anomalous diffusion and anomalous mobility. This model will
be studied in a forthcoming paper, in continuity with \cite{lang}.

\subsection{Filling factor: Fermi-Dirac entropy}
\label{sec_ff}

If we take $h(\rho)=1$ and $g(\rho)=\rho(1-\rho/\sigma_{0})$, we obtain
\begin{eqnarray}
\label{ff1}
\frac{\partial\rho}{\partial t}=\nabla\cdot \left ( D \nabla\rho+
\chi \rho(1-\rho/\sigma_{0}) \nabla \Phi\right ).
\end{eqnarray}
This corresponds to a normal diffusion $D(\rho)=D$ and a mobility
$\chi(\rho)=\chi(1-\rho/\sigma_{0})$ vanishing linearly when the
density reaches the maximum value $\rho_{max}=\sigma_{0}$. The
associated stochastic process is
\begin{eqnarray}
\label{ff2}
\frac{d{\bf r}}{dt}=-\chi(1-\rho/\sigma_{0})\nabla \Phi+\sqrt{2D}{\bf R}(t).
\end{eqnarray}
The generalized entropy associated with Eq. (\ref{ff1}) is the
Fermi-Dirac entropy in position space
\begin{eqnarray}
\label{ff3}
S=-\sigma_{0}\int \left\lbrace \frac{\rho}{\sigma_{0}}\ln\frac{\rho}{\sigma_{0}}+\left (1-\frac{\rho}{\sigma_{0}}\right)\ln \left (1-\frac{\rho}{\sigma_{0}}\right)\right\rbrace  d{\bf r},\nonumber\\
\end{eqnarray}
and the stationary solution is the Fermi-Dirac distribution in
position space
\begin{eqnarray}
\label{ff4}
\rho=\frac{\sigma_{0}}{1+ e^{\beta \Phi+\alpha}}.
\end{eqnarray}
From Eq. (\ref{ff4}), we see that, in the stationary state,
$\rho\le \sigma_{0}$. This bound is similar to the Pauli exclusion
principle in quantum mechanics. In fact, we can show that $\rho({\bf
r},t)$ remains bounded by $\sigma_{0}$ during the whole evolution.  For
$\sigma_{0}\rightarrow +\infty$, we recover the standard model (\ref{sm1}).

An alternative model, with the same entropy and the same equilibrium
states, is obtained by taking $h(\rho)=1/(1-\rho/\sigma_{0})$ and
$g(\rho)=\rho$. This leads to
\begin{eqnarray}
\label{ff5}
\frac{\partial\rho}{\partial t}=\nabla\cdot \left (-D\sigma_{0} \nabla\ln(1-\rho/\sigma_{0})+
\chi \rho \nabla \Phi\right ).
\end{eqnarray}
This corresponds to a nonlinear diffusion with $D(\rho)=-
\sigma_{0}(D/\rho)\ln (1-\rho/\sigma_{0})$ and a constant mobility
$\chi(\rho)=\chi$. Equation (\ref{ff5}) can be put in the form of a
generalized Smoluchowski equation (\ref{gs1})  with a pressure law
\begin{eqnarray}
\label{ff6}
p(\rho)=-T\sigma_{0} \ln(1-\rho/\sigma_{0}).
\end{eqnarray}
For $\rho\ll\sigma_{0}$, we recover the ``isothermal'' equation of
state $p=\rho T$ leading to the standard model (\ref{sm1}). However,
for higher densities, the equation of state is modified and the
pressure diverges when $\rho\rightarrow \sigma_{0}$. This
prevents the density from exceeding the maximum value $\sigma_{0}$.

In the context of chemotaxis, the model (\ref{ff1}) has been introduced by
Hillen \& Painter \cite{hp} and, independently, by Chavanis
\cite{gen,crrs,degrad}.  It provides a regularization of the standard
Keller-Segel model preventing overcrowding, blow-up and unphysical
singularities. The filling factor $(1-\rho/\sigma_{0})$ takes into
account the fact that the particles cannot interpenetrate because of
their finite size $a$. Therefore, the maximum allowable density is
$\sigma_{0}\sim 1/a^{d}$. It is achieved when all the cells are packed
together. In the model (\ref{ff1}), it is assumed that the mobility vanishes
when the density reaches the close packing value ($\rho\rightarrow
\sigma_{0}$) while the diffusion is not affected. The alternative model (\ref{ff5}) has been introduced in Chavanis \cite{gen,degrad}. 
In that case, the mobility is assumed to be constant and the
regularization preventing overcrowding is taken into account in the
pressure law (\ref{ff6}). As explained in Sec. \ref{sec_func}, we can
multiply the diffusion term and the mobility term by the {\it same}
positive function $\chi({\bf r},t)$ in order to obtain a more general
model with the same entropy and the same equilibrium states. Note
finally that an equation similar to Eq. (\ref{ff1}) has been
introduced by Robert
\& Sommeria \cite{rs} (see also \cite{csr}) in the statistical mechanics of
two-dimensional turbulence for two vorticity levels $0$ and
$\sigma_{0}$. In that case, $\rho$ represents the coarse-grained
vorticity $\overline{\omega}$ and $\Phi$ plays the role of the stream
function $\psi$. The ``exclusion principle'' leading to the
Fermi-Dirac entropy (\ref{ff3}) is a consequence of the 2D Euler
equation implying that the vorticity levels cannot overlap so that
$\overline{\omega}({\bf r},t)\le \sigma_{0}$. These analogies between
chemotaxis and 2D turbulence are further discussed in \cite{degrad}.

\subsection{Fermi, Bose and intermediate statistics}
\label{sec_fbi}

If we take $h(\rho)=1$ and $g(\rho)=\rho(1+K\rho)$, we obtain
\begin{eqnarray}
\label{fbi1}
\frac{\partial\rho}{\partial t}=\nabla\cdot \left ( D \nabla\rho+
\chi \rho(1+K\rho) \nabla \Phi\right ).
\end{eqnarray}
This corresponds to a normal diffusion $D(\rho)=D$ and a variable
mobility $\chi(\rho)=\chi(1+K\rho)$.  The associated stochastic process
is
\begin{eqnarray}
\label{fbi2}
\frac{d{\bf r}}{dt}=-\chi(1+K\rho)\nabla \Phi+\sqrt{2D}{\bf R}(t).
\end{eqnarray}
Here, $K$ is a real number taking positive or negative values.  When
$K>0$ the mobility is enhanced in regions of large densities and when
$K<0$, it is reduced. This takes into account inclusion ($K>0$) or
exclusion ($K<0$) principles. For $K=0$, we recover the standard model
(\ref{sm1}). The generalized entropy associated with Eq. (\ref{fbi1}) is
\begin{eqnarray}
\label{fbi3}
S=-\int \left\lbrack {\rho}\ln{\rho}-\frac{1}{K}\left (1+K \rho\right)\ln \left (1+K\rho\right)\right\rbrack  d{\bf r},\nonumber\\
\end{eqnarray}
and the stationary solution is 
\begin{eqnarray}
\label{fbi4}
\rho=\frac{1}{e^{\beta \Phi+\alpha}-K}.
\end{eqnarray}
For $K=+1$ we obtain the Bose-Einstein statistics and for $K=-1$ we
obtain the Fermi-Dirac statistics. For other values of $K$, we obtain
intermediate statistics (quons) interpolating between fermions and
bosons. For $K=0$, we recover the Boltzmann statistics.

An alternative model with the same entropy and the same equilibrium
states is obtained by taking $h(\rho)=1/(1+K\rho)$ and $g(\rho)=\rho$. This leads to
\begin{eqnarray}
\label{fbi5}
\frac{\partial\rho}{\partial t}=\nabla\cdot \left (\frac{1}{K} \nabla\ln(1+K\rho)+
\chi \rho \nabla \Phi\right ).
\end{eqnarray}
This corresponds to a nonlinear diffusion such that $D(\rho)=
(D/K\rho)\ln(1+K\rho) $ and a constant mobility $\chi(\rho)=\chi$.
 The pressure law is
\begin{eqnarray}
\label{fbi6}
p(\rho)=\frac{T}{K} \ln(1+K\rho).
\end{eqnarray}
We recall that Eq. (\ref{fbi1}) can be obtained from the master
equation (\ref{kin2}) when the transition probabilities are of the
form (\ref{kin1}) with $a(\rho)=\rho$ and
$b(\rho)=1+K\rho$. Alternatively, the model (\ref{fbi5}) corresponds to
$a(\rho)=\rho/\sqrt{1+K\rho}$ and $b(\rho)=\sqrt{1+K\rho}$.

The NFP equations (\ref{fbi1}) and (\ref{fbi5}) have been introduced
by Kaniadakis \& Quarati \cite{kq} (see also \cite{csr} in the context of
the violent relaxation of 2D vortices and stellar systems). For $K=+1$,
they can provide a dynamical model of the Bose-Einstein condensation
in the canonical ensemble which has been studied in detail in
\cite{bose}. When coupled to the field equation (\ref{gle3}), the NFP
equations (\ref{fbi1}) and (\ref{fbi5}) could also provide generalized
Keller-Segel models of chemotaxis.

\subsection{Mixed model: anomalous diffusion and filling factor}
\label{sec_mixed}

The previous models focus individually on two important effects:
anomalous diffusion (see Secs. \ref{sec_pld}-\ref{sec_pldd}) and
exclusion constraints when the density becomes too large (see
Sec. \ref{sec_ff}). Here we introduce a mixed model which combines these two
effects in a single equation. If we take $h(\rho)=q\rho^{q+\mu-1}$ and
$g(\rho)=\rho^{\mu+1}(1-\rho/\sigma_{0})$, we obtain
\begin{eqnarray}
\label{mixed1}
\frac{\partial\rho}{\partial t}=\nabla\cdot \left ( Dq \rho^{q+\mu-1}\nabla\rho+
\chi \rho^{\mu+1}(1-\rho/\sigma_{0}) \nabla \Phi\right ).\nonumber\\
\end{eqnarray}
This corresponds to a power law diffusion such that
$D(\rho)=\lbrack{Dq}/({q+\mu})\rbrack \rho^{q+\mu-1}$ and a mobility
$\chi(\rho)=\chi\rho^{\mu}(1-\rho/\sigma_{0})$.  The associated
stochastic process is
\begin{eqnarray}
\label{mixed2}
\frac{d{\bf r}}{dt}=-\chi \rho^{\mu}(1-\rho/\sigma_{0})\nabla \Phi+\sqrt{\frac{2Dq}{q+\mu}}\rho^{\frac{q+\mu-1}{2}}{\bf R}(t).\nonumber\\
\end{eqnarray}
The generalized entropy corresponding to Eq. (\ref{mixed1}) is obtained by integrating twice the relation 
\begin{eqnarray}
\label{mixed3}
C''(\rho)=\frac{q\rho^{q-2}}{1-\rho/\sigma_{0}}.
\end{eqnarray}
A first integration gives
\begin{eqnarray}
\label{mixed4}
C'(\rho)=q\sigma_{0}^{q-1}\Phi_{q-2}\left(\frac{\rho}{\sigma_{0}}\right ),
\end{eqnarray}
where 
\begin{eqnarray}
\label{mixed5}
\Phi_{m}(t)=\int_{0}^{t}\frac{x^{m}}{1-x}dx.
\end{eqnarray}
Therefore, the generalized entropy can be expressed as
\begin{eqnarray}
\label{mixed6}
C(\rho)=q\sigma_{0}^{q}\int_{0}^{\rho/\sigma_{0}}\Phi_{q-2}(t)dt.
\end{eqnarray}
Note that it does not depend on $\mu$ since the term $\rho^{\mu}$ can
be put in factor of the diffusion current in Eq. (\ref{mixed1}); see
the discussion in Sec. \ref{sec_func}.

Let us consider some particular cases. (i) For $q=1$,
Eq. (\ref{mixed1}) has the same entropy and the same equilibrium
states as Eq. (\ref{ff1}). (ii) For $\sigma_{0}\rightarrow +\infty$,
we recover Eq. (\ref{pldd1}). (iii) For $\mu=0$ and $q=2$, we have
\begin{eqnarray}
\label{mixed7}
\frac{\partial\rho}{\partial t}=\nabla\cdot \left ( D \nabla\rho^{2}+
\chi \rho (1-\rho/\sigma_{0}) \nabla \Phi\right ).
\end{eqnarray}
The generalized entropy is 
\begin{eqnarray}
\label{mixed8}
S=-2\sigma_{0}^{2}\int  \left (1-\frac{\rho}{\sigma_{0}}\right)\ln \left (1-\frac{\rho}{\sigma_{0}}\right) d{\bf r},
\end{eqnarray}
and the stationary solution is 
\begin{eqnarray}
\label{mixed9}
\rho=\sigma_{0}\left\lbrack 1-e^{(\beta \Phi+\alpha)/2\sigma_{0}}\right\rbrack_{+}.
\end{eqnarray}
For $\sigma_{0}\rightarrow +\infty$, we recover Eq. (\ref{pld12}).
We can also consider the alternative model 
\begin{eqnarray}
\label{mixed10}
\frac{\partial\rho}{\partial t}=\nabla\cdot \left (\frac{2\rho D}{1-\rho/\sigma_{0}}\nabla\rho+
\chi \rho  \nabla\Phi\right ),
\end{eqnarray}
which has the same entropy and the same equilibrium states as Eq. (\ref{mixed7}). The pressure law is
\begin{eqnarray}
\label{mixed11}
p(\rho)=-2T\sigma_{0}^{2}\left\lbrack \ln(1-\rho/\sigma_{0})-\rho/\sigma_{0}\right \rbrack.
\end{eqnarray}
(iv) For $(\mu,q)=(0,0)$ and performing the transformation
$qD\rightarrow D$, or directly taking $h(\rho)=1/\rho$ and
$g(\rho)=\rho(1-\rho/\sigma_{0})$, we  obtain
\begin{eqnarray}
\label{mixed12}
\frac{\partial\rho}{\partial t}=\nabla\cdot \left ( D \nabla\ln\rho+
\chi \rho (1-\rho/\sigma_{0}) \nabla \Phi\right ).
\end{eqnarray}
This corresponds to a logarithmic diffusion and a modified mobility
taking into account an exclusion principle through the filling factor. The generalized entropy is obtained from the relation
\begin{eqnarray}
\label{mixed13}
C''(\rho)=\frac{1}{\rho^{2}(1-\rho/\sigma_{0})}, 
\end{eqnarray}
leading to
\begin{eqnarray}
\label{mixed14}
C'(\rho)=-\frac{1}{\sigma_{0}}\left\lbrace \ln\left (\frac{\sigma_{0}}{\rho}-1\right )+\frac{\sigma_{0}}{\rho}\right\rbrace,
\end{eqnarray}
and finally to the explicit expression
\begin{eqnarray}
\label{mixed15}
S=-\int \left (1-\frac{\rho}{\sigma_{0}}\right )\ln \left (\frac{\sigma_{0}}{\rho}-1\right ) d{\bf r}.
\end{eqnarray}
We can consider the alternative model 
\begin{eqnarray}
\label{mixed16}
\frac{\partial\rho}{\partial t}=\nabla\cdot \left \lbrack \frac{D}{\rho(1-\rho/\sigma_{0})}\nabla\rho+
\chi \rho  \nabla \Phi\right \rbrack,
\end{eqnarray}
with the same entropy and the same equilibrium states. The associated pressure law is
\begin{eqnarray}
\label{mixed17}
p(\rho)=-T\ln\left (\frac{\sigma_{0}}{\rho}-1\right ).
\end{eqnarray}

\subsection{$\kappa$-entropy}
\label{sec_kappa}

We consider the $\kappa$-entropy
\begin{eqnarray}
\label{kappa1}
S_{\kappa}=-\frac{1}{2\kappa}\int (\rho^{1+\kappa}-\rho^{1-\kappa})d{\bf r}.
\end{eqnarray}
This entropy was  introduced by Kaniadakis \cite{k1}. It can be written
\begin{eqnarray}
\label{kappa2}
S_{\kappa}=-\int \rho \ln_{(\kappa)}\rho \, d{\bf r},
\end{eqnarray}
with the $\kappa$-logarithm
\begin{eqnarray}
\label{kappa3}
\ln_{(\kappa)}(x)=\frac{1}{2\kappa}(x^{\kappa}-x^{-\kappa}).
\end{eqnarray}
We have 
\begin{eqnarray}
\label{kappa4}
C'(\rho)=\frac{1}{2\kappa}\left\lbrack (1+\kappa)\rho^{\kappa}-(1-\kappa)\rho^{-\kappa}\right\rbrack,
\end{eqnarray}
\begin{eqnarray}
\label{kappa5}
C''(\rho)=\frac{1}{2\rho}\left\lbrack (1+\kappa)\rho^{\kappa}+(1-\kappa)\rho^{-\kappa}\right\rbrack.
\end{eqnarray}
If we take $g(\rho)=\rho$ and $h(\rho)=\rho C''(\rho)$, we obtain the NFP equation
\begin{eqnarray}
\label{kappa6}
\frac{\partial\rho}{\partial t}=\nabla\cdot \left ( \frac{D}{2} \nabla (\rho^{1+\kappa}+\rho^{1-\kappa})+
\chi \rho \nabla \Phi\right ).
\end{eqnarray}
This corresponds to a power law diffusion $D(\rho)=\frac{D}{2}(\rho^{\kappa}+\rho^{-\kappa})$ and a constant
mobility $\chi(\rho)=\chi$.  The associated stochastic process is
\begin{eqnarray}
\label{kappa7}
\frac{d{\bf r}}{dt}=-\chi\nabla \Phi+\sqrt{D}(\rho^{\kappa}+\rho^{-\kappa})^{1/2} {\bf R}(t),
\end{eqnarray}
and the stationary solution of Eq. (\ref{kappa6}) can be written
\begin{eqnarray}
\label{kappa8}
\rho=\left (\frac{1-\kappa}{1+\kappa}\right )^{\frac{1}{2\kappa}}e_{(\kappa)}^{-(\beta \Phi+\alpha)/\sqrt{1-\kappa^{2}}},
\end{eqnarray}
with the $\kappa$-exponential
\begin{eqnarray}
\label{kappa9}
e_{(\kappa)}(x)=(\kappa x+\sqrt{1+\kappa^{2}x^{2}})^{1/\kappa}.
\end{eqnarray}
Finally, the pressure law is
\begin{eqnarray}
\label{kappa10}
p(\rho)=\frac{1}{2}T(\rho^{1+\kappa}+\rho^{1-\kappa}).
\end{eqnarray}
For $\kappa=0$, we recover
the standard model (\ref{sm1}). These results can be generalized to the $(\kappa,r)$ entropy 
\begin{eqnarray}
\label{kappa11}
S_{\kappa,r}=-\frac{1}{2\kappa}\int \rho^{r}(\rho^{1+\kappa}-\rho^{1-\kappa})d{\bf r},
\end{eqnarray}
which reduces in some special cases to the Tsallis \cite{tsallis}, Abe
\cite{abe} and Kaniadakis \cite{k1} entropies. The corresponding NFP
equation can be written
\begin{eqnarray}
\label{kappa12}
\frac{\partial\rho}{\partial t}=\nabla\cdot \left \lbrack \frac{D}{2\kappa} \nabla \left (a\rho^{1+a}
+b\rho^{1-b}\right )
+\chi \rho \nabla \Phi\right \rbrack,
\end{eqnarray}
with $a=\kappa+r$ and $b=\kappa -r$. Of course, we could give many
other examples of generalized Fokker-Planck equations since there
exists an infinite number of distributions and entropic
functionals. Therefore, we found it more convenient in \cite{gen} to
formulate the problem in a general setting, using an arbitrary
entropic functional of the form (\ref{h5}).

\section{Nonlinear mean field Fokker-Planck equations in phase space}
\label{sec_phase}

We now describe nonlinear mean field Fokker-Planck equations in phase
space taking into account the inertia of the particles. Overdamped
models will be recovered in a limit of strong friction.

\subsection{Generalized Kramers equation}
\label{sec_gk}

We consider a system of $N$  particles in interaction whose dynamics
is described by the stochastic Ito-Langevin equations
\begin{eqnarray}
\label{gk1} \frac{d{\bf r}_{i}}{dt}={\bf v}_{i},
\end{eqnarray}
\begin{eqnarray}
\label{gk2} \frac{d{\bf v}_{i}}{dt}=-\xi(f_i){\bf
v}_{i}-\nabla\Phi_i+\sqrt{2D(f_i)}{\bf R}_{i}(t),
\end{eqnarray}
where $\Phi({\bf r},t)$ is a self-consistent potential given by the
mean field Eq. (\ref{gle2}). In ordinary models, the friction $\xi$
and the diffusion coefficient $D$ are constant. In that case, the
statistical equilibrium state is the Maxwell-Boltzmann distribution $f
\sim e^{-\beta \epsilon}$ where $\epsilon=v^{2}/2+\Phi({\bf r})$ is
the individual energy and the temperature $T=1/\beta$ is given by the
Einstein relation $T=D/\xi$. Here, for sake of generality, the
friction coefficient $\xi(f)$ and the diffusion coefficient $D(f)$ are
allowed to depend on the distribution function $f({\bf r},{\bf
v},t)=\langle\sum_{i=1}^{N}\delta({\bf r}-{\bf r}_{i}(t))\delta({\bf
v}-{\bf v}_{i}(t))\rangle$. This can take into account microscopic
constraints that affect the dynamics and modify the equilibrium
distribution. The evolution of the distribution function $f({\bf
r},{\bf v},t)$ is governed by the nonlinear mean field Fokker-Planck
equation
\begin{eqnarray}
\label{gk3}
\frac{\partial f}{\partial t}+{\bf v}\cdot \frac{\partial f}{\partial {\bf r}}-\nabla\Phi\cdot \frac{\partial f}{\partial {\bf v}} =\frac{\partial}{\partial {\bf v}}\cdot \left\lbrack \frac{\partial}{\partial {\bf v}}(D(f)f)+\xi(f)f{\bf v}\right\rbrack,\nonumber\\
\end{eqnarray}
coupled to Eq. (\ref{gle2}). We introduce the notations
\begin{eqnarray}
\label{gk4} D h(f)=\frac{d}{df}(f D(f)), \qquad \xi g(f)=f\xi(f),
\end{eqnarray}
where $D$ and $\xi$ are positive constants and $h(f)$ and $g(f)$ are
positive functions. The ordinary model with constant diffusion
$D(f)=D$ and constant friction $\xi(f)=\xi$ is recovered for $h(f)=1$
and $g(f)=f$. With these notations, the NFP equation (\ref{gk3}) can be
rewritten in the form of a generalized Kramers (GK) equation
\begin{eqnarray}
\label{gk5} \frac{\partial f}{\partial t}+{\bf v}\cdot
\frac{\partial f}{\partial {\bf r}}-\nabla\Phi\cdot \frac{\partial
f} {\partial {\bf v}} =\frac{\partial}{\partial {\bf v}}\cdot \left
( D h(f)\frac{\partial f}{\partial {\bf v}}+\xi g(f){\bf v}\right
).\nonumber\\
\end{eqnarray}
It can be put in the conservative form
\begin{eqnarray}
\label{gk6}
\frac{df}{dt}=-\frac{\partial}{\partial {\bf v}}\cdot {\bf J},
\end{eqnarray}
where
\begin{eqnarray}
\label{gk7} {\bf J}=- \left \lbrack D h(f)\frac{\partial f}{\partial
{\bf v}}+\xi g(f){\bf v}\right \rbrack,
\end{eqnarray}
is a diffusion current. This structure guarantees the conservation
of mass $M=\int f d{\bf r}d{\bf v}$.

\subsection{Generalized free energy and H-theorem}
\label{sec_nh}

We define the energy by
\begin{eqnarray}
\label{nh1}
E=\frac{1}{2}\int f v^{2}\, d{\bf r}d{\bf v}+\frac{1}{2}\int \rho \Phi\, d{\bf r}=K+W,
\end{eqnarray} 
where $K$ is the kinetic energy and $W$ is the potential energy. We
define the temperature by
\begin{eqnarray}
\label{nh2}
T=\frac{D}{\xi}.
\end{eqnarray} 
The Einstein relation is preserved in the generalized thermodynamical
framework.  We introduce the generalized
entropic functional
\begin{eqnarray}
\label{nh3}
S=-\int C(f)\, d{\bf r}d{\bf v},
\end{eqnarray}
where $C(f)$ is a convex function defined by
\begin{eqnarray}
\label{nh4}
C''(f)=\frac{h(f)}{g(f)}.
\end{eqnarray}
Finally, we introduce the generalized free energy
\begin{eqnarray}
\label{nh5}
F=E-TS.
\end{eqnarray}
The definition of the free energy (Legendre transform) is preserved in
the generalized thermodynamical framework. Explicitly,
\begin{eqnarray}
\label{nh5b}
F[f]=\frac{1}{2}\int f v^{2}\, d{\bf r}d{\bf v}+\frac{1}{2}\int \rho \Phi\, d{\bf r}+T\int C(f)\, d{\bf r}d{\bf v}.\nonumber\\
\end{eqnarray}
A straightforward
calculation (see Appendix \ref{sec_ht}) shows that
\begin{eqnarray}
\label{nh6}
\dot F=
-\int \frac{1}{\xi g(f)}\left (Dh(f)\frac{\partial f}{\partial {\bf v}}+\xi g(f){\bf v}\right )^{2}d{\bf r}d{\bf v}.
\end{eqnarray}
Therefore, $\dot F\le 0$. We can also introduce the Massieu function
$J=S-\beta E$ which is related to the free energy by $J=-\beta F$. It
satisfies $\dot J\ge 0$. If $D=0$ (leading to $T=0$), we get $F=E$ so
that $\dot E\le 0$. If $\chi=0$ (leading to $\beta=0$), we get $J=S$
so that $\dot S\ge 0$.

\subsection{Stationary solution}
\label{sec_nss}

The steady states of Eq. (\ref{gk5}) must satisfy  $\dot F=0$. According to Eq. (\ref{nh6}), this  implies ${\bf J}={\bf 0}$ or explicitly
\begin{eqnarray}
\label{nss1}
Dh(f)\frac{\partial f}{\partial {\bf v}}+\xi g(f){\bf v}={\bf 0}.
\end{eqnarray}
Using Eqs. (\ref{nh2}) and (\ref{nh4}), we get
\begin{eqnarray}
\label{nss2}
C''(f)\frac{\partial f}{\partial {\bf v}}+\beta {\bf v}={\bf 0},
\end{eqnarray}
which can be integrated into
\begin{eqnarray}
\label{nss3} C'(f)=-\beta \left \lbrack \frac{v^{2}}{2}+\lambda({\bf
r})\right \rbrack,
\end{eqnarray}
where $\lambda({\bf r})$ is a function of the position. Since $C$ is
convex, this relation can be reversed to give
\begin{eqnarray}
\label{nss4} f({\bf r},{\bf v})=F\left \lbrack \beta \left (
\frac{v^{2}}{2}+\lambda({\bf r})\right )\right \rbrack,
\end{eqnarray}
where $F(x)=(C')^{-1}(-x)$ is a decreasing function. Since ${\bf J}={\bf 0}$ and $\partial f/\partial t=0$, the steady  solution of Eq. (\ref{gk5}) 
must also cancel the advective term
\begin{eqnarray}
\label{nss5}
{\bf v}\cdot \frac{\partial f}{\partial {\bf r}}-\nabla\Phi\cdot  \frac{\partial f}{\partial {\bf v}}=0.
\end{eqnarray}
In other words, the steady solution of Eq. (\ref{gk5}) is a particular
stationary solution of the Vlasov equation (l.h.s.) whose form is
selected by the ``collision'' term (r.h.s.). Substituting
Eq. (\ref{nss4}) in Eq. (\ref{nss5}), we get
\begin{eqnarray}
\label{nss6} (\nabla\lambda-\nabla\Phi)\cdot {\bf v}=0,
\end{eqnarray}
which must be true for all ${\bf v}$. 
This yields 
\begin{eqnarray}
\label{nss7} \lambda({\bf
r})=\Phi({\bf r})+\alpha/\beta,
\end{eqnarray}
where $\alpha$ is a constant of
integration. Therefore, the stationary solution of Eq. (\ref{gk5}) is given
by
\begin{eqnarray}
\label{nss8} C'(f)=-\beta \epsilon -\alpha,
\end{eqnarray}
or, equivalently,
\begin{eqnarray}
\label{nss9} f({\bf r},{\bf v})=F\lbrack \beta\epsilon({\bf r},{\bf v})+\alpha\rbrack,
\end{eqnarray}
where $\epsilon={v^{2}}/{2}+\Phi({\bf r})$ is the energy of a
particle. Thus, in the steady state, the distribution function
$f=f(\epsilon)$ is a function of the individual energy. The potential
$\Phi$ is determined by an integrodifferential equation obtained by
substituting Eq. (\ref{nss9}) in Eq.  (\ref{gle2}), using $\rho=\int f
d{\bf v}$. The constant $\alpha$ is determined by the conservation of
mass.  On the other hand, differentiating Eq. (\ref{nss8}), we obtain
\begin{eqnarray}
\label{nss10} \frac{df}{d\epsilon}=-\frac{\beta}{C''(f)}.
\end{eqnarray}
Since $C$ is convex, i.e. $C''>0$, the preceding relation implies that
$f'(\epsilon)<0$. Therefore, $f(\epsilon)$ is a decreasing function of
the energy.

\subsection{Minimum of free energy}
\label{sec_nmin}

The critical points of free energy at fixed mass are determined by the
variational problem
\begin{eqnarray}
\label{nmin1}
\delta F+T\alpha\delta M=0,
\end{eqnarray}
where $\alpha$ is a Lagrange multiplier. 
These variations give 
\begin{eqnarray}
\label{nmin2}
C'(f)=-\beta\epsilon-\alpha.
\end{eqnarray}
Therefore, comparing with Eq. (\ref{nss8}), we find that a stationary
solution of Eq. (\ref{gk5}) is a critical point of $F$ at fixed
mass. Furthermore, it is shown in \cite{gen} that a steady state of
Eq. (\ref{gk5}) is linearly dynamically stable iff it is a {\it
minimum} (at least local) of $F$ at fixed mass. In this sense,
dynamical and generalized thermodynamical stability in the canonical
ensemble coincide. This property also results from Lyapunov's direct
method \cite{frank}. Finally, if $F$ is bounded from below, we
conclude from the above properties that the system will converge to a
stable steady state for $t\rightarrow +\infty$ which is a (local)
minimum of $F[f]$ at fixed mass.  If several local minima exist, the
choice of the final steady state will depend on a complicated notion
of basin of attraction. In conclusion, we have the important
result: {\it a steady solution of the generalized Kramers equation
(\ref{gk5}) is linearly dynamically stable iff it is a (local) minimum
of the free energy $F[f]$ at fixed mass $M[f]=M$.} This corresponds to
the minimization problem:
\begin{eqnarray}
\label{nmin3}
\min_{f}\quad \lbrace F[f]\quad |\quad M[f]=M\rbrace.
\end{eqnarray}
Taking the second variations of $F$ and using Eq. (\ref{nss10}),  the
condition of dynamical stability is
\begin{eqnarray}
\label{nmin4} \delta^{2}F[\delta f]=-{1\over 2}\biggl\lbrace \int {(\delta
f)^{2}\over f'(\epsilon)}d{\bf  r}d{\bf v}-\int \delta\rho\delta\Phi
d{\bf  r}\biggr\rbrace \ge 0,\quad
\end{eqnarray}
for all perturbations $\delta f$ that conserve mass.

\subsection{Particular cases}
\label{sec_npc}

If we take $h(f)=1$ and $g(f)=1/C''(f)$ we get
\begin{eqnarray}
\label{npc1}
\frac{\partial f}{\partial t}+{\bf v}\cdot \frac{\partial f}{\partial {\bf r}}-\nabla\Phi\cdot  \frac{\partial f}{\partial {\bf v}}=\frac{\partial}{\partial {\bf v}}\cdot \left (D\frac{\partial f}{\partial {\bf v}}+\frac{\xi}{C''(f)}{\bf v}
\right ).\nonumber\\
\end{eqnarray}
In that case, we have a constant diffusion $D(f)=D$  and a variable friction $\xi(f)=\xi/[fC''(f)]$. If we take $g(f)=f$ and $h(f)=fC''(f)$ we get
\begin{eqnarray}
\label{npc2}
\frac{\partial f}{\partial t}+{\bf v}\cdot \frac{\partial f}{\partial {\bf r}}-\nabla\Phi\cdot  \frac{\partial f}{\partial {\bf v}}=\frac{\partial}{\partial {\bf v}}\cdot \left (DfC''(f)\frac{\partial f}{\partial {\bf v}}+\xi f{\bf v}
\right ).\nonumber\\
\end{eqnarray}
In that case, we have a constant friction $\xi(f)=\xi$  and a variable diffusion $D(f)=Df \lbrack C(f)/f\rbrack'$.

\subsection{Functional derivative}
\label{sec_nfun}

For a given free energy functional $F[f]$, we can introduce
phenomenologically a dynamical model by writing the evolution of the
distribution function as a continuity equation
$d_{t}f=-\frac{\partial}{\partial {\bf v}} \cdot {\bf J}$ where the
current is proportional to the gradient in velocity space of the
functional derivative of the free energy, i.e.
\begin{eqnarray}
\label{nfun1}
\frac{\partial f}{\partial t}+{\bf v}\cdot \frac{\partial f}{\partial {\bf r}}-\nabla\Phi\cdot  \frac{\partial f}{\partial {\bf v}}=\frac{\partial}{\partial {\bf v}}\cdot \left \lbrack \xi({\bf r},{\bf v},t)f \frac{\partial}{\partial {\bf v}}\frac{\delta F}{\delta f}\right \rbrack.\nonumber\\
\end{eqnarray}
For the free energy (\ref{nh5b}), we have  
\begin{eqnarray}
\label{nfun2}
\frac{\delta F}{\delta f}=TC'(f)+\frac{v^{2}}{2}+\Phi,
\end{eqnarray}
so that
\begin{eqnarray}
\label{nfun3}
\frac{df}{dt}=
\frac{\partial}{\partial {\bf v}}\cdot \left \lbrack \xi({\bf r},{\bf v},t)\left (T f C''(f) \frac{\partial f}{\partial {\bf v}}+f {\bf v}\right )\right\rbrack,
\end{eqnarray}
where we have introduced the material derivative
$d/dt=\partial/\partial t+{\bf v}\cdot \partial/\partial {\bf
r}-\nabla\Phi\cdot \partial/\partial {\bf v}$ in the advective term.
This equation is more general than Eq. (\ref{gk5}). It shows that, for
a given free energy, we can introduce an infinite class of NFP
equations where $\xi({\bf r},{\bf v},t)$ is an {\it arbitrary}
positive function of position, velocity and time. In particular, it
can be a function of $f({\bf r},{\bf v},t)$. If we set $\xi({\bf
r},{\bf v},t)=\xi g(f)/f$ we recover Eq. (\ref{gk5}). We can also
write Eq. (\ref{nfun3}) in the alternative form
\begin{eqnarray}
\label{nfun4}
\frac{df}{dt}=\frac{\partial}{\partial {\bf v}}\cdot \left \lbrack \tilde{\xi}({\bf r},{\bf v},t)\left (T \frac{\partial f}{\partial {\bf v}}+\frac{1}{C''(f)} {\bf v}\right )\right\rbrack, 
\end{eqnarray}
where $\tilde{\xi}({\bf r},{\bf v},t)$ is an {\it arbitrary} positive
function of position, velocity and time. If we set $\tilde{\xi}({\bf
r},{\bf v},t)=\xi h(f)$ we recover Eq. (\ref{gk5}). These two
alternative forms (\ref{nfun3}) and (\ref{nfun4}) were given in 
\cite{gen}. On the other
hand, the general structure of Eq. (\ref{nfun1}) implies an
$H$-theorem for the free energy ($\dot F\le 0$). The derivation is similar to that given in Sec. \ref{sec_func}.

\subsection{Equation of state}
\label{sec_eos}

The stationary solutions of the nonlinear Kramers equation (\ref{gk5})
are of the form $f=f(\epsilon)$ with $f'(\epsilon)<0$ where
$\epsilon={v^{2}}/{2}+\Phi({\bf r})$ is the energy of a particle. The
function $f$ is determined by the convex function $C$ according to
Eq. (\ref{nss8}). Therefore, at equilibrium, the density $\rho=\int f
d{\bf v}$ and the pressure $p=\frac{1}{d}\int f v^2 d{\bf v}$ can be
expressed as $\rho=\rho(\Phi({\bf r}))$ and $p=p(\Phi({\bf
r}))$. Eliminating the potential $\Phi({\bf r})$ between these
expressions, we obtain a barotropic equation of state $p=p(\rho)$
where the function $p(\rho)$ is entirely determined by the convex
function $C(f)$.  Furthermore, the condition that the distribution
function is a function $f=f(\epsilon)$ of the energy alone implies the
condition of hydrostatic balance. Indeed, we have
\begin{eqnarray}
\label{edd1} \nabla p={1\over d}\int f'(\epsilon)\nabla\Phi v^{2}d{\bf v}={1\over d}\nabla\Phi \int {\partial f\over\partial {\bf v}}\cdot {\bf v}d{\bf v}\nonumber\\
=-\nabla\Phi\int f d{\bf v}=-\rho\nabla\Phi.
\end{eqnarray}
The condition of
hydrostatic equilibrium can also be written $p'(\Phi)=-\rho(\Phi)$ or
\begin{eqnarray}
\label{edd2} p'(\rho)=-\frac{\rho}{\rho'(\Phi)}.
\end{eqnarray}
Let us introduce the free energy functional
\begin{eqnarray}
\label{strong12ng} F[\rho]=\int \rho\int^{\rho}{p(\rho')\over
\rho'^{2}} \,d \rho'd{\bf r}+{1\over 2}\int\rho\Phi d{\bf r}.
\end{eqnarray}
In Appendix \ref{sec_passage}, we show that this functional of $\rho$
can be deduced from the free energy functional $F[f]$ given by
Eq. (\ref{nh5}) by using the relation (\ref{nss9}) valid at
equilibrium. Furthermore, we
show in Appendix \ref{sec_equi} that the minimization problem
(\ref{nmin3}) is equivalent to the minimization problem
\begin{eqnarray}
\label{min3ng}
\min_{\rho}\quad \lbrace F[\rho]\quad |\quad M[\rho]=M\rbrace,
\end{eqnarray}
where $\rho({\bf r})$ is the density profile corresponding to the
distribution function $f({\bf r},{\bf v})$. This equivalence
considerably simplifies the study of the stability of a steady state
of the NFP equation (\ref{gk5}).  A critical point of $F[\rho]$ at
fixed mass satisfies the condition of hydrostatic balance. Indeed,
writing $\delta F-\alpha\delta M=0$, we have $\int^{\rho}[p'(\rho')/\rho']d\rho'+\Phi-\alpha=0$ implying $\nabla p+\rho\nabla\Phi={\bf 0}$. On the
other hand, taking the second variations of $F$ and using
Eq. (\ref{edd2}), the condition of stability can be written
\begin{eqnarray}
\label{min4ng} \delta^{2}F[\delta\rho]=-{1\over 2}\biggl\lbrace \int {(\delta
\rho)^{2}\over \rho'(\Phi)}d{\bf  r}-\int \delta\rho\delta\Phi
d{\bf  r}\biggr\rbrace \ge 0,
\end{eqnarray}
for all perturbations $\delta\rho$ that conserve mass.

\subsection{The strong friction limit}
\label{sec_strong}

In this section, we shall derive the generalized Smoluchowski equation
from the generalized Kramers equation in the strong friction limit
$\xi\rightarrow +\infty$. The general case where both the diffusion
coefficient and the friction coefficient depend on the distribution
function is treated in
\cite{lemou} by using a Chapman-Enskog expansion. Here, we restrict
ourselves to the generalized Kramers equation with constant friction
coefficient
\begin{eqnarray}
\label{strong1}
\frac{\partial f}{\partial t}+{\bf v}\cdot \frac{\partial f}{\partial {\bf r}}-\nabla\Phi\cdot  \frac{\partial f}{\partial {\bf v}}=\frac{\partial}{\partial {\bf v}}\cdot \left\lbrack \xi \left (TfC''(f)\frac{\partial f}{\partial {\bf v}}+f{\bf v}
\right )\right\rbrack.\nonumber\\
\end{eqnarray}
In that case, it is possible to develop a procedure simpler that the
Chapman-Enskog expansion (see \cite{banach}). Let us derive the
hierarchy of hydrodynamic equations associated with
Eq. (\ref{strong1}). Defining the density and the local velocity by
\begin{eqnarray}
\label{strong2} \rho=\int f\,d{\bf v}, \qquad \rho{\bf u}=\int f{\bf
v}\,d{\bf v},
\end{eqnarray}
and integrating Eq.~(\ref{strong1}) on velocity, we get the continuity equation
\begin{eqnarray}
\label{strong3} {\partial\rho\over\partial t}+\nabla\cdot (\rho{\bf u})=0.
\end{eqnarray}
Next, multiplying Eq.~(\ref{strong1}) by ${\bf v}$ and integrating on
velocity, we obtain the momentum equation
\begin{eqnarray}
\label{strong4} {\partial\over\partial t}(\rho
u_{i})+{\partial\over\partial x_{j}}(\rho u_{i}u_{j})= -{\partial
P_{ij}\over\partial x_{j}}-\rho{\partial\Phi\over\partial
x_{i}}-\xi\rho u_i,\nonumber\\
\end{eqnarray}
where we have defined the pressure tensor
\begin{eqnarray}
\label{strong5}P_{ij}=\int fw_{i}w_{j}\,d{\bf v},
\end{eqnarray}
where ${\bf w}={\bf v}-{\bf u}$ is the relative velocity. Using Eq. (\ref{strong3}), the momentum equation can be rewritten in the form
\begin{eqnarray}
\label{strong6} \rho \left ({\partial u_{i}\over\partial t}+u_{j}{\partial u_{i}\over\partial x_{j}}\right )= -{\partial
P_{ij}\over\partial x_{j}}-\rho{\partial\Phi\over\partial
x_{i}}-\xi\rho u_i.\nonumber\\
\end{eqnarray}

We now consider the strong friction limit $\xi\rightarrow +\infty$
with fixed $T$. Since the term in parenthesis in Eq. (\ref{strong1})
must vanish to leading order, we find that the out of equilibrium
distribution function $f_0({\bf r},{\bf v},t)$ is given by
\begin{eqnarray}
\label{strong7} C'(f_0)=-\beta\left\lbrack \frac{v^{2}}{2}+\lambda({\bf
r},t)\right\rbrack+O(\xi^{-1}),
\end{eqnarray}
where $\lambda({\bf r},t)$ is a constant of integration that is
determined by the density according to
\begin{eqnarray}
\label{strong8} \rho({\bf r},t)=\int f_{0}d{\bf v}=\rho[\lambda({\bf
r},t)].
\end{eqnarray}
Note that the distribution function $f_0$ is {\it isotropic} so that
the velocity ${\bf u}({\bf r},t)=O(\xi^{-1})$ and the pressure
tensor  $P_{ij}=p\delta_{ij}+O(\xi^{-1})$ where $p$ is given by
\begin{eqnarray}
\label{strong9} p({\bf r},t)=\frac{1}{d}\int f_{0}v^{2}d{\bf
v}=p[\lambda({\bf r},t)].
\end{eqnarray}
Eliminating $\lambda({\bf r},t)$ between the two expressions
(\ref{strong8}) and (\ref{strong9}), we find that the fluid is {\it
barotropic} with an equation of state $p=p(\rho)$ entirely determined
by the function $C(f)$. Of course, this is the same equation of state
as the one obtained at equilibrium (see Sec.
\ref{sec_eos}). Now, considering the momentum equation (\ref{strong4}) in
the limit $\xi\rightarrow +\infty$, we find that
\begin{eqnarray}
\label{strong10} \rho{\bf u}=-\frac{1}{\xi}(\nabla
p+\rho\nabla\Phi)+O(\xi^{-2}).
\end{eqnarray}
Inserting this relation in the continuity equation  (\ref{strong3}), we obtain
the generalized Smoluchowski equation \cite{gen,banach,lemou}:
\begin{eqnarray}
\label{strong11}\frac{\partial\rho}{\partial t}=\nabla\cdot
\left\lbrack \frac{1}{\xi}(\nabla p+\rho\nabla\Phi)\right\rbrack.
\end{eqnarray}
The free energy associated to this equation is
\begin{eqnarray}
\label{strong12} F[\rho]=\int \rho\int^{\rho}{p(\rho')\over
\rho'^{2}} \,d \rho'd{\bf r}+{1\over 2}\int\rho\Phi d{\bf r}.
\end{eqnarray}
It can be deduced from the free energy (\ref{nh5}) by using
Eq. (\ref{strong7}) to express $F[f]$ as a functional $F[\rho]=F[f_0]$
of the density (see Appendix \ref{sec_passage}).  A direct calculation leads to the
$H$-theorem
\begin{eqnarray}
\label{strong13} \dot F=
-\int \frac{1}{\xi\rho}(\nabla p+\rho\nabla \Phi)^{2}d{\bf r}\le 0.
\end{eqnarray}
The stationary solutions of the generalized Smoluchowski equation
(\ref{strong11}) are critical points of free energy at fixed
mass. They satisfy the condition of hydrostatic balance
\begin{eqnarray}
\label{strong14}\nabla p+\rho\nabla\Phi={\bf 0}.
\end{eqnarray}
From Lyapunov's direct method, we conclude that a steady state of the
generalized Smoluchowski equation (\ref{strong11}) is linearly
dynamically stable iff it is a (local) minimum of $F$ at fixed mass
$M$. This corresponds to the minimization problem (\ref{min3ng}).

The condition of hydrostatic balance (\ref{strong14}) only holds at
equilibrium.  In the strong friction limit $\xi\rightarrow +\infty$,
the out-of-equilibrium distribution is of the form
$f_{0}=f_{0}(\epsilon')$ with $\epsilon'={v^{2}}/{2}+\lambda({\bf
r},t)$. Taking the gradient of Eq. (\ref{strong9}) and using a
procedure similar to that followed in Eq. (\ref{edd1}) with $\lambda({\bf
r},t)$ in place of $\Phi({\bf r})$, we obtain
\begin{eqnarray}
\label{strong15} \nabla p=-\rho\nabla\lambda.
\end{eqnarray}
Since $p=p(\rho)$ and $\lambda=\lambda(\rho)$,  this can be
rewritten
\begin{eqnarray}
\label{strong16} \lambda'(\rho)=-\frac{p'(\rho)}{\rho},
\end{eqnarray}
so that the out-of-equilibrium chemical potential $\lambda({\bf r},t)$
is given by
\begin{eqnarray}
\label{strong17} \lambda(\rho)=-\int^{\rho}\frac{p'(x)}{x}dx.
\end{eqnarray}
At equilibrium, comparing Eqs. (\ref{strong15}) and (\ref{strong14}),
we have $\lambda({\bf r})=\Phi({\bf r})+\alpha/\beta$ and Eq. (\ref{strong7}) leads to Eq. (\ref{nss8}).

\subsection{The damped Euler equations}
\label{sec_damped}

The generalized Smoluchowski equation (\ref{strong11}) can also be
obtained formally from the damped Euler \footnote{Hydrodynamical Euler
equations involving a friction force $-\xi(|{\bf u}|) {\bf u}$ have appeared in
various contexts for different reasons. We may mention, for example,
bottom-wall friction in 2D turbulence \cite{pt}, frictional force of
air on a turbulent soap film \cite{rw}, effective dynamical friction
in the process of violent relaxation for collisionless stellar systems
\cite{csr} and Epstein or Stokes friction laws for the dynamics of dust
particles in the solar nebula \cite{aa}.}  equations \cite{gen}:
\begin{eqnarray}
\label{damped1} {\partial\rho\over\partial t}+\nabla\cdot (\rho{\bf u})=0,
\end{eqnarray}
\begin{eqnarray}
\label{damped2}\frac{\partial {\bf u}}{\partial t}+({\bf u}\cdot \nabla){\bf u}=-\frac{1}{\rho}\nabla p-\nabla\Phi-\xi {\bf u}.
\end{eqnarray}
The Lyapunov functional associated with the damped Euler equations is the generalized free energy
\begin{eqnarray}
\label{ht1grt}
F[\rho,{\bf u}]=\int \rho\int^{\rho}{p(\rho')\over
\rho'^{2}} \,d \rho'd{\bf r}+{1\over 2}\int\rho\Phi d{\bf r}+\int\rho \frac{{\bf u}^{2}}{2}d{\bf r}.\nonumber\\
\end{eqnarray}
It satisfies an $H$-theorem of the form (see Appendix \ref{sec_ht}):
\begin{eqnarray}
\label{ht1ff}
\dot F=-\int \xi\rho {\bf u}^2  d{\bf r}\le 0.
\end{eqnarray}
A steady state of the damped Euler equations (\ref{damped1})-(\ref{damped2}) satisfies the condition of hydrostatic equilibrium (\ref{strong14}). Furthermore, from Lyapunov's direct method, it is linearly dynamically stable iff it is a minimum of the free energy (\ref{ht1grt}) at fixed mass. This corresponds to the minimization problem 
\begin{eqnarray}
\label{fdr}
\min_{\rho,{\bf u}}\quad \lbrace F[\rho,{\bf u}]\quad |\quad M[\rho]=M\rbrace,
\end{eqnarray}
The friction coefficient in Eq. (\ref{damped2}) measures the
importance of inertial effects. For $\xi=0$, we recover the usual
barotropic Euler equations. Alternatively, if we consider the
strong friction limit $\xi\rightarrow +\infty$, we can neglect the
inertial term in Eq. (\ref{damped2}) and we get
\begin{eqnarray}
\label{damped3}\xi {\bf u}=-\frac{1}{\rho}\nabla p-\nabla\Phi+O(\xi^{-1}).
\end{eqnarray}
Substituting this relation in the continuity equation (\ref{damped1}),
we obtain the generalized Smoluchowski equation
\begin{eqnarray}
\label{damped4}\frac{\partial\rho}{\partial t}=\nabla\cdot
\left\lbrack \frac{1}{\xi}(\nabla p+\rho\nabla\Phi)\right\rbrack.
\end{eqnarray}
The physical justification of the damped Euler equations
(\ref{damped1})-(\ref{damped2}) is not clear. They can be obtained
from Eqs. (\ref{strong3})-(\ref{strong4}) if we close the hierarchy by
invoking a local thermodynamic equilibrium (L.T.E.)  condition
\cite{gen}. However, the rigorous justification of this
L.T.E. condition is not established, so this approach remains
heuristic. Nevertheless, hydrodynamic equations (hyperbolic models) of
the form (\ref{damped1})-(\ref{damped2}) have been proposed in the
context of chemotaxis to describe the organization of endothelial
cells
\cite{gamba,filbet,csbio}. They lead to the formation of filaments
that are interpreted as the beginning of a vasculature. These
filaments are not obtained in the Keller-Segel model (parabolic model)
which leads to point-wise blow up or round aggregates
\cite{horstmann,sc}.

\section{Explicit examples}
\label{sec_ee}

In this section, we give explicit examples showing the passage from
the generalized Kramers equation to the generalized Smoluchowski
equation in the strong friction limit. 

\subsection{Isothermal systems: Boltzmann entropy}
\label{sec_is}

If we consider the Boltzmann entropy
\begin{equation}
\label{is1} S_{B}[f]=-\int f\ln f d{\bf r}d{\bf v},
\end{equation}
we get the ordinary Kramers equation
\begin{eqnarray}
\label{is2}
\frac{df}{dt}=\frac{\partial}{\partial {\bf v}}\cdot\left \lbrack \xi \left (T\frac{\partial f}{\partial {\bf v}}+f{\bf v}
\right )\right\rbrack.
\end{eqnarray}
The stationary state is the isothermal (Maxwell-Boltzmann)
distribution
\begin{equation}
\label{is3} f=A e^{-\beta\epsilon},
\end{equation}
where $A$ is determined by the conservation of mass. The equation
of state is the isothermal one
\begin{equation}
\label{is4} p=\rho T.
\end{equation}
In the strong
friction limit, we obtain the ordinary Smoluchowski equation
\begin{eqnarray}
\label{is5} {\partial\rho\over\partial t}=\nabla \cdot \biggl\lbrack
{1\over\xi}(T\nabla \rho+\rho\nabla\Phi)\biggr\rbrack.
\end{eqnarray}
The corresponding free energy is the Boltzmann free energy 
\begin{eqnarray}
{F}[\rho]=T\int \rho \ln\rho\ d{\bf r}+{1\over
  2}\int\rho\Phi \ d{\bf r},
\label{is6}
\end{eqnarray}
and the stationary solution
is the Boltzmann distribution
\begin{equation}
\label{is7} \rho=A' e^{-\beta\Phi}
\end{equation}
where $A'=(2\pi/\beta)^{d/2} A$.

\subsection{Polytropes: Tsallis entropy}
\label{sec_poly}

If we consider the Tsallis $q$-entropy
\begin{equation}
\label{poly1} S_{q}[f]=-{1\over q-1}\int (f^{q}-f)  d{\bf r} d{\bf
v},
\end{equation}
we get the polytropic Kramers equation
\begin{eqnarray}
\label{poly2}
\frac{df}{dt}=\frac{\partial}{\partial {\bf v}}\cdot\left \lbrack \xi \left (T\frac{\partial f^{q}}{\partial {\bf v}}+f{\bf v}
\right )\right\rbrack.
\end{eqnarray}
The stationary state is the 
polytropic distribution
\begin{equation}
\label{poly3} f=\biggl\lbrack \mu-{(q-1)\beta\over
q}\epsilon\biggr\rbrack_{+}^{1\over q-1},
\end{equation}
where $\mu$ is determined by the conservation of mass.  The index $n$
of the polytrope is related to the parameter $q$ by the relation
\begin{equation}
\label{poly4} n=\frac{d}{2}+\frac{1}{q-1}.
\end{equation}
Isothermal distribution functions are recovered in the limit
$q\rightarrow 1$ (i.e. $n\rightarrow +\infty$). We shall consider
$q>0$ so that $C$ is convex. We have to distinguish two cases. (i) For
$q>1$, i.e. $n>d/2$, the distribution has a compact support since $f$
is defined only for $\epsilon\le \epsilon_{m}\equiv \mu
q/[|q-1|\beta]$ (it vanishes at $\epsilon=\epsilon_{m}$). For
$\epsilon\ge \epsilon_{m}$, we set $f=0$. For $q\rightarrow +\infty$,
i.e. $n=d/2$, $f$ is the Heaviside function. (ii) For $q<1$, the
distribution is defined for all energies. For large velocities, it
behaves like $f\sim v^{-(d-2n)}$. Therefore, the density and the pressure
are finite only for $n<-1$, i.e. $d/(d+2)<q<1$. Therefore the range of
allowed parameters are
\begin{equation}
\label{poly5} q>1, \qquad n>\frac{d}{2}\quad ({\rm case}\ 1),
\end{equation}
\begin{equation}
\label{poly6} \frac{d}{d+2}<q<1, \qquad n<-1 \quad ({\rm case}\ 2).
\end{equation}
The distribution function (\ref{poly3}) leads to the polytropic
equation of state (see Appendix \ref{sec_pol})
\begin{equation}
\label{poly7} p=K\rho^{\gamma}, \qquad \gamma=1+{1\over n}.
\end{equation}
For $n>d/2$ the polytropic
constant is
\begin{equation}
\label{poly8} K=\frac{1}{n+1}\left\lbrack A S_{d}
2^{\frac{d}{2}-1}\frac{\Gamma\left (d/2\right )\Gamma\left
(1-d/2+n\right )}{\Gamma(1+n)}\right \rbrack^{-1/n},
\end{equation}
and for $n<-1$, we have
\begin{equation}
\label{poly9} K=-\frac{1}{n+1}\left\lbrack A S_{d}
2^{\frac{d}{2}-1}\frac{\Gamma\left (d/2\right )\Gamma\left (-n\right
)}{\Gamma(d/2-n)}\right \rbrack^{-1/n},
\end{equation}
where $A=(\beta |q-1|/q)^{1/(q-1)}$. In the strong friction limit, we get the polytropic
Smoluchowski equation
\begin{eqnarray}
\label{poly10} {\partial\rho\over\partial t}=\nabla \cdot \biggl\lbrack
{1\over\xi}(K\nabla \rho^{\gamma}+\rho\nabla\Phi)\biggr\rbrack.
\end{eqnarray}
The generalized free energy is the Tsallis free energy
\begin{eqnarray}
{F}[\rho]={K\over\gamma -1}\int (\rho^{\gamma}-\rho) \ d{\bf
r}+{1\over
  2}\int\rho\Phi \ d{\bf r},
\label{poly11}
\end{eqnarray}
and the stationary solution is the polytropic distribution
\begin{equation}
\label{poly12} \rho=\biggl\lbrack \lambda-{\gamma-1\over
K\gamma}\Phi\biggr\rbrack^{1\over\gamma-1}.
\end{equation}
Other useful relations valid for polytropic distributions are given in
\cite{cst}.  We note that a polytropic distribution with index $q$ in phase
space yields a polytropic distribution with index
$\gamma=1+2(q-1)/\lbrack 2+d(q-1)\rbrack$ in physical space. In this
sense, Tsallis distributions are stable laws since the functions
$f(\epsilon)$ and $\rho(\Phi)$ have a similar structure. By comparing
Eq. (\ref{poly3}) to Eq. (\ref{poly12}) or Eq. (\ref{nh5}) with
Eq. (\ref{poly1}) to Eq. (\ref{poly11}), we note that $K$ plays the
same role in physical space as the temperature $T=1/\beta$ in phase
space. It is sometimes called a ``polytropic temperature''.

Quite generally, we define the local kinetic temperature $T({\bf r})$
by $\frac{d}{2}T({\bf r})=\frac{1}{2}\langle v^{2}\rangle$ or
equivalently $p({\bf r})=\rho({\bf r})T({\bf r})$. It is proportional
to the velocity dispersion. Since, at equilibrium, $\rho=\rho(\Phi)$
and $p=p(\Phi)$, we conclude that $T({\bf r})=T[\Phi({\bf r})]$ is a
function of the potential $\Phi$. This is true for any barotropic
fluid at equilibrium. Now, for a polytropic distribution, using
Eq. (\ref{poly7}), we have $T({\bf r})=K\rho^{\gamma-1}$. Then, using
Eq. (\ref{poly12}) we obtain $T({\bf
r})=K\lambda-\frac{\gamma-1}{\gamma}\Phi({\bf r})$ so that $T$ is a
linear function of $\Phi$ with a gradient
$(\gamma-1)/\gamma=1/(n+1)=2(q-1)/[(d+2)q-d]$ \cite{cst,du}.

\subsection{Fermions: Fermi-Dirac entropy}
\label{sec_fermi}

If we consider the Fermi-Dirac entropy
\begin{equation}
\label{fermi1} S_{FD}[f]=-\eta_{0}\int \biggl\lbrace {f\over \eta_{0}}\ln
{f\over \eta_{0}}+\biggl (1-{f\over \eta_{0}}\biggr )\ln \biggl
(1-{f\over \eta_{0}}\biggr ) \biggr\rbrace d{\bf r} d{\bf v},
\end{equation}
we get the fermionic Kramers equation
\begin{eqnarray}
\label{fermi2}
\frac{df}{dt}
=\frac{\partial}{\partial {\bf v}}\cdot\left \lbrack \xi \left
(-T\eta_{0}\frac{\partial}{\partial {\bf v}}\ln\left
(1-\frac{f}{\eta_{0}}\right )+f{\bf v}
\right )\right\rbrack.
\end{eqnarray}
The corresponding equation with normal diffusion and nonlinear friction is
\begin{eqnarray}
\label{fermi2bis}
\frac{df}{dt}
=\frac{\partial}{\partial {\bf v}}\cdot\left \lbrack D \left
(\frac{\partial f}{\partial {\bf v}}+\beta f(1-f/\eta_0){\bf v}
\right )\right\rbrack.
\end{eqnarray} 
The stationary state is the Fermi-Dirac distribution function
\begin{equation}
\label{fermi3} f={\eta_{0}\over 1+\lambda e^{\beta\epsilon}},
\end{equation}
where $\lambda>0$ is determined by the conservation of mass.  The
Fermi-Dirac distribution function (\ref{fermi3}) satisfies the constraint
$f\le \eta_{0}$ which is related to the Pauli exclusion principle in
quantum mechanics. The isothermal distribution function (\ref{is3}) is
recovered in the non-degenerate limit $f\ll \eta_{0}$. This is the
case when $\lambda\rightarrow +\infty$, valid at high temperatures
$T\rightarrow +\infty$. On the other hand, in the completely
degenerate limit, the distribution is a step function
$f=\eta_{0}H(\epsilon-\epsilon_{F})$ where
$\epsilon_{F}=-\frac{1}{\beta}\ln\lambda$ is the Fermi energy. This is the
case when $\lambda\rightarrow 0$, valid at low temperatures
$T\rightarrow 0$. This limiting distribution corresponds to a polytrope with
index $n=d/2$.  The distribution in physical space, obtained by
integrating the Fermi-Dirac statistics (\ref{fermi3}) on the velocity, can be
written
\begin{equation}
\label{fermi4} \rho={\eta_{0}S_{d}2^{{d\over
2}-1}\over\beta^{d/2}}I_{{d\over 2}-1}(\lambda e^{\beta\Phi}),
\end{equation}
where $I_{n}$ is the Fermi integral
\begin{equation}
\label{fermi5} I_{n}(t)=\int_{0}^{+\infty}{x^{n}\over 1+te^{x}}dx.
\end{equation}
The  quantum equation of state for fermions is given in parametric form by
\begin{equation}
\label{fermi6} \rho={\eta_{0}S_{d}2^{{d\over
2}-1}\over\beta^{d/2}}I_{{d\over 2}-1}(t),\qquad
p={\eta_{0}S_{d}2^{{d\over 2}}\over d\beta^{{d\over 2}+1}}I_{{d\over
2}}(t).
\end{equation}
At high temperatures ($t\rightarrow +\infty$) we recover the classical
isothermal law $p=\rho T$ and at low temperatures ($t\rightarrow 0$)
we get a polytropic equation of state $p=K\rho^{\gamma}$ with
$\gamma=(d+2)/2$ (i.e. $n=d/2$) and
$K=\frac{1}{d+2}(\frac{d}{\eta_{0}S_{d}})^{2/d}$
\cite{fermid}.  In the strong friction limit, we get the fermionic
Smoluchowski equation (\ref{strong11}) where the equation of state is
given by (\ref{fermi6}). The fermionic Smoluchowski-Poisson system has been
studied in \cite{csr,crrs,bln}.

\section{Conclusion}
\label{sec_conc}

In this paper, we have studied a general class of nonlinear mean field
Fokker-Planck equations \cite{gen} associated with a formalism of
effective generalized thermodynamics (E.G.T.).  We have given several
physical examples of application and we have shown that NFP equations
can provide generalizations of the standard Keller-Segel model
describing the chemotaxis of biological populations.  The main
properties of these NFP equations are valid for a large class of
entropic functionals, encompassing the Boltzmann, the Fermi-Dirac, the
Bose-Einstein and the Tsallis statistics. Indeed, the rich
mathematical structure of these equations is almost independent on the
precise form of the entropy. These results should therefore evidence
which properties in statistical mechanics are specific to the standard
Boltzmann entropy and which properties are valid for a larger class of
entropies. The distinguished feature of the Boltzmann entropy is that
it can be obtained from a combinatorial analysis, assuming that all
the microstates are equiprobable. However, equiprobability of the
microstates is a strong postulate and it is not clear whether it has a
universal scope. The universality of the Boltzmann entropy has been
criticized long ago by Einstein \cite{ein10} who argued that the
statistics applicable on a system depends on its underlying dynamics
(see discussion in \cite{cohen}). For example, the Boltzmann
distribution can be obtained from a stochastic process describing a
classical random walk where the kicks have uniform sizes. However,
different distributions emerge when the stochastic process describes a
{\it biased} random walk where the kicks depend on the region where
the particle happens to be. We think that this is the case in many
physical systems. This results in a very complex geometrical structure
of phase space (fractal, multi-fractal,...) leading to
non-Boltzmannian distributions at equilibrium. Indeed, in such
circumstances, the microstates are {\it not} equiprobable since the
system prefers certain regions of phase space rather than others. It
would be interesting to derive the corresponding generalized entropies
(Tsallis, Abe, Kaniadakis,...) directly from a combinatorial
analysis. For example, Tsallis entropy could be the natural entropy on
a fractal phase-space. The selection of the entropy demands a complete
specification of the microdynamics of the system in agreement with the
statement given long ago by Einstein
\cite{ein10}.

\appendix

\section{The isotropic BGK operator}
\label{sec_iso}

Let us consider a simple kinetic equation where the generalized Fokker-Planck
operator in Eq. (\ref{gk5}) is replaced by a generalized isotropic BGK
operator. This equation has been introduced in Appendix A of
\cite{lemou} and we provide here some complements. The kinetic
equation is
\begin{eqnarray}
\label{iso1} \frac{\partial f}{\partial t}+{\bf v}\cdot
\frac{\partial f}{\partial {\bf r}}-\nabla\Phi\cdot \frac{\partial
f}{\partial {\bf v}} =-\frac{f-f_0}{\tau},
\end{eqnarray}
where $f_{0}({\bf r},{\bf v},t)$ is given by
\begin{eqnarray}
\label{iso2} C'(f_0)=-\beta\left\lbrack \frac{v^{2}}{2}+\lambda({\bf
r},t)\right\rbrack,
\end{eqnarray}
or
\begin{eqnarray}
\label{iso3} f_0=F\left\lbrace \beta\left\lbrack
\frac{v^{2}}{2}+\lambda({\bf r},t)\right\rbrack\right\rbrace,
\end{eqnarray}
with $F(x)=(C')^{-1}(-x)$. The function $\lambda({\bf r},t)$ is
determined by the density by writing $\rho=\int f_{0}d{\bf
v}=\rho(\lambda)$. First, we show that Eq. (\ref{iso1}) admits an
H-theorem for the free energy (\ref{nh5}). Recalling that the left hand side
(Vlasov term) conserves the energy and the Casimirs, hence $F$, we find that
\begin{eqnarray}
\label{iso4} \dot F=-\int \left (TC'(f)+\frac{v^{2}}{2}\right
)\frac{f-f_0}{\tau}d{\bf r}d{\bf v}.
\end{eqnarray}
Using Eq. (\ref{iso2}) and the fact that $\int f_{0}d{\bf v}=\int f d{\bf v}$,
we have the identity
\begin{eqnarray}
\label{iso5}
\int \left (TC'(f_0)+\frac{v^{2}}{2}\right )\frac{f-f_0}{\tau}d{\bf r}d{\bf v}\nonumber\\
=-\beta\int \lambda({\bf r},t)\frac{f-f_0}{\tau}d{\bf r}d{\bf v}=0.
\end{eqnarray}
Therefore, we can rewrite Eq. (\ref{iso4}) in the form
\begin{eqnarray}
\label{iso6} \dot F=-T\int \left \lbrack C'(f)-C'(f_0)\right \rbrack
\frac{f-f_0}{\tau}d{\bf r}d{\bf v}.
\end{eqnarray}
Since $C$ is convex, we have 
\begin{eqnarray}
\label{iso7} \left \lbrack C'(f)-C'(f_0)\right \rbrack ({f-f_0})\ge 0,
\end{eqnarray}
so that $\dot F\le 0$. On the other hand, using a procedure similar to
that described in Sec. \ref{sec_nss}, it is straightforward to prove
that the steady states of Eq. (\ref{iso1}) are given by
Eq. (\ref{nss8}) and that a steady state of Eq. (\ref{iso1}) is
dynamically stable iff it is a (local) minimum of $F$ at fixed mass.
The first two hydrodynamic equations associated with Eq. (\ref{iso1})
are
\begin{eqnarray}
\label{iso8} {\partial\rho\over\partial t}+\nabla\cdot (\rho{\bf
u})=0,
\end{eqnarray}
\begin{eqnarray}
\label{iso9} {\partial\over\partial t}(\rho
u_{i})+{\partial\over\partial x_{j}}(\rho u_{i}u_{j})= -{\partial
P_{ij}\over\partial x_{j}}-\rho{\partial\Phi\over\partial
x_{i}}-\frac{1}{\tau}\rho u_{i}.\nonumber\\
\end{eqnarray}
In the limit $\tau\rightarrow 0$, we can repeat the same arguments as
in Sec. \ref{sec_strong} and we obtain the generalized Smoluchowski equation
\begin{eqnarray}
\label{iso10}\frac{\partial\rho}{\partial t}=\nabla\cdot
\left\lbrack \tau (\nabla p+\rho\nabla\Phi)\right\rbrack.
\end{eqnarray}
This equation can also be obtained from a Chapman-Enskog
expansion \cite{lemou}. Rigorous mathematical results have been obtained
recently in \cite{dolbeault}.

\section{Connection between dynamical and thermodynamical stability for nonlinear mean field Fokker-Planck equations}
\label{sec_conn}

Let us consider a small perturbation $\delta\rho$ around a stationary
solution of the nonlinear mean field Fokker-Planck equation
(\ref{dde3}). We write the time dependence of the perturbation as
$\delta\rho\sim e^{\lambda t}$. It can be shown that $\lambda$ is real,
so that the perturbation is damped exponentially for $\lambda<0$
(stable) or increases exponentially for $\lambda>0$ (unstable). The
linearized Fokker-Planck equation can be written
\begin{eqnarray}
\lambda\delta\rho=-\nabla\cdot \delta{\bf J}.
\label{conn1}
\end{eqnarray}
If we multiply Eq. (\ref{conn1}) by $C''(\rho)\delta\rho$ and integrate the resulting expression over the volume, we get
\begin{eqnarray}
\lambda\int C''(\rho)(\delta\rho)^{2}d{\bf r}=\int \delta{\bf J}\cdot
 \left\lbrack C''(\rho)\nabla\delta\rho+C'''(\rho)\delta\rho\nabla\rho\right\rbrack d{\bf r},\nonumber\\
\label{conn2}
\end{eqnarray}
where we have used an integration by parts in the r.h.s. Now, the
linear variation of the current (\ref{dde5}) around equilibrium can be written
\begin{eqnarray}
-\frac{\delta {\bf J}}{D}=h'(\rho)\delta\rho\nabla\rho+h(\rho)\nabla\delta\rho+\beta g'(\rho)\delta\rho\nabla\Phi+\beta g(\rho)\nabla\delta\Phi.\nonumber\\
\label{conn3}
\end{eqnarray}
Using $h(\rho)=C''(\rho)g(\rho)$ and the relation
\begin{eqnarray}
C''(\rho)\nabla\rho=-\beta\nabla\Phi,
\label{conn4}
\end{eqnarray}
resulting from Eq. (\ref{sta3}), we obtain 
\begin{eqnarray}
-\frac{\delta {\bf J}}{Dg(\rho)}=C'''(\rho)\delta\rho\nabla\rho+C''(\rho)\nabla\delta\rho+\beta\nabla\delta\Phi.\nonumber\\
\label{conn5}
\end{eqnarray}
Then, Eq. (\ref{conn2}) can be rewritten
\begin{eqnarray}
\lambda\int C''(\rho)(\delta\rho)^{2}d{\bf r}+\beta\int \delta{\bf J}\cdot \nabla\delta\Phi d{\bf r}=-\int \frac{(\delta{\bf J})^{2}}{Dg(\rho)}d{\bf r}.\nonumber\\
\label{conn6}
\end{eqnarray}
Now, multiplying Eq. (\ref{conn1}) by $\delta\Phi$ and integrating
over the volume we find that
\begin{eqnarray}
\lambda\int \delta\rho \delta\Phi d{\bf r}=\int \delta{\bf J}\cdot \nabla\delta\Phi d{\bf r},
\label{conn7}
\end{eqnarray}
where we have used an integration by parts in the r.h.s. Inserting  Eq. (\ref{conn7}) in Eq. (\ref{conn6}), we obtain
\begin{eqnarray}
-\lambda\beta \int \frac{(\delta\rho)^{2}}{\rho'(\Phi)}d{\bf r}+\lambda\beta\int \delta\rho \delta\Phi d{\bf r}=-\int \frac{(\delta{\bf J})^{2}}{Dg(\rho)}d{\bf r}.\nonumber\\
\label{conn8}
\end{eqnarray}
On the other hand, the second variations of the rate of free energy
dissipation (\ref{h8}) around equilibrium are
\begin{eqnarray}
\delta^{2}{\dot F}=-\int \frac{(\delta{\bf J})^{2}}{\beta D g(\rho)}d{\bf r},
\label{conn9}
\end{eqnarray}
and they are clearly negative.  Inserting Eqs. (\ref{conn9}) and
(\ref{min4}) in Eq. (\ref{conn8}), we finally obtain the relation
\begin{eqnarray}
2\lambda \delta^{2}{F}=\delta^{2}{\dot F}\le 0.
\label{conn10}
\end{eqnarray}
This relation shows that a steady state of the nonlinear mean field
Fokker-Planck equation (\ref{dde3}) is linearly dynamically stable
($\lambda<0$) iff it is a minimum (at least local) of the free energy
at fixed mass ($\delta^{2}F>0$). Therefore, dynamical and
generalized thermodynamical stability coincide.

\section{Stability of the homogeneous phase and critical point}
\label{sec_bif}

We consider a homogeneous stationary solution $\rho({\bf r})=\rho$ of
the nonlinear mean field Fokker-Planck equation (\ref{dde3}). The
corresponding potential is $\Phi({\bf r})=\Phi=U\rho$ where $U\equiv
\int u({\bf x})d{\bf x}$. The dynamical evolution of a small
perturbation around equilibrium is given by the linearized equation
\begin{eqnarray}
\label{bif1}
\frac{\partial\delta\rho}{\partial t}=\nabla\cdot \left\lbrack Dh(\rho)\nabla
\delta\rho+\chi g(\rho)\nabla\delta\Phi\right\rbrack\nonumber\\
=Dg(\rho)\left\lbrack C''(\rho)\Delta\delta\rho+\beta\Delta\delta\Phi\right\rbrack,
\end{eqnarray}
where we have used Eqs. (\ref{h4}) and (\ref{h6}) to get the second
line. The perturbations $\delta\rho$ and $\delta\Phi$ can be
decomposed in Fourier modes of the form $f({\bf r},t)=\int
\hat{f}({\bf k})e^{i{\bf k}\cdot {\bf r}}e^{\lambda({\bf k})t}d{\bf
k}$ leading to the dispertion relation
\begin{eqnarray}
\label{bif2}
\lambda({\bf k})=-Dg(\rho)k^{2}\left\lbrack C''(\rho)-\beta \hat{v}(k)\right\rbrack,
\end{eqnarray} 
where we have used
$\delta\hat{\Phi}=(2\pi)^{d}\hat{u}(k)\delta\hat{\rho}$ (convolution)
and set $\hat{v}(k)=-(2\pi)^{d}\hat{u}(k)$. The system is stable
($\lambda<0$) if
\begin{eqnarray}
\label{bif3}
C''(\rho)-\beta \hat{v}(k)>0,
\end{eqnarray} 
for all $k$ and unstable (to some wavenumbers) otherwise. For a
potential satisfying $\hat{v}<0$, the homogeneous phase is always
stable. Otherwise, there exists a critical point in the problem. The
homogeneous phase is stable for
\begin{eqnarray}
\label{bif4}
T>T_{c}\equiv \frac{\hat{v}(k)_{max}}{C''(\rho)},
\end{eqnarray} 
where $\hat{v}(k)_{max}$ is the largest value of $\hat{v}(k)$ achieved
for $k=k_{*}$. On the other hand, for $T<T_{c}$ the homogeneous phase is unstable to
wavenumbers satisfying $\hat{v}(k)>TC''(\rho)$. The growth rate of the
mode $k$ is given by Eq. (\ref{bif2}). This stability analysis has
been explicited for particular potentials of the form (\ref{gle2}) in
\cite{hb,csjeans}. It has also been generalized to potentials of the form
(\ref{gle3}) in \cite{chemojeans}.

The condition of generalized thermodynamical stability demands that
$\rho$ is a minimum of the free energy $F[\rho]$, given by Eq. (\ref{min6}), at
fixed mass. Using Eq. (\ref{sta5}), the stability criterion
(\ref{min4}) can be rewritten
\begin{eqnarray}
\label{bif5}
\delta^{2}F=\frac{1}{2\beta}\left\lbrace \int  C''(\rho)(\delta\rho)^{2}
d{\bf r}+\beta\int \delta\rho\delta\Phi d{\bf r}\right\rbrace \ge 0,\nonumber\\
\end{eqnarray} 
for all perturbations $\delta\rho$ that conserve mass. We need therefore to investigate the eigenvalue equation
\begin{eqnarray}
\label{bif6}
C''(\rho)\delta\rho+\beta\delta\Phi=\lambda\delta\rho.
\end{eqnarray} 
The homogeneous phase is a minimum of free energy at fixed mass (stable) iff all the eigenvalues $\lambda$ are positive. If at least one eigenvalue is negative, the homogeneous phase is an unstable saddle point of free energy at fixed mass. Solving the eigenvalue Eq. (\ref{bif6})  in Fourier space, we get
\begin{eqnarray}
\label{bif7}
C''(\rho)-\beta\hat{v}(k)=\lambda.
\end{eqnarray} 
The spectrum of eiganvalues is continuous and Eq. (\ref{bif7})
determines the eigenmode $k$ corresponding to the eigenvalue
$\lambda$.  Therefore, the homogeneous phase is a minimum of free
energy at fixed mass (stable) iff the l.h.s. of Eq. (\ref{bif7}) is
positive for all $k$. This returns the condition (\ref{bif3}),
i.e. $T>T_{c}$. Alternatively, for $T<T_{c}$, there exists modes $k$
such that $\lambda<0$ implying $\delta^{2}F<0$ for these modes. In
that case, the homogeneous phase is a maximum or a saddle point of
free energy at fixed mass (unstable). Using the Parseval theorem, the
second variations of free energy (\ref{bif5}) can be written
\begin{eqnarray}
\label{bif8}
\delta^{2}F=\frac{1}{2\beta}(2\pi)^{d}\int \left\lbrack  C''(\rho)-\beta \hat{v}(k)\right\rbrack |\delta\hat{\rho}|^{2}d{\bf k},
\end{eqnarray} 
leading directly to the preceding results. We also check on this
explicit example (homogeneous state) that the conditions of dynamical
and generalized thermodynamical stability coincide. As shown in Appendix
\ref{sec_conn}, this is also true for inhomogeneous equilibrium distributions.

Finally, let us show that $T_{c}$ corresponds to a bifurcation
point. The general steady state of the nonlinear mean field
Fokker-Planck equation (\ref{dde3}) is given by
Eq. (\ref{sta3}). Close to the bifurcation point, the inhomogeneous
equilibrium density profile can be written $\rho({\bf
r})=\rho+\epsilon({\bf r})$ where $\rho$ is the homogeneous solution
and $\epsilon({\bf r})\ll \rho$. Substituting this relation in
Eq. (\ref{sta3}) and expanding the equation to first order in
$\epsilon$, we get
\begin{eqnarray}
\label{bif9}
C''(\rho)\epsilon({\bf r})=-\beta\int\epsilon({\bf r}')u(|{\bf r}-{\bf
r}'|)d{\bf r}'.
\end{eqnarray} 
In Fourier space, this relation becomes
\begin{eqnarray}
\label{bif10}
C''(\rho)\hat{\epsilon}({\bf k})=\beta \hat{v}(k)\hat{\epsilon}({\bf k}).
\end{eqnarray} 
This equation has a non-zero solution $\hat{\epsilon}({\bf k})\neq 0$
provided that there exists a mode $k=k_*$ such that $C''(\rho)=\beta
\hat{v}(k_*)$. This precisely corresponds to the 
condition $T=T_{c}$. Therefore, stable stationary inhomogeneous
solutions appear for $T<T_{c}$ precisely when the homogeneous phase
becomes unstable. For the BMF model
\cite{cvb}, the stability analysis can be carried out explicitly and we find that the phase transition at $T=T_c$ is second order. More generally, there
may exist other systems where inhomogeneous solutions are stable for
$T>T_{c}$. In that case, they are in ``competition'' with the
homogeneous solution. One of these solutions is stable (global minimum
of free energy) and the other metastable (local minimum of free
energy). This is usually associated with a first order phase
transition.

\section{Passage from $F[f]$ to $F[\rho]$}
\label{sec_passage}

We assume that the distribution function $f({\bf r},{\bf v},t)$ is given by an expression of the form
\begin{equation}
\label{passage1} C'(f)=-\beta\biggl\lbrack {v^{2}\over
2}+\lambda({\bf r},t)\biggr\rbrack.
\end{equation}
This expression appeared at several occasions in our analysis (see
Secs. \ref{sec_nss}, \ref{sec_eos}, \ref{sec_strong} and Appendix
\ref{sec_equi}).  Since $C$ is convex, the foregoing relation can be
reversed to give
\begin{equation}
\label{passage2} f=F\biggl \lbrack \beta\biggl ({v^{2}\over
2}+\lambda({\bf r},t)\biggr )\biggr\rbrack,
\end{equation}
where $F(x)=(C')^{-1}(-x)$. Since the distribution function is isotropic, the
local velocity vanishes: ${\bf u}={\bf 0}$. The density and the
pressure are then given by
\begin{equation}
\label{passage3}\rho=\int f d{\bf v}=\rho[\lambda],\qquad p={1\over
d}\int f v^{2}d{\bf v}=p[\lambda].
\end{equation}
The first relation determines $\lambda({\bf r},t)$ as a function of
the density $\rho({\bf r},t)$.  Substituting Eq. (\ref{passage2}) in
Eq. (\ref{passage3}), we find that
\begin{equation}
\label{passage4}\rho={1\over \beta^{d/2}}g(\beta\lambda), \qquad p={1\over \beta^{d+2\over 2}}h(\beta\lambda),
\end{equation}
with
\begin{equation}
\label{passage5} g(x)=2^{d-2\over 2}S_{d} \int_{0}^{+\infty} F(x+t)\ t^{d-2\over 2} dt,
\end{equation}
\begin{equation}
\label{passage6} h(x)={1\over d}2^{d\over 2}S_{d} \int_{0}^{+\infty} F(x+t)\ t^{d\over 2} dt,
\end{equation}
where $S_{d}$ is the surface of a unit sphere in $d$
dimensions. Eliminating $\lambda$ between the foregoing expressions, we find
that the fluid is {\it barotropic}, in the sense that $p=p(\rho)$
where the equation of state is entirely specified by $C(f)$. We can
now express the free energy (\ref{nh5}) as a functional of $\rho$ by writing
$F[\rho]\equiv F[f]$. The energy (\ref{nh1}) is simply given by
\begin{equation}
\label{passage7}
E={d\over 2}\int p\ d{\bf r}+{1\over 2}\int\rho\Phi d{\bf
  r}.
\end{equation}
On the other hand, the entropy (\ref{nh3}) can be written
\begin{eqnarray}
{S}=-{2^{d-2\over 2}S_{d}\over\beta^{d/2}}\int d{\bf
  r}\int_{0}^{+\infty}C\lbrack F(t+\beta\lambda)\rbrack \ t^{d-2\over 2}dt.
\label{passage8}
\end{eqnarray}
Integrating by parts  and using $C'\lbrack F(x)\rbrack=-x$, we find that
\begin{eqnarray}
{S}=-{2^{d/2}S_{d}\over d\beta^{d/2}}\int d{\bf r}\int_{0}^{+\infty}
F'(t+\beta\lambda)(t+\beta\lambda)t^{d/2}dt.\nonumber\\
\label{passage9}
\end{eqnarray}
Integrating by parts one more time and using Eqs. (\ref{passage4}),
(\ref{passage5}) and (\ref{passage6}), we finally obtain
\begin{eqnarray}
{S}={d+2\over 2}\beta\int p d{\bf r}+\beta\int \lambda\rho d{\bf r}.
\label{passage10}
\end{eqnarray}
Collecting all the previous results, the free energy (\ref{nh5}) becomes
\begin{eqnarray}
{F}[\rho]=-\int\rho \biggl (\lambda+{p\over\rho}\biggr )d{\bf
r} +{1\over 2}\int\rho\Phi d{\bf r}. \label{passage11}
\end{eqnarray}
Finally, using the relation $h'(x)=-g(x)$ obtained from
Eqs. (\ref{passage5}) and (\ref{passage6}) by a simple integration by parts, it
is easy to check that Eq. (\ref{passage4}) implies
\begin{eqnarray}
\lambda'(\rho)=-\frac{p'(\rho)}{\rho},
\label{passage12}
\end{eqnarray}
so that
\begin{eqnarray}
\lambda+{p\over\rho}=-\int^{\rho}{p(\rho')\over\rho'^{2}}d\rho'.
\label{passage13}
\end{eqnarray}
Hence, the free energy can be written more explicitly as
\begin{eqnarray}
{F}[\rho]=\int \rho \int^{\rho}{p(\rho')\over\rho'^{2}}d\rho'
d{\bf r}+{1\over
  2}\int\rho\Phi d{\bf r}.
\label{passage14}
\end{eqnarray}

\section{Equivalence between the stability criteria of the generalized Kramers and Smoluchowski equations}
\label{sec_equi}

Let us consider the minimization problem (\ref{nmin3}). We want to
determine the distribution $f_*({\bf r},{\bf v})$ which minimizes the
free energy $F[f]$ at fixed mass $M[f]=M$. To solve this minimization
problem, we can proceed in two steps. {\it First step:} we determine
the distribution $f_1({\bf r},{\bf v})$ which minimizes $F[f]$ at
fixed density profile $\rho({\bf r})=\int fd{\bf v}$.  This gives a
distribution $f_1[\rho({\bf r}),{\bf v}]$ depending on $\rho({\bf r})$
and ${\bf v}$. Substituting this distribution in the functional
$F[f]$, we obtain a functional $F[\rho]\equiv F[f_1]$ of the
density. {\it Second step:} we determine the density $\rho_*({\bf r})$
which minimizes $F[\rho]$ at fixed mass $M[\rho]=M$. Finally, we have
$f_*({\bf r},{\bf v})=f_1[\rho_*({\bf r}),{\bf v}]$.
 
Let us be more explicit. If we fix the density profile $\rho({\bf
r})$, the potential energy $W[\rho]$ is automatically
determined. Therefore, minimizing $F[f]=E[f]-TS[f]$ at fixed density
profile is equivalent to minimizing $\tilde{F}[f]=K[f]-TS[f]$ at fixed
density profile, where $K[f]$ is the kinetic energy. The distribution
$f_1({\bf r},{\bf v})$ that extremizes $\tilde{F}[f]$ with the
constraint $\int f \, d{\bf v} =\rho({\bf r})$ satisfies the first
order variations $\delta F+\int \lambda({\bf r}) \delta (\int f d{\bf
v}) d{\bf r}=0$, where $\lambda({\bf r})$ is a Lagrange
multiplier. This leads to
\begin{equation}
\label{equi1} C'(f_{1})=-\beta\biggl\lbrack {v^{2}\over
2}+\lambda({\bf r})\biggr\rbrack,
\end{equation}
where $\lambda({\bf r})$ is related to $\rho({\bf r})$ by writing
$\rho=\int f_{1}d{\bf v}$. Since $\delta^{2}F=-T\delta^{2}S=\frac{1}{2}T\int C''(f_{1})(\delta f)^{2}d{\bf r}d{\bf v}\ge 0$
the distribution $f_{1}$ is a {\it minimum} of $F[f]$ at fixed density
profile. Now, we remark that Eq. (\ref{equi1}) has the form
(\ref{passage1}) so that the functional $F[\rho]\equiv F[f_{1}]$ is
explicitly given by
\begin{eqnarray}
{F}[\rho]=\int \rho \int^{\rho}{p(\rho')\over\rho'^{2}}d\rho'
d{\bf r}+{1\over
  2}\int\rho\Phi d{\bf r}.
\label{equi3}
\end{eqnarray}
Therefore, we conclude that {\it $f_{*}({\bf r},{\bf
v})=f_1[\rho_*({\bf r}),{\bf v}]$ is a minimum of $F[f]$ at fixed mass
iff $\rho_{*}({\bf r})$ is a minimum of $F[\rho]$ at fixed
mass}. Thus, the variational problems (\ref{nmin3}) and
(\ref{min3ng}) are equivalent for global minimization. 

We shall now show that they are equivalent for local minimization.  A
critical point of (\ref{nmin3}) is a local minimum of $F[{f}]$ at
fixed mass iff inequality (\ref{nmin4}) is satisfied for all
perturbations $\delta{f}$ that conserve mass. A critical point of
(\ref{min3ng}) is a local minimum of $F[\rho]$ at fixed mass iff
inequality (\ref{min4ng}) is satisfied for all perturbations
$\delta{\rho}$ that conserve mass. In order to make the connection
between the second order variations (\ref{nmin4}) and (\ref{min4ng}),
the idea is to project the perturbation $\delta f$ on a suitable space
and write $\delta f=\delta f_{\|}+\delta f_{\perp}$ where $\delta
f_{\perp}$ is the orthogonal perturbation (this is a relatively
general method that has been applied in different contexts; see
\cite{frank,bouchet,assise}). We can always write the perturbation in the form
\begin{eqnarray}
\delta f=\delta f_{\|}+\delta f_{\perp}=\frac{\delta\rho}{\int f'(\epsilon)d{\bf v}}f'(\epsilon)+\delta f_{\perp}, 
\label{app9}
\end{eqnarray}
where $\delta f_{\perp}\equiv \delta f-\delta f_{\|}$ ensures that all the
perturbations are considered.  By construction, we have $\int \delta f
d{\bf v}=\int \delta f_{\|} d{\bf v}=\delta\rho$ so that $\int
\delta f_{\perp} d{\bf v}=0$. Therefore, $\delta f_{\perp}$ is orthogonal to $\delta f_{\|}$ in the sense that
\begin{eqnarray}
\int \delta f_{\|} \delta f_{\perp} \frac{1}{f'(\epsilon)}d{\bf v} \propto \int \delta f_{\perp} d{\bf v}=0.
\label{app10}
\end{eqnarray}
Then, we readily obtain
\begin{eqnarray}
\int \frac{(\delta f)^{2}}{f'(\epsilon)}d{\bf v}=\int \frac{(\delta f_{\perp})^{2}}{f'(\epsilon)} d{\bf v}+\int \frac{(\delta f_{\|})^{2}}{f'(\epsilon)}d{\bf v}\nonumber\\
=\int \frac{(\delta f_{\perp})^{2}}{f'(\epsilon)} d{\bf v}
+\frac{(\delta\rho)^2}{\int f'(\epsilon)d{\bf v}}.
\label{app11}
\end{eqnarray}
Now, a critical point of (\ref{nmin3}) is of the form $f=f(\epsilon)$
with $\epsilon=v^2/2+\Phi({\bf r})$ and $f'(\epsilon)<0$. This implies
that $\rho=\int f(\epsilon)d{\bf v}=\rho(\Phi)$ and $\rho'(\Phi)=\int
f'(\epsilon)d{\bf v}$.  Therefore, Eq. (\ref{app11}) can be rewritten
\begin{eqnarray}
\int \frac{(\delta f)^{2}}{f'(\epsilon)}d{\bf v}
=\int \frac{(\delta f_{\perp})^{2}}{f'(\epsilon)} d{\bf v}+\frac{(\delta\rho)^2}{\rho'(\Phi)}.
\label{app11bis}
\end{eqnarray}
Combining Eqs. (\ref{app11bis}), (\ref{nmin4}) and (\ref{min4ng}) we
finally obtain
\begin{eqnarray}
\delta^{2}F[\delta f]=-\frac{1}{2}\int \frac{(\delta f_{\perp})^{2}}{f'(\epsilon)} d{\bf r}d{\bf v} +\delta^{2}F[\delta \rho].
\label{app14}
\end{eqnarray}
If $\delta^{2}F[\delta \rho]\ge 0$ for all perturbations $\delta \rho$
that conserve mass, then $\delta^{2}F[\delta f]\ge 0$ for all
perturbations $\delta f$ that conserve mass. Alternatively, if there
exists a perturbation $\delta\rho_{*}$ such that
$\delta^{2}F[\delta\rho_{*}]< 0$, by taking $\delta f_*$ in the form
(\ref{app9}) with $\delta\rho=\delta\rho_{*}$ and $\delta
f_{\perp}=0$, we get $\delta^{2}F[\delta f_{*}]=\delta^{2}F[\delta
\rho_{*}]< 0$. We conclude that $f({\bf r},{\bf v})$ is a local
minimum of $F[f]$ at fixed $M$ iff $\rho({\bf r})$ is a local minimum
of $F[\rho]$ at fixed $M$.  Thus: (\ref{nmin3}) $\Leftrightarrow$
(\ref{min3ng}) for local and global minimizations.

There are several consequences to this result (see also the more
detailed discussion in \cite{assise}):

(i) We have seen that $f$ is a linearly dynamically stable steady
state of the generalized Kramers equation (\ref{gk5}) iff it is a
(local) minimum of $F[f]$ at fixed mass $M[f]=M$. On the other hand,
we have seen that $\rho$ is a linearly dynamically stable steady state
of the generalized Smoluchowski equation (\ref{gs1}) iff it is a
(local) minimum of $F[\rho]$ at fixed mass $M[\rho]=M$. According to
the above-mentioned result (\ref{nmin3}) $\Leftrightarrow$
(\ref{min3ng}), we conclude that: {\it $f_{*}({\bf r},{\bf
v})=f_1[\rho_*({\bf r}),{\bf v}]$ is a linearly dynamically stable
steady state of the generalized Kramers equation iff $\rho_{*}({\bf
r})$ is a linearly dynamically stable steady state of the generalized
Smoluchowski equation}.

(ii) It can be shown that a distribution function which minimizes a
functional of the form $F[f]=E[f]-TS[f]$ (where $T$ is a constant and
$S[f]$ is given by Eq. (\ref{nh3}) where $C(f)$ is an arbitrary convex
function) at fixed mass $M[f]=M$ is a nonlinearly \footnote{We
implicitly consider here the {\it formal} nonlinear dynamical stability in the
sense of Holm {\it et al.} \cite{holm}.} dynamically stable stationary
solution of the Vlasov equation (see, e.g., the case of stellar
systems
\cite{canto}). Therefore, the generalized mean field Kramers equation
(\ref{gk5}), which precisely solves this minimization problem, can be
used as a {\it numerical algorithm} to construct nonlinearly
dynamically stable stationary solutions of the Vlasov
equation. Indeed, if a distribution is linearly dynamically stable
with respect to the generalized Kramers equation then it is
nonlinearly dynamically stable with respect to the Vlasov equation
(but the converse may be wrong, see below).

(iii) It can be shown that a density profile is a nonlinearly
dynamically stable stationary solution of the Euler equation with a
barotropic equation of state iff it minimizes a functional of the form
$F[\rho]$ (where $F[\rho]$ is given by Eq. (\ref{gs4})) at fixed mass
$M[\rho]=M$ (see, e.g., the case of barotropic stars
\cite{canto}). Therefore, the generalized mean field Smoluchowski
equation (\ref{gs1}), which precisely solves this minimization
problem, can be used as a numerical algorithm to construct nonlinearly
dynamically stable stationary solutions of the Euler equation.
Indeed, a density profile is linearly dynamically stable with respect
to the generalized Smoluchowski equation iff it is nonlinearly
dynamically stable with respect to the barotropic Euler equation.

(iv) According to the above mentioned results, a distribution function
$f=f(\epsilon)$ with $f'(\epsilon)<0$ is a nonlinearly dynamically
stable stationary solution of the Vlasov equation if the corresponding
barotropic gas with equation of state $p=p(\rho)$ is a nonlinearly
dynamically stable stationary solution of the Euler equation. In
astrophysics, this corresponds to the nonlinear Antonov first law
\cite{canto}.  However, the minimization of $F[f]=E[f]-TS[f]$ at fixed
mass $M[f]=M$ (problem with one constraint) is just a {\it sufficient}
condition of nonlinear dynamical stability with respect to the Vlasov
equation. Thus, this minimization problem does not allow to construct
all the nonlinearly dynamically stable stationary solutions of the
Vlasov equation.  A larger class of nonlinearly dynamically stable
stationary solutions is obtained by maximizing a Casimir functional of
the form $S[f]$ (where $S[f]$ is given by Eq. (\ref{nh3})) at fixed
mass $M[f]=M$ and energy $E[f]=E$ (problem with two constraints)
\cite{canto}. A numerical algorithm solving this maximization problem
is proposed in \cite{gen}. In case of ``ensemble inequivalence'',
these solutions cannot be obtained by minimizing $F[f]=E[f]-TS[f]$ at
fixed mass $M[f]=M$. Therefore, the problem with two constraints
(``microcanonical'') provides a {\it refined} condition of nonlinear
dynamical stability with respect to the problem with one constraint
(``canonical'').

\section{Extension of the Eddington formula}
\label{sec_eddington}

In Sec. \ref{sec_eos}, we have seen that a distribution function
$f=f(\epsilon)$ determines a barotropic equation of state $p=p(\rho)$
and we have explained how to obtain it (some explicit examples have
been given in Sec. \ref{sec_ee}).  We shall now consider the inverse
problem: find the distribution function $f=f(\epsilon)$ leading to the
equation of state $p(\rho)$.  This problem was first encountered in
astrophysics and solved by Eddington \cite{eddig}. In astrophysics, a
distribution function of the form $f=f(\epsilon)$ describes a
particular class of spherical stellar systems that are stationary
solutions of the Vlasov-Poisson system. To any such stellar system, we
can associate a corresponding barotropic star with an equation of
state $p=p(\rho)$ which is a stationary solution of the Euler-Poisson
system. The problem is to find the equation of state $p=p(\rho)$
corresponding to the distribution function $f=f(\epsilon)$ and {\it
vice et versa}. A similar problem arises in the context of nonlinear
mean field Fokker-Planck equations. Thus, we can adapt many results of
astrophysics to the present situation. For sake of generality, we
shall consider the Eddington inverse problem in a space of arbitrary
dimension $d$.

Knowing the equation of state $p=p(\rho)$, we can obtain the
equilibrium density $\rho(\Phi)$ by integrating the condition of
hydrostatic equilibrium (\ref{edd2}). The problem now is to determine
$f(\epsilon)$ from the knowledge of $\rho(\Phi)$.  Let us rewrite the
density in the form
\begin{eqnarray}
\label{edd3}
\rho(\Phi)=\int_{0}^{+\infty}f(\epsilon)S_{d}v^{d-1}dv.
\end{eqnarray}
We shall consider two cases:

(i) We first assume that the distribution function has a compact
support so that $f=0$ if $\epsilon\ge \epsilon_{m}$. In that case,
the range of integration in Eq. (\ref{edd3}) is restricted to $v\le
\sqrt{2(\epsilon_{m}-\Phi)}$ so that
\begin{eqnarray}
\label{edd4}
\rho(\Phi)=\int_{0}^{\sqrt{2(\epsilon_{m}-\Phi)}}f(\epsilon)S_{d}v^{d-1}dv.
\end{eqnarray}
Taking $\epsilon=\frac{v^{2}}{2}+\Phi$ as a variable of integration instead of $v$ we obtain
\begin{eqnarray}
\label{edd5}
\rho(\Phi)=\int_{\Phi}^{\epsilon_{m}}f(\epsilon)S_{d}\left\lbrack
2(\epsilon-\Phi)\right\rbrack^{\frac{d-2}{2}}d\epsilon.
\end{eqnarray}
It is convenient at this stage  to define $\psi=\epsilon_{m}-\Phi$
and $x=\epsilon_{m}-\epsilon$. In terms of these variables, we get
\begin{eqnarray}
\label{edd6} \rho(\psi)=2^{\frac{d-2}{2}}S_{d}\int_{0}^{\psi}f(x)(\psi-x)^{\frac{d-2}{2}}dx.
\end{eqnarray}
In $d=3$, taking the derivative of Eq. (\ref{edd6}) we find
\begin{eqnarray}
\label{edd7} \frac{1}{\sqrt{8}\pi}\frac{d\rho}{d\psi}=\int_{0}^{\psi}\frac{f(x)}{\sqrt{\psi-x}}dx.
\end{eqnarray}
This is an Abel integral whose solution is
\begin{eqnarray}
\label{edd8} f(x)=\frac{1}{\sqrt{8}\pi^{2}}\frac{d}{dx}\int_{0}^{x}\frac{d\rho}{d\psi} \frac{d\psi}{\sqrt{x-\psi}}\nonumber\\
=\frac{1}{\sqrt{8}\pi^{2}}\left\lbrack
\int_{0}^{x}\frac{d^{2}\rho}{d\psi^{2}}
\frac{d\psi}{\sqrt{x-\psi}}+\frac{1}{\sqrt{x}}\left
(\frac{d\rho}{d\psi}\right )_{\psi=0}\right\rbrack.
\end{eqnarray}
This is the Eddington formula \cite{eddig}. In $d=1$, Eq. (\ref{edd6}) becomes
\begin{eqnarray}
\label{edd9} \frac{1}{\sqrt{2}}\rho(\psi)=\int_{0}^{\psi}\frac{f(x)}{\sqrt{\psi-x}}dx.
\end{eqnarray}
Comparing with the previous case, we immediately have
\begin{eqnarray}
\label{edd10} f(x)=\frac{1}{\sqrt{2}\pi}\frac{d}{dx}\int_{0}^{x}\rho(\psi) \frac{d\psi}{\sqrt{x-\psi}}\nonumber\\
=\frac{1}{\sqrt{2}\pi}\left\lbrack \int_{0}^{x}\frac{d\rho}{d\psi} \frac{d\psi}{\sqrt{x-\psi}}+\frac{1}{\sqrt{x}}\rho(\psi=0)\right\rbrack.
\end{eqnarray}
Finally, in $d=2$, Eq. (\ref{edd5}) reduces to
\begin{eqnarray}
\label{edd11}
\rho(\Phi)=2\pi\int_{\Phi}^{\epsilon_{m}}f(\epsilon)d\epsilon,
\end{eqnarray}
and we get the very simple result
\begin{eqnarray}
\label{edd12}
f(\epsilon)=-\frac{1}{2\pi}\frac{d\rho}{d\Phi}(\epsilon).
\end{eqnarray}

(ii) We now consider the case where the distribution function takes
strictly positive  values for all energies so that
$\epsilon_{m}\rightarrow +\infty$. Then Eq. (\ref{edd3}) becomes
\begin{eqnarray}
\label{edd13}
\rho(\Phi)=\int_{\Phi}^{+\infty}f(\epsilon)S_{d}\left\lbrack
2(\epsilon-\Phi)\right\rbrack^{\frac{d-2}{2}}d\epsilon.
\end{eqnarray}
In $d=3$, taking the derivative of Eq. (\ref{edd13}) we obtain
\begin{eqnarray}
\label{edd14}
-\frac{1}{\sqrt{8}\pi}\frac{d\rho}{d\Phi}=\int_{\Phi}^{+\infty}\frac{f(\epsilon)}{\sqrt{\epsilon-\Phi}}d\epsilon.
\end{eqnarray}
This is an Abel integral whose  solution is
\begin{eqnarray}
\label{edd15} f(\epsilon)=\frac{1}{\sqrt{8}\pi^{2}}\frac{d}{d\epsilon}\int_{\epsilon}^{+\infty}\frac{d\rho}{d\Phi} \frac{d\Phi}{\sqrt{\Phi-\epsilon}}\nonumber\\
=\frac{1}{\sqrt{8}\pi^{2}}\left\lbrack
\int_{\epsilon}^{+\infty}\frac{d^{2}\rho}{d\Phi^{2}}
\frac{d\Phi}{\sqrt{\Phi-\epsilon}}-\lim_{\Phi\rightarrow
+\infty}\frac{\frac{d\rho}{d\Phi}}{\sqrt{\Phi-\epsilon}}\right\rbrack.
\end{eqnarray}
In $d=1$, Eq. (\ref{edd13}) becomes
\begin{eqnarray}
\label{edd16}
\frac{1}{\sqrt{2}}\rho(\Phi)=\int_{\Phi}^{+\infty}\frac{f(\epsilon)}{\sqrt{\epsilon-\Phi}}d\epsilon,
\end{eqnarray}
and we get
\begin{eqnarray}
\label{edd17} f(\epsilon)=-\frac{1}{\sqrt{2}\pi}\frac{d}{d\epsilon}\int_{\epsilon}^{+\infty}\rho(\Phi) \frac{d\Phi}{\sqrt{\Phi-\epsilon}}\nonumber\\
=-\frac{1}{\sqrt{2}\pi}\left\lbrack
\int_{\epsilon}^{+\infty}\frac{d\rho}{d\Phi}\frac{d\Phi}{\sqrt{\Phi-\epsilon}}-\lim_{\Phi\rightarrow
+\infty}\frac{\rho}{\sqrt{\Phi-\epsilon}}\right\rbrack.
\end{eqnarray}
For $d=2$, Eq. (\ref{edd12}) remains unchanged. For example, the distribution function associated with the Fermi-Dirac statistics in
physical space
\begin{eqnarray}
\label{edd18}\rho(\Phi)=\frac{\sigma_{0}}{1+ e^{\beta\Phi+\alpha}},
\end{eqnarray}
is
\begin{eqnarray}
\label{edd19}f(\epsilon)=\frac{\sigma_{0}\beta}{8\pi\cosh^{2}\left\lbrack\frac{1}{2}(\beta\epsilon+\alpha)\right\rbrack}, \qquad (d=2).
\end{eqnarray}

\section{Derivation of the $H$-theorems}
\label{sec_ht}

Let us derive the $H$-theorem (\ref{h8}) for the NFP equation (\ref{dde3})-(\ref{gle2}). The time variations of the entropy (\ref{h5}) and of the energy  (\ref{h1}) associated with an external potential are
\begin{eqnarray}
\label{ht1}
\dot S=-\int C'(\rho)\frac{\partial\rho}{\partial t}d{\bf r}, \qquad \dot E=\int\Phi_{ext}\frac{\partial\rho}{\partial t}d{\bf r}.
\end{eqnarray}
On the other hand, using Eqs. (\ref{h2}) and (\ref{gle2}), the energy
associated with a binary potential of interaction can be written
\begin{eqnarray}
\label{ht2}
E=\frac{1}{2}\int \rho({\bf r},t)u(|{\bf r}-{\bf r}'|)\rho({\bf r}',t)d{\bf r}d{\bf r}'.
\end{eqnarray}
Its time derivative is 
\begin{eqnarray}
\label{ht3}
\dot E=\frac{1}{2}\int \frac{\partial\rho}{\partial t}({\bf r},t)u(|{\bf r}-{\bf r}'|)\rho({\bf r}',t)d{\bf r}d{\bf r}'\nonumber\\
+\frac{1}{2}\int \rho({\bf r},t)u(|{\bf r}-{\bf r}'|)\frac{\partial\rho}{\partial t}({\bf r}',t)d{\bf r}d{\bf r}'.
\end{eqnarray}
Interchanging the dummy variables ${\bf r}$ and ${\bf r}'$ and using
Eq.  (\ref{gle2}), we finally obtain
\begin{eqnarray}
\label{ht4}
\dot E=\int\Phi\frac{\partial\rho}{\partial t}d{\bf r}.
\end{eqnarray}
Therefore, the time variation of the free energy (\ref{h7}) is given by
\begin{eqnarray}
\label{ht5}
\dot F=\dot E-T\dot S=\int (\Phi+TC'(\rho))\frac{\partial\rho}{\partial t} d{\bf r}.
\end{eqnarray}
Using Eq. (\ref{dde4}) and integrating by parts, we obtain
\begin{eqnarray}
\label{ht6}
\dot F
=\int {\bf J}\cdot  (\nabla\Phi+TC''(\rho)\nabla\rho ) d{\bf r}. 
\end{eqnarray}
Inserting Eqs. (\ref{dde5}), (\ref{h4}) and (\ref{h6}) in
Eq. (\ref{ht6}), we finally obtain the $H$-theorem (\ref{h8}).

Let us derive the $H$-theorem (\ref{h9}) for the NFP equation
(\ref{dde3})-(\ref{gle3}). The time variation of the energy given by
Eq. (\ref{h3}) is
\begin{eqnarray}
\label{ht7}
\dot E=\frac{1}{\lambda}\int \left (\nabla\Phi\cdot \nabla\frac{\partial\Phi}{\partial t}+k^{2}\Phi\frac{\partial\Phi}{\partial t}\right )d{\bf r}\nonumber\\
+\int \frac{\partial\rho}{\partial t}\Phi d{\bf r}+\int\rho\frac{\partial\Phi}{\partial t}d{\bf r}.
\end{eqnarray}
Integrating the first term by parts and using Eq. (\ref{gle3}), it can be rewritten
\begin{eqnarray}
\label{ht8}
\dot E=-\frac{1}{\lambda\epsilon}\int (\Delta\Phi-k^{2}\Phi-\lambda\rho)^{2} d{\bf r}+\int \frac{\partial\rho}{\partial t}\Phi d{\bf r}.
\end{eqnarray}
Inserting Eq. (\ref{dde3}) and proceeding as above, we obtain the
$H$-theorem (\ref{h9}).

Consider now the generalized Smoluchowski equation (\ref{gs1}). The
time variation of the free energy (\ref{gs4}) is
\begin{eqnarray}
\label{ht9}
\dot F=\int \left (\Phi+\int^{\rho}\frac{p(\rho')}{\rho^{'2}}d\rho' d{\bf r}+\frac{p}{\rho}\right )\frac{\partial\rho}{\partial t}d{\bf r}
\end{eqnarray}
Inserting Eq. (\ref{gs1}) in Eq. (\ref{ht9}) and integrating by parts
we obtain the $H$-theorem (\ref{gs5}).

Let us derive the $H$-theorem (\ref{nh6}) for the NFP equation
(\ref{gk5})-(\ref{gle2}). The time variations of the entropy (\ref{nh3}) and
of the energy (\ref{nh1}) are
\begin{eqnarray}
\label{ht10}
\dot S=-\int C'(f)\frac{\partial f}{\partial t}d{\bf r}d{\bf v}, \quad \dot E=\int \left (\frac{v^{2}}{2}+\Phi\right )\frac{\partial f}{\partial t}d{\bf r}d{\bf v}.\nonumber\\
\end{eqnarray}
For $D=\xi=0$, the NFP equation (\ref{gk5})-(\ref{gle2}) reduces to
the Vlasov equation. The Vlasov equation conserves the energy and the
Casimirs. Indeed, using integrations by parts, we have
\begin{eqnarray}
\label{ht11}
\dot E=\int \left (\frac{v^{2}}{2}+\Phi\right )\left (-{\bf v}\cdot \frac{\partial f}{\partial {\bf r}}+\nabla\Phi\cdot \frac{\partial f}{\partial {\bf v}}\right )d{\bf r}d{\bf v} \nonumber\\
=\int (\nabla\Phi\cdot {\bf v}-{\bf v}\cdot \nabla\Phi)f d{\bf r}d{\bf v}=0, \qquad
\end{eqnarray}
and 
\begin{eqnarray}
\label{ht12}
\dot S=-\int C'(f)\left (-{\bf v}\cdot \frac{\partial f}{\partial {\bf r}}+\nabla\Phi\cdot \frac{\partial f}{\partial {\bf v}}\right )d{\bf r}d{\bf v} \nonumber\\
=\int \left\lbrack {\bf v}\cdot \frac{\partial C(f)}{\partial {\bf r}}-\nabla\Phi\cdot\frac{\partial C(f)}{\partial {\bf v}}\right\rbrack  d{\bf r}d{\bf v}\nonumber\\
=\int \left\lbrack \frac{\partial}{\partial {\bf r}}\cdot (C(f){\bf v})-\frac{\partial}{\partial {\bf v}}\cdot (C(f)\nabla\Phi)\right\rbrack d{\bf r}d{\bf v}=0. \qquad
\end{eqnarray}
Therefore,  coming back to the NFP equation  (\ref{gk5})-(\ref{gle2}), the only contribution to the time variation of the free energy (\ref{nh5}) comes from the Fokker-Planck current. Using Eqs. (\ref{ht10}) and (\ref{gk6}), we have
\begin{eqnarray}
\label{ht13}
\dot F=\int \left (\frac{v^{2}}{2}+\Phi+TC'(f)\right )\frac{\partial f}{\partial t} d{\bf r}d{\bf v}\nonumber\\
 =-\int \left (\frac{v^{2}}{2}+\Phi+TC'(f)\right )\cdot \frac{\partial {\bf J}}{\partial {\bf v}} d{\bf r}d{\bf v}\nonumber\\
=\int \left ({\bf v}+TC''(f)\frac{\partial f}{\partial {\bf v}}\right )\cdot {\bf J}  d{\bf r}d{\bf v}. \qquad
\end{eqnarray}
Inserting Eqs. (\ref{gk7}), (\ref{nh2}) and (\ref{nh4}) in Eq. (\ref{ht13}) and integrating by parts
we obtain the $H$-theorem (\ref{nh6}).

Let us finally derive the $H$-theorem (\ref{ht1ff}) for the damped
Euler equations (\ref{damped1}), (\ref{damped2}) and (\ref{gle2}). The
time variation of the free energy (\ref{ht1grt}) is
\begin{eqnarray}
\label{ht14}
\dot F=\int \left (\Phi+\int^{\rho}\frac{p(\rho')}{\rho^{'2}}d\rho' d{\bf r}+\frac{p}{\rho}+\frac{{\bf u}^{2}}{2}\right )\frac{\partial\rho}{\partial t}d{\bf r}\nonumber\\
+\int \rho {\bf u}\cdot \frac{\partial {\bf u}}{\partial t}d{\bf r}.
\end{eqnarray}
Substituting Eqs. (\ref{damped1}) and (\ref{damped2}) in Eq. (\ref{ht14}) and integrating by parts, we get
\begin{eqnarray}
\label{ht15}
\dot F=\int \rho {\bf u}\cdot \left\lbrack \nabla\left (\frac{{\bf u}^{2}}{2}\right )-\xi {\bf u}-({\bf u}\cdot \nabla){\bf u}\right \rbrack d{\bf r}.
\end{eqnarray}
Using $({\bf u}\cdot\nabla){\bf u}=\nabla({\bf u}^{2}/2)-{\bf u}\times (\nabla\times {\bf u})$ we finally obtain the result (\ref{ht1ff}).

\section{Polytropic equation of state}
\label{sec_pol}

For $n>d/2$ (case 1), the polytropic DF can we written
\begin{eqnarray}
\label{pol1}
f=A(\epsilon_{m}-\epsilon)_{+}^{n-d/2}.
\end{eqnarray}
The density and the pressure can be expressed as
\begin{eqnarray}
\label{pol2}
\rho=AS_{d}Q_{0}(\Phi), \qquad p=\frac{1}{d}AS_{d}Q_{2}(\Phi),
\end{eqnarray}
with
\begin{eqnarray}
\label{pol3}
Q_{k}=\int_{0}^{\sqrt{2(\epsilon_{m}-\Phi)}}\left (\epsilon_{m}-\Phi-\frac{v^{2}}{2}\right )^{n-d/2}v^{k+d-1}dv.\quad 
\end{eqnarray}
Setting $x=v^{2}/\lbrack 2(\epsilon_{m}-\Phi)\rbrack$, we obtain
\begin{eqnarray}
\label{pol4}
Q_{k}=2^{(k+d-2)/2}(\epsilon_{m}-\Phi)^{n+k/2}\nonumber\\
\times\int_{0}^{1}(1-x)^{n-d/2}x^{(k+d-2)/2}dx. 
\end{eqnarray}
The integral can be expressed in terms of Gamma functions leading to 
\begin{eqnarray}
\label{pol5}
Q_{k}=2^{(k+d-2)/2}(\epsilon_{m}-\Phi)^{n+k/2}\nonumber\\
\times\frac{\Gamma((d+k)/2)\Gamma(1-d/2+n)}{\Gamma(1+k/2+n)}.
\end{eqnarray}
Then, the density and the pressure can be expressed in terms of the potential $\Phi$ as 
\begin{eqnarray}
\label{pol6}
\rho=A S_{d}(\epsilon_{m}-\Phi)^{n}2^{d/2-1}\frac{\Gamma(d/2)\Gamma(1-d/2+n)}{\Gamma(1+n)},\qquad 
\end{eqnarray}
\begin{eqnarray}
\label{pol7}
p=\frac{A S_{d}}{n+1}(\epsilon_{m}-\Phi)^{n+1}2^{d/2-1}\frac{\Gamma(d/2)\Gamma(1-d/2+n)}{\Gamma(1+n)},\qquad 
\end{eqnarray}
where we have used the identity $\Gamma(n+1)=n\Gamma(n)$ to simplify
the second expression. Eliminating the potential $\Phi$ between these
equations, we obtain the polytropic equation of state (\ref{poly7}) with $K$
given by Eq. (\ref{poly8}).

For $n<-1$ (case 2), the polytropic DF can we written
\begin{eqnarray}
\label{pol8}
f=A(\epsilon_{m}+\epsilon)^{n-d/2}.
\end{eqnarray}
The density and the pressure can be expressed as
\begin{eqnarray}
\label{pol9}
\rho=AS_{d}R_{0}(\Phi), \qquad p=\frac{1}{d}AS_{d}R_{2}(\Phi),
\end{eqnarray}
with
\begin{eqnarray}
\label{pol10}
R_{k}=\int_{0}^{+\infty}\left (\epsilon_{m}+\Phi+\frac{v^{2}}{2}\right )^{n-d/2}v^{k+d-1}dv.\quad 
\end{eqnarray}
Setting $x=v^{2}/\lbrack 2(\epsilon_{m}+\Phi)\rbrack$, we obtain
\begin{eqnarray}
\label{pol11}
R_{k}=2^{(k+d-2)/2}(\epsilon_{m}+\Phi)^{n+k/2}\nonumber\\
\times\int_{0}^{+\infty}(1+x)^{n-d/2}x^{(k+d-2)/2}dx. 
\end{eqnarray}
The integral can be expressed in terms of Gamma functions leading to 
\begin{eqnarray}
\label{pol12}
R_{k}=2^{(k+d-2)/2}(\epsilon_{m}+\Phi)^{n+k/2}\nonumber\\
\times\frac{\Gamma((d+k)/2)\Gamma(-k/2-n)}{\Gamma(d/2-n)}.
\end{eqnarray}
Then, the density and the pressure can be expressed in terms of the potential $\Phi$ as 
\begin{eqnarray}
\label{pol13}
\rho=A S_{d}(\epsilon_{m}+\Phi)^{n}2^{d/2-1}\frac{\Gamma(d/2)\Gamma(-n)}{\Gamma(d/2-n)},\qquad 
\end{eqnarray}
\begin{eqnarray}
\label{pol14}
p=-\frac{A S_{d}}{n+1}(\epsilon_{m}+\Phi)^{n+1}2^{d/2-1}\frac{\Gamma(d/2)\Gamma(-n)}{\Gamma(d/2-n)},\qquad 
\end{eqnarray}
where we have used the identity $\Gamma(n+1)=n\Gamma(n)$ to simplify
the second expression. Eliminating the potential $\Phi$ between these
equations, we obtain the polytropic equation of state (\ref{poly7}) with $K$
given by Eq. (\ref{poly9}). 

Note, as a final remark, that spatially homogeneous polytropic
distributions are obtained by taking $\Phi({\bf r})=0$ in the above
expressions.


\begin{thebibliography}{99}

\bibitem{fokker}  {\small A.D. Fokker, Ann. Physik  {\bf 43}, 810 (1914).}

\bibitem{planck}  {\small M. Planck, Sitzber. Preuss. Akad. Wiss., p. 324 (1917).}

\bibitem{einstein}  {\small A. Einstein, Ann. Physik  {\bf 17}, 549 (1905).}

\bibitem{smoluchowski}  {\small M. von Smoluchowski, Ann. Physik  {\bf 48}, 1103 (1915).}

\bibitem{kramers}  {\small H.A. Kramers, Physica A  {\bf 7}, 284 (1940).}

\bibitem{klein}  {\small O. Klein, Arkiv for Mathematik, Astronomi, och Fysik {\bf 16}, No 5 (1921).}

\bibitem{risken}  {\small H. Risken, {\it The Fokker-Planck equation} (Springer, 1989).}

\bibitem{cras}  {\small P.H. Chavanis, C. R. Physique  {\bf 7}, 318 (2006).}

\bibitem{tsallis}  {\small  C. Tsallis, J. Stat. Phys.  {\bf 52}, 479  (1988).}

\bibitem{abe}  {\small  S. Abe, Phys. Lett. A {\bf 224}, 326 (1997).}

\bibitem{br}  {\small  E.P. Borges, I. Roditi, Phys. Lett. A {\bf 246}, 399 (1998).}

\bibitem{k1}  {\small G. Kaniadakis, Physica A  {\bf 296}, 405 (2001).}

\bibitem{naudts}  {\small J. Naudts, Physica A  {\bf 340}, 32 (2004).}

\bibitem{k2}  {\small G. Kaniadakis, M. Lissia, A.M. Scarfone, Phys. Rev. E  {\bf 71}, 046128 (2005).}

\bibitem{hc}  {\small J. Harvda, F. Charvat, Kybernetica {\bf 3}, 30 (1967).}

\bibitem{mittal}  {\small D.P. Mittal, Metrika {\bf 22}, 35 (1975).}

\bibitem{st}  {\small B.D. Sharma, I.J. Taneja, Metrika {\bf 22}, 205 (1975).}

\bibitem{reyni}  {\small A. Reyni, {\it Probability Theory} (North-Holland, Amsterdam, 1970).}

\bibitem{sm}  {\small B.D. Sharma, D.P. Mittal, J. Math. Sci.  {\bf 1}, 28 (1975).}

\bibitem{kl}  {\small G. Kaniadakis, M. Lissia, Physica A  {\bf 340}, xv-xix (2004).}

\bibitem{pp}  {\small  A.R. Plastino, A. Plastino, Physica A  {\bf 222}, 347 (1995).}

\bibitem{spohn}  {\small  H. Spohn, J. Physique  {\bf 3}, 69 (1993).}

\bibitem{bukman}  {\small  C. Tsallis, D.J. Bukman, Phys. Rev. E  {\bf 54}, R2197 (1996).}

\bibitem{stariolo}  {\small  D. Stariolo,  Phys. Rev. E  {\bf 55}, 4806 (1997).}

\bibitem{borland}  {\small  L. Borland,  Phys. Rev. E  {\bf 57}, 6634 (1998).}

\bibitem{nobre}  {\small  F. Nobre, E. Curado, G. Rowlands,   Physica A  {\bf 334}, 109 (2004).}

\bibitem{kq}  {\small  G. Kaniadakis, P. Quarati  Phys. Rev. E  {\bf 49}, 5103 (1994).}

\bibitem{csr} {\small P.H. Chavanis, J. Sommeria, R. Robert, Astrophys. J.  {\bf 471}, 385 (1996).}

\bibitem{lb}  {\small D. Lynden-Bell, Mon. Not. Roy. Astr. Soc.  {\bf 136}, 101 (1967).}

\bibitem{bose}  {\small J. Sopik, C. Sire, P.H. Chavanis, Phys. Rev. E {\bf 74}, 011112 (2006).}

\bibitem{martinez}  {\small  S. Martinez, A.R. Plastino, A. Plastino, Physica A  {\bf 259}, 183 (1998).}

\bibitem{frank1}  {\small T.D. Frank, Physica A {\bf 310}, 397 (2002).}

\bibitem{gen}  {\small P.H. Chavanis, Phys. Rev. E {\bf 68}, 036108 (2003).}

\bibitem{gen2}  {\small P.H. Chavanis, Physica A {\bf 332}, 89 (2004).}

\bibitem{cn}  {\small E. Curado, F. Nobre, Phys. Rev. E {\bf 67}, 021107 (2003).}

\bibitem{frank}  {\small  T.D. Frank, {\it Nonlinear Fokker-Planck Equations: Fundamentals and Applications} (Springer-Verlag, 2005). }

\bibitem{houches}  {\small {\it Dynamics and thermodynamics of systems with long range interactions}, edited by Dauxois, T., Ruffo, S., Arimondo, E. and  Wilkens, M. Lect. Not. in Phys. {\bf 602}  (Springer, 2002)}

\bibitem{kuramoto}  {\small  Y. Kuramoto, {\it Chemical oscillations, waves, and turbulence} (Springer, Berlin, 1984). }

\bibitem{ma}  {\small N. Martzel, C. Aslangul, J. Phys. A {\bf 34}, 11225 (2001).}

\bibitem{hb}  {\small P.H. Chavanis, Physica A {\bf 361}, 55 (2006); Physica A {\bf 361}, 81 (2006).}

\bibitem{crs}  {\small P.H. Chavanis, C. Rosier, C. Sire, Phys. Rev. E {\bf  66}, 036105 (2002).}

\bibitem{sc}  {\small C. Sire, P.H. Chavanis, Phys. Rev. E {\bf 66}, 046133 (2002).}

\bibitem{cvb}  {\small P.H. Chavanis, J. Vatteville, F. Bouchet, Eur. Phys. J. B  {\bf  46}, 61 (2005).}

\bibitem{antoni}  {\small M. Antoni, S. Ruffo, Phys. Rev. E  {\bf  52}, 2361 (1995).}

\bibitem{lang}  {\small  P.H. Chavanis, C. Sire, Phys. Rev. E {\bf 69}, 016116 (2004). }

\bibitem{shiino}  {\small  M. Shiino, Phys. Rev. E {\bf 67}, 056118 (2003). }

\bibitem{murray}  {\small J.D. Murray, {\it Mathematical Biology} (Springer, Berlin, 1991).}

\bibitem{patlak}  {\small C.S. Patlak, Bull. of Math. Biophys.  {\bf 15}, 311 (1953).}

\bibitem{ks}  {\small E.F. Keller, L.A. Segel, J. Theor. Biol.  {\bf 30}, 225 (1971).}

\bibitem{os}  {\small H. Othmer, A. Stevens, SIAM J. Appl. Math.  {\bf 57}, 1044 (1997).}

\bibitem{csbio}  {\small  P.H. Chavanis, C. Sire, Physica A {\bf 384}, 199 (2007). }

\bibitem{horstmann}  {\small D. Horstmann, {\it From 1970 until present: the Keller-Segel model in chemotaxis and its consequences,} Jahresber. Deutsch. Math. Verein.   {\bf 106}, 51 (2004).}

\bibitem{csmass}  {\small  P.H. Chavanis, C. Sire, Physica A {\bf 387}, 1999 (2008). }

\bibitem{lemou}  {\small P.H. Chavanis, P. Lauren\c cot, M. Lemou,  Physica A {\bf 341}, 145 (2004).}

\bibitem{new}  {\small P.H. Chavanis,  [arXiv:0803.0263]}

\bibitem{post}  {\small C. Sire, P.H. Chavanis, Phys. Rev. E {\bf 69}, 066109 (2004).}

\bibitem{tcoll}  {\small P.H. Chavanis, C. Sire, Phys. Rev. E {\bf 70}, 026115 (2004).}

\bibitem{virial1}  {\small P.H. Chavanis, C. Sire, Phys. Rev. E {\bf 73}, 066103 (2006).}

\bibitem{virial2}  {\small P.H. Chavanis, C. Sire, Phys. Rev. E {\bf 73}, 066104 (2006).}

\bibitem{dh}  {\small P. Debye, E. H\"uckel, Phys. Z. {\bf 24}, 305 (1923).}

\bibitem{rs}  {\small R. Robert, J. Sommeria, Phys. Rev. Lett. {\bf 69}, 2776 (1992).}

\bibitem{physicaD}  {\small P.H. Chavanis, Physica D  {\bf 200}, 257  (2005).}

\bibitem{burgers}  {\small J.M. Burgers, {\it The Nonlinear Diffusion Equation} (Riedel, Boston, 1974).}

\bibitem{stevens}  {\small A. Stevens, SIAM J. Appl. Math.  {\bf 61}, 183 (2000).}

\bibitem{ng}  {\small T.J. Newman, R. Grima, Phys. Rev. E  {\bf 70}, 051916 (2004).}

\bibitem{crrs}  {\small P.H. Chavanis, M. Ribot, C. Rosier, C. Sire, Banach Center Publ. {\bf 66}, 103 (2004)}

\bibitem{kinbio}  {\small  P.H. Chavanis, C. Sire, Physica A {\bf 384}, 199 (2007).}

\bibitem{grima}  {\small R. Grima, Curr. Topics. Dev. Bio  {\bf 81}, 435 (2008).}

\bibitem{super}  {\small P.H. Chavanis,  Physica A {\bf 359}, 177 (2006).}


\bibitem{pudritz}  {\small  D.E. McLaughlin, R.E. Pudritz, Astrophys. J. {\bf 476}, 750 (1997).}

\bibitem{logotrope}  {\small  P.H. Chavanis, C. Sire, Physica A {\bf 375}, 140 (2007).}

\bibitem{hp}  {\small T. Hillen, K. Painter, Adv. Appl. Math.  {\bf 26}, 280  (2001).}

\bibitem{degrad}  {\small P.H. Chavanis, Eur. Phys. J. B   {\bf 54}, 525  (2006).}

\bibitem{banach}  {\small P.H. Chavanis,  Banach Center Publ. {\bf 66}, 79 (2004).}

\bibitem{pt}  {\small J. Paret, P. Tabeling, Phys. Fluids    {\bf 10}, 3126  (1998).}

\bibitem{rw}  {\small M. Rivera, X.L. Wu, Phys. Rev. Lett.    {\bf 85}, 976  (2000).}

\bibitem{aa}  {\small P.H. Chavanis, A\&A    {\bf 356}, 1089  (2000).}	

\bibitem{gamba}  {\small A. Gamba, D. Ambrosi, A. Coniglio, A. de Candia, S. di Talia, E. Giraudo, G. Serini, L. Preziosi, F. Bussolino, Phys. Rev. Lett. {\bf 90}, 118101 (2003).}

\bibitem{filbet}  {\small F. Filbet, P. Lauren\c cot, B. Perthame, J. Math. Biol.   {\bf 50}, 189  (2005).}

\bibitem{cst}  {\small  P.H. Chavanis, C. Sire, Physica A {\bf 356}, 419 (2005).}

\bibitem{du}  {\small  Du Jiulin, Astrophys. Space Sci. {\bf 306}, 247 (2006).}

\bibitem{fermid}  {\small  P.H. Chavanis,  Phys. Rev. E {\bf 69}, 066126 (2004).}

\bibitem{bln}  {\small P. Biler, P. Lauren\c cot, T. Nadzieja, Adv. Differential Equations {\bf 9}, 563 (2004).}
 
\bibitem{ein10}  {\small A. Einstein, Ann. Physik  {\bf 33}, 1275 (1910).}
 
\bibitem{cohen}  {\small  E.G.D. Cohen, Physica A {\bf 305}, 19 (2002).}



\bibitem{dolbeault}  {\small J. Dolbeault, P. Markowich, D. \"Olz, C. Schmeiser, Arch. Rational Mech. Anal. {\bf 1}, 133 (2007).}

\bibitem{csjeans}  {\small  P.H. Chavanis, C. Sire, [arXiv:0708.3163] }

\bibitem{chemojeans}  {\small P.H. Chavanis,  Eur. Phys. J. B {\bf 52}, 433 (2006).}

\bibitem{bouchet}  {\small  F. Bouchet,  [arXiv:0710.5094]}

\bibitem{assise}  {\small  P.H. Chavanis, AIP Conf. Proc. {\bf 970}, 39 (2008).}

\bibitem{holm}  {\small  D.D. Holm, J.E. Marsden, T. Ratiu, A. Weinstein, Phys. Rep. {\bf 123}, 1 (1985).}

\bibitem{canto}  {\small P.H. Chavanis,  A\&A {\bf 451}, 109 (2006).}

\bibitem{eddig}  {\small A.S. Eddington,  MNRAS {\bf 76}, 572 (1916).}


\end{thebibliography}
\end{document}